\documentclass[trackchanges]{aastex701}

\usepackage{graphicx}
\usepackage{longtable}
\usepackage[dvipsnames]{xcolor}
\usepackage{gensymb}
\usepackage{mathtools}
\usepackage[version=4]{mhchem}
\usepackage{plimsoll}

\begin{document}

\title{Chemical Tracers for 3D Atmospheric Asymmetries on WASP-69~b}

\author[orcid=0009-0008-7545-5022,sname='Bangera']{Nidhi Bangera}
\affiliation{Space Research Institute, Austrian Academy of Sciences, Schmiedlstrasse 6, 8042 Graz, Austria}
\affiliation{Institute for Theoretical and Computational Physics, Graz University of Technology, Petersgasse 16, 8010 Graz, Austria}
\email[show]{nidhirohit.bangera@oeaw.ac.at}  

\author[orcid=0000-0001-9355-3752,sname='Carone']{Ludmila Carone} 
\affiliation{Space Research Institute, Austrian Academy of Sciences, Schmiedlstrasse 6, 8042 Graz, Austria}
\email{ludmila.carone@oeaw.ac.at}

\author[orcid=0000-0001-9273-9694,sname='Soni']{Vikas Soni}
\affiliation{Space Research Institute, Austrian Academy of Sciences, Schmiedlstrasse 6, 8042 Graz, Austria}
\email{vikas.soni@oeaw.ac.at}

\author[orcid=0000-0002-4258-6703,sname='Goodis Gordon']{Kenneth Goodis Gordon}
\affiliation{Space Research Institute, Austrian Academy of Sciences, Schmiedlstrasse 6, 8042 Graz, Austria}
\email{KennethEverett.GoodisGordon@oeaw.ac.at}

\author[orcid=0000-0003-4426-9530, sname='Fossati' ]{Luca Fossati}
\affiliation{Space Research Institute, Austrian Academy of Sciences, Schmiedlstrasse 6, 8042 Graz, Austria}
\email{luca.fossati@oeaw.ac.at}

\author[orcid=0000-0002-0469-5191,sname='Lecoq-Molinos']{Helena Lecoq-Molinos}
\affiliation{Space Research and Planetary Sciences, Physics Institute, University of Bern, Gesellschaftsstrasse 6, 3012 Bern, Switzerland}
\email{helena.lecoqmolinos@unibe.ch}

\author[orcid=0000-0002-7180-081X, sname='Rimmer' ]{Paul B. Rimmer}
\affiliation{ Cavendish Laboratory, University of Cambridge, United Kingdom}
\email{pbr27@cam.ac.uk}

\author[orcid=0000-0002-8900-3667,sname='Woitke']{Peter Woitke}
\affiliation{Space Research Institute, Austrian Academy of Sciences, Schmiedlstrasse 6, 8042 Graz, Austria}
\email{peter.woitke@oeaw.ac.at}

\author[orcid=0000-0002-8275-1371,sname='Helling']{Christiane Helling}
\affiliation{Space Research Institute, Austrian Academy of Sciences, Schmiedlstrasse 6, 8042 Graz, Austria}
\affiliation{Institute for Theoretical and Computational Physics, Graz University of Technology, Petersgasse 16, 8010 Graz, Austria}
\email{christiane.helling@oeaw.ac.at}

\begin{abstract}

Warm giant exoplanets exhibit strong three-dimensional temperature contrasts that can significantly alter atmospheric chemistry through quenching and photochemistry, yet transmission spectra are commonly interpreted using one-dimensional, limb-averaged models. Such simplifications may bias inferred atmospheric properties, particularly metallicity and C/O ratio. In this work, we investigate the relative influence of atmospheric composition and three-dimensional thermal structure on atmospheric chemistry and transmission spectra using WASP-69~b as a test case. WASP-69~b is a $\sim$900 K warm Saturn, residing in a thermal regime especially sensitive to disequilibrium chemistry. We use three-dimensional general circulation model derived pressure–temperature profiles as inputs for a one-dimensional photochemical-kinetics model to resolve longitudinal and latitudinal chemical asymmetries across the atmosphere. We find that \ce{CH4} exhibits strong latitudinal variations linked to deep quench temperatures, while \ce{SO2} shows longitudinal asymmetries driven by upper-atmospheric photochemistry and irradiation geometry. In contrast, \ce{CO2} remains comparatively insensitive to spatial thermal variations and emerges as a robust tracer of atmospheric metallicity. Synthetic transmission spectra reveal that three-dimensional chemical asymmetries can produce spectral variations comparable to those induced by metallicity itself, demonstrating that limb-averaged interpretations can mask substantial spatial structure in warm giant exoplanet atmospheres.

\end{abstract}

\keywords{\uat{Exoplanet atmospheres}{487} --- \uat{Planetary Atmospheres}{1244}}

\section{Introduction} 

Over the past decade, transmission and emission spectroscopy of hot Jupiters and warm Neptunes have revealed a diverse inventory of molecular species, including \ce{H2O}, CO, and \ce{CO2} in their atmospheres \citep[e.g.][]{Brogi2018,JWST2023,Schlawin2024}.
These observations have demonstrated that many irradiated gaseous planets possess atmospheric compositions that cannot be fully explained by thermochemical equilibrium alone. In particular, planets with equilibrium temperatures below 1000~K occupy a transitional regime where chemical timescales become comparable to atmospheric transport timescales, allowing disequilibrium processes to strongly influence the observable atmospheric composition \citep[e.g.][]{Visscher2011, Moses2011, Venot2014, Marley2015}. As a result, molecules such as \ce{CH4}, \ce{NH3}, \ce{HCN}, and \ce{C2H2} have emerged as key tracers of atmospheric kinetics and transport processes in hot hydrogen-dominated atmospheres. Recent observations with the James Webb Space Telescope (JWST) have expanded the range of detected disequilibrium species with the detections of photochemically produced \ce{SO2} \citep[e.g.][]{Tsai2023, Dyrek2024}, demonstrating that sulfur chemistry may also provide an important diagnostic of disequilibrium processes on irradiated exoplanets. 

WASP-69~b ($T_{\mathrm{eq}}\approx 900$~K, \citet{Anderson2014}) is a particularly compelling target for studying the effect of disequilibrium processes on the atmospheric composition. Its intermediate equilibrium temperature places it in a regime where \ce{CH4}–CO interconversion is highly sensitive to vertical mixing efficiency and temperature variations, making its carbon chemistry especially dynamic \citep{Bangera2025}. Observations of \ce{CH4}, \ce{C2H2}, and HCN have suggested that WASP-69~b may possess an unusually carbon-rich atmosphere \citep{Guilluy2022}; however, these interpretations remain uncertain because the inferred molecular abundances depend sensitively on the assumed atmospheric temperature structure. In particular, in \citet{Bangera2025} we demonstrated that enhanced \ce{C2H2} concentrations can arise from a high C/O ratio or from photochemical processes under different thermal conditions. As a consequence, it remains unclear whether the observed molecular signatures primarily reflect the atmospheric composition or the effects of disequilibrium chemistry in a spatially varying atmosphere.

Most atmospheric retrievals and chemistry models used to interpret transmission spectra assume one-dimensional (1D), chemically homogeneous atmospheres \citep[see e.g.,][]{barstow2020}. While such approaches capture the dominant spectral features, they neglect the inherently three-dimensional (3D) nature of irradiated giant exoplanets. Tidally locked hot and warm Jupiters exhibit strong latitudinal and longitudinal temperature gradients driven by permanent day-night irradiation contrasts and global atmospheric circulation \citep[e.g.,][]{Showman2009, 2016MNRAS.460..855H, Parmentier2021, 2023A&A...671A.122H}. These circulation patterns produce equatorial jets, hotspot offsets, and high-latitude Rossby gyres, resulting in temperature variations of several hundred kelvin throughout the observable atmosphere. Since chemical kinetics depend sensitively on local gas temperature, these thermal gradients naturally give rise to three-dimensional chemical asymmetries. Thermal asymmetries are particularly important for disequilibrium chemistry, since the relevant chemical timescales are often comparable to, or longer than, atmospheric transport timescales \citep{Baeyens2021}.

The impact of 3D atmospheric structure is particularly relevant for transmission spectroscopy. During transit, observations probe the planetary limb, where the measured signal integrates over regions with potentially very different thermal and chemical properties. As a result, the retrieved atmospheric properties represent a weighted average of these diverse regions rather than a single atmospheric state. Previous studies have investigated atmospheric asymmetries at the planetary limb, particularly in the context of inhomogeneous cloud coverage \citep[e.g.,][]{Kempton2017,Powell2019} and spatially varying equilibrium chemistry \citep[e.g.,][]{Fortney2010}. 
In particular, \cite{Carone2025} demonstrated that latitudinal variations across the terminator can produce observable asymmetries in transit spectra, highlighting that transmission spectroscopy probes a three-dimensional atmospheric structure rather than a single averaged limb. More recently, asymmetries have begun to be directly inferred from transmission spectra of hot and ultra-hot Jupiters \citep[e.g.,][]{Eherenreich2020,Prinoth2022,Espinoza2024}.

This raises the question: which observable molecules trace the atmospheric composition, and which instead trace the 3D thermal structure and atmospheric dynamics? Species such as \ce{CH4}, \ce{CO}, \ce{NH3}, and \ce{N2} are largely controlled by quench conditions in the deep atmosphere and are therefore sensitive to vertical mixing \citep[e.g.][]{Moses2011}. In contrast, photochemically produced species such as \ce{SO2}, \ce{C2H2} and HCN are formed in the upper atmosphere and depend strongly on local irradiation and temperature conditions \citep[e.g.][]{Hobbs2021, Tsai2023}. These species therefore probe distinct atmospheric layers and may respond differently to spatial temperature asymmetries. 

In this work, we investigate how 3D temperature structure shapes disequilibrium chemistry in the atmosphere of WASP-69~b. Building on the 1D disequilibrium chemistry study in \cite{Bangera2025}, we extend this analysis into three-dimensions by using pressure-temperature profiles extracted from the \texttt{ExoRad} general circulation model (GCM) as inputs for the ARGO photochemical-kinetics framework. In addition to the previously studied carbon- and nitrogen-bearing species, we include a sulfur chemistry investigation to explore the formation and spatial distribution of \ce{SO2}. We examine a range of metallicities and C/O ratios motivated by observational constrains from emission spectroscopy \citep{Schlawin2024}, while using a baseline model with 10 $\times$ solar metallicity and C/O = 0.55 to isolate the effects of atmospheric temperature asymmetries. 

Specifically, we address the following questions:
\begin{itemize}
    \item What molecules exhibit the largest variation in abundance within the possible atmospheric metallicity and carbon-to-oxygen (C/O) ratios for WASP-69~b, as constrained by its emission spectra?
    \item How do longitudinal and latitudinal temperature asymmetries modify the global distribution of disequilibrium species?
    \item Which observable molecules are primarily controlled by metallicity, and which are dominated by three-dimensional atmospheric structure and are thus signatures of 3D dynamics and disequilibrium processes?
\end{itemize}

This paper proceeds as follows: Section \ref{sec: Approach} describes the modeling framework, including the \texttt{ExoRad} GCM-derived P–T structure and the \textsc{ARGO} photochemical–kinetics model. Section \ref{sec: Results} presents the atmospheric composition of WASP-69~b across a range of metallicities and C/O ratios. Section \ref{sec: Asymmetry} then focuses on the baseline model to quantify the impact of spatial thermal asymmetries on atmospheric chemistry, while Section \ref{sec: Discussion} discusses the observational implications and future directions. Finally, Section \ref{sec: Conclusions} summarizes the main findings.  

\section{Approach} \label{sec: Approach}

A combined framework consisting of a 3D GCM and a 1D photochemical-kinetics model is employed to explore chemistry in the atmosphere of WASP-69~b. The framework is applied to WASP-69~b ($T_\mathrm{eq} \approx$ 900~K), an inflated warm Saturn orbiting a K-type host star \citep{Faedi2011,Anderson2014}. The input parameters adopted in this study are listed in Table \ref{Table:Input}. 

   \begin{table}[ht!]
      \caption{Input parameters for WASP-69~b \textsc{Argo} simulations.
      }
      \centering 
      \def\arraystretch{1.3}
         \label{Table:Input}
        \begin{tabular}{|c|c|}
             \hline
            \textbf{Parameter}  & \textbf{Value} \\
           \hline
           \textbf{WASP-69~b} & \\
           \hline
           Surface gravity g (m s$^{-2}$) & 5.73\\
           Radius [R$_{\rm{Jup}}$] & 1.06\\
           Mass [M$_{\rm{Jup}}$] & 0.26\\
           T$_{\rm{global}}$ & 900~K\\
           p$_{\rm{gas}}$-T$_{\rm{gas}}$ profile & This work\\
           Metallicity & 6, 10, 14 $\times$ Solar\\
           C/O & 0.55, 0.64, 0.95\\
           Host star & K5\\
           Incident stellar spectrum & From \cite{Bangera2025}\\
           Kzz profile & From \cite{Moses2022}\\
           Gas-pressure range & 1.3$\times 10^{-8}$ to 650 bar\\
           Vertical levels & 67\\
            
            \hline 
            
         \end{tabular}

        \tablecomments{Mass and radius estimate for WASP-69~b are taken from \cite{Anderson2014}.}  
         
   \end{table}
   

\subsection{3D Atmosphere Modeling}
The 3D thermodynamics of WASP-69~b are provided for different metallicities and C/O ratios using the planetary parameters as listed in Table~\ref{Table:Input} via the 3D General Circulation Model (GCM) framework \texttt{ExoRad} \citep{Carone2020,Baeyens2021,Schneider2022a}. Four representative cases are considered, spanning metallicities of 6×, 10×, and 14× solar and C/O ratios of 0.55, 0.64, and 0.95 for WASP-69~b, in line with the spectroscopic constraints by \citet{Schlawin2024}.

The 3D GCM \texttt{ExoRad} uses the hydrodynamical core \texttt{MITgcm} \citep{Adcroft2004} to solve the Navier-Stokes equations on a rotating sphere. The dynamical core uses the finite volume method for spatial discretization and the second order Adams-Bashforth method for explicit time stepping in solving the HPE on a staggered Arakawa C grid \citep{ArakawaLamb1977,Showman2009}. The core is complemented with physical parametrizations suitable for externally irradiated tidally locked gas giants of a wide range of global temperatures $T_{\rm global}=600$~K$,\ldots 2600$~K \citep{Plaschzug2025}. The latest implementation includes full radiative transfer  (\texttt{expeRT/MITgcm}, \citealt{Schneider2022b,Schneider2022a}).

\texttt{ExoRad} solves the hydrostatic primitive equations (HPE, \citealt{Showman2009}) that assume a hydrostatic atmosphere and the ideal gas law in a C32 cubed-sphere grid, corresponding to a horizontal resolution of 128 $\times$ 64 in longitude ($\phi$) and latitude ($\theta$) ($2.8^\circ\times 2.8^\circ$). As the vertical coordinate, the local gas pressure is adopted, $p_\mathrm{gas}$~[Pa], which covers  $10^{-5}$ bar to 700 bar. A logarithmic spacing with 41 grid cells is applied between $10^{-5}$ bar and 100 bar to resolve the radiative and dynamically active regions, while linear spacing with six grid cells is used between 100 bar and 700 bar with a step size of 100 bar. Thus, a total of 47 vertical layers is implemented. In this work,  the standard dynamical timestep of $\Delta t =25~$s is adopted.  Each simulation is run for $1000\,{\rm days}$ and time averaged over the last 100 days.  

To ensure numerical stability against unphysical gravity wave reflection in the upper atmosphere, a sponge layer is applied between $10^{-4}-10^{-5}$ bar, in which the zonal horizontal velocity $u$ is damped by a Rayleigh friction
term towards its longitudinal mean $\overline{u}$ \citep{Carone2020}. To avoid shear flow instability at the bottom, basal drag \citep{LiuShowman2013ApJ} is applied between 500 and 700~bar \citep{Carone2020}.

In this work, the following gas-phase opacity species are considered: H$_2$O (from ExoMol\footnote{\url{https://www.exomol.com}} -- \citealt{Tennyson2016_ExoMol, TennysonEtal2020jqsrtExomol2020}), Na \citep{Allard19_Na_K}, K \citep{Allard19_Na_K}, the latter two including pressure broadening, CO$_2$, CH$_4$, NH$_3$, CO, H$_2$S, HCN, SiO, PH$_3$ and FeH as listed in \citet[][Table. 1]{Schneider2022a}. In addition, collision-induced absorption for H$_2$--H$_2$ \citep{BorysowEtal2001jqsrtH2H2highT, Borysow2002jqsrtH2H2lowT, Richard2012} and H$_2$--He \citep{BorysowEtal1988apjH2HeRT, BorysowFrommhold1989apjH2HeOvertones, BorysowEtal1989apjH2HeRVRT} and Rayleigh scattering for H$_2$ \citep{Dalgarno1962}, and He \citep{Chan1965} are considered. The gas opacities are binned and tabulated via the correlated-k-method \citep{Goody1989} into 11 spectral bins, spanning 0.26 -300~$\mu$m \citep{Schneider2022a}. The gas is assumed to be in local thermal equilibrium (LTE), that is, the number densities of the opacity species are calculated in chemical equilibrium for the specified metallicities and C/O ratios (Table.~\ref{Table:Input}). An interior temperature of 195~K is assumed based on the parametric fit of \cite{Thorngren2019}.

\subsection{Postprocessing 3D Thermodynamics}
Four characteristic points from the GCM grid are used to represent planet-wide features: the substellar point, evening terminator, morning terminator and antistellar point, at (latitude $\theta$,longitude $\phi$) coordinates of (0,0), (0,90), (0,-90), and (0,-180) respectively.

\begin{figure}[ht!]
\centering
\includegraphics[width = 0.5\linewidth]{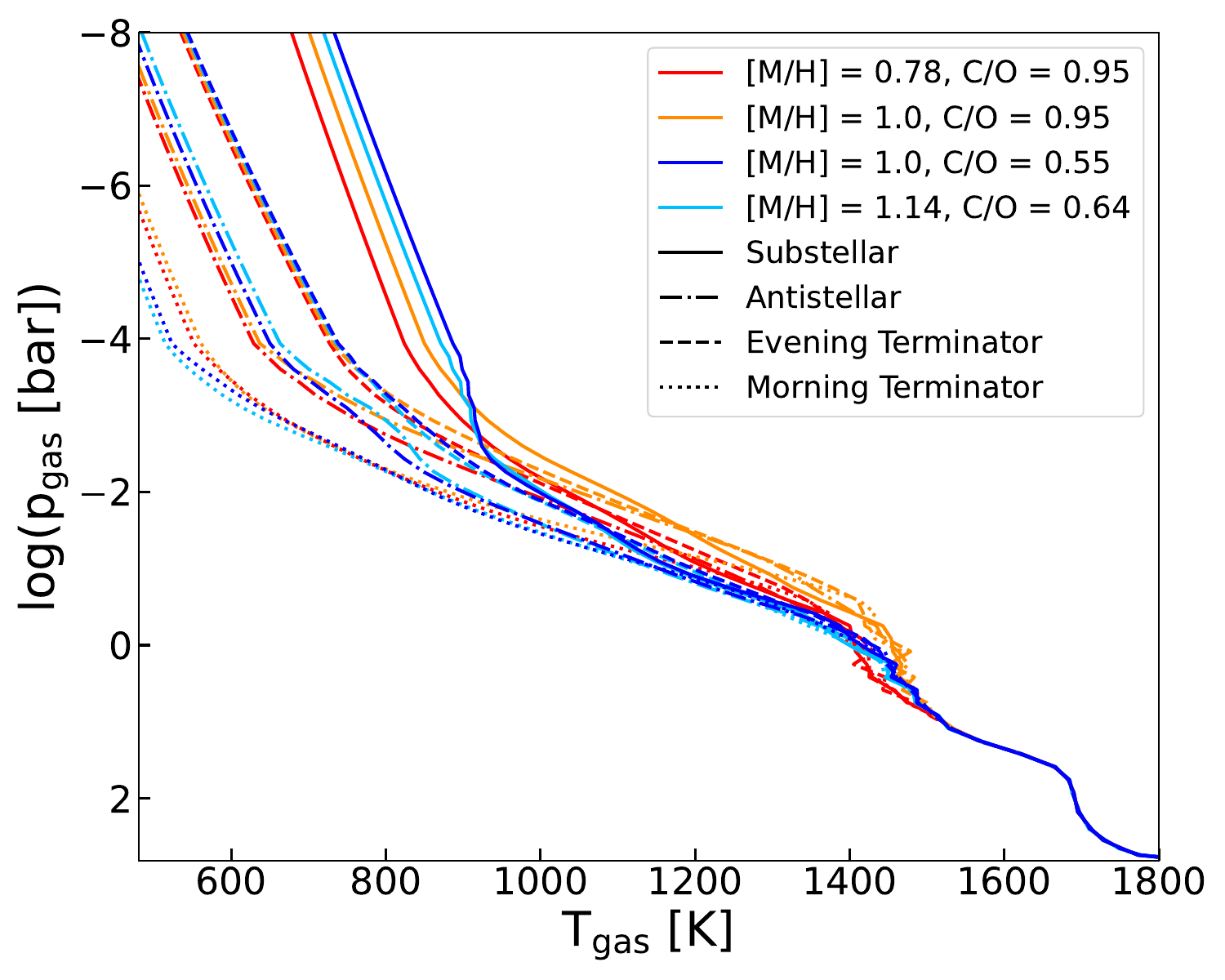}
\caption{1D ($\rm{T_{gas}}$-$\rm{p_{gas}}$) profiles for different [M/H] and C/O combinations for WASP-69~b. }
\label{GCM WASP69 b}
\end{figure}

To probe the upper photochemically active pressure layers, each GCM profile is extended from its top boundary of 10$^{-4}$ bar to 10$^{-8}$ bar using a generalized exponential extrapolation \citep{Madhusudhan2009}:

\begin{equation}
    \rm{p_{gas}} = p_0\times \rm{exp}(\alpha(T_{gas}-T_0)^\beta)
\end{equation}
where, $p_{\rm{gas}}$ is the pressure in bars, $T_{\rm{gas}}$ is the temperature in K, and $p_0$ and $T_0$ are the pressure and temperature at the top of the atmosphere, and $\alpha$ and $\beta$ are free parameters. Throughout this work, we adopt $p_0 = 10^{-8}$ bars, $T_0 = T(p_{\rm{gas}} = 10^{-4} ~\mathrm{bar}) - 200~\mathrm{K}$, $\alpha = 0.65$ and $\beta = 0.5$ such that the temperature decreases with height. The resulting  (${T_{\rm{gas}}}-{p_{\rm{gas}}}$) profiles for WASP-69~b are plotted in Figure \ref{GCM WASP69 b}. In the pressure range -4$\leq$ log$_{10}$ (${p_{gas}}$ [bar]) $\leq$1, a difference in C/O causes a difference of approximately 100~K in temperature. The effect of differing metallicity on the local gas temperature is smaller. 

\subsection{Atmospheric Chemistry} \label{sec:ARGO}

The 1D kinetic simulations are performed using the \textsc{ARGO} code \citep{Rimmer&Helling,ErratumRimmer_2019}, which solves the coupled continuity equations for all included chemical species under stellar insolation and vertical diffusion. The chemical kinetics are based on the STAND network \citep{Hobbs2021}, an H/C/N/O/S chemical network containing over 6600 reactions valid across temperatures from 100--30000~K. STAND2020 includes reactions involving species containing up to six H, two C, two N, three O, and two S atoms. This network has been extended to include new ion-neutral, photochemical and thermal ionization reactions for sulfur bearing species, hereafter referred to as STAND2025 (Bangera et al., in prep). Further details of the disequilibrium chemistry modeling with \textsc{ARGO} are provided in Appendix~\ref{appendix: argo}. Although the chemical network includes ion-neutral chemistry and thermal ionization processes, we find that ion chemistry does not significantly affect the dominant observable neutral species in the atmosphere of WASP-69~b under the temperature regime explored here. We therefore focus this work on the neutral disequilibrium chemistry relevant for transmission spectroscopy. Representative ion abundance profiles are included in Appendix~\ref{appendix: molecules} for completeness. A more detailed investigation of ion chemistry in gaseous exoplanet atmospheres will be presented in forthcoming work.

In Section~\ref{sec: Results}, a comparison is made between the atmospheric composition in chemical equilibrium and disequilibrium. The equilibrium abundances here are calculated using \texttt{GGchem} \citep{Woitke2018}.

Four combinations of metallicity and C/O ratios are modeled for the atmosphere of WASP-69~b under disequilibrium conditions:

\begin{enumerate}
    \item [$\blacksquare$][M/H] = 0.78, C/O = 0.95
    \item [$\blacksquare$][M/H] = 1.0, C/O = 0.95
    \item [$\blacksquare$][M/H] = 1.0, C/O = 0.55
    \item [$\blacksquare$][M/H] = 1.14, C/O = 0.64
\end{enumerate}
   
\noindent The low ([M/H] = 0.78) and high ([M/H] = 1.14) metallicity cases are motivated by constraints from emission spectra \citep{Schlawin2024}, while the [M/H] = 1.0  cases represent intermediate scenarios. The model with [M/H] = 1.0 and C/O = 0.55 serves as our baseline. Each model of WASP-69~b is initialized with elemental concentrations at the lower boundary of $p_{\rm{gas}}$ = 650~bar. To vary the C/O ratio for a given metallicity, the C and O concentrations are scaled inversely such that the total C+O remains unaltered. 

For the baseline model, we further extract pressure-temperature profiles for a grid of latitude and longitude points to capture the thermal asymmetries relevant to transmission spectroscopy. Since the thermal structure is approximately symmetric about the equator, only positive latitudes are modeled and are assumed to be representative of their corresponding negative latitudes.

In Section \ref{sec: Asymmetry}, we evaluate the cumulative effect of the individual 1D atmospheric columns by constructing global maps of molecular abundances. These maps are of two types. The first is a vertically integrated column density ($N_i(\theta,\phi)$ (in molecules/cm$^2$)) map, obtained by integrating the concentration at a given latitude $\theta$ and longitude $\phi$, from the top of the modeled atmosphere ($p_{\rm{gas}}$ = 10$^{-8}$~bar) down to pressure levels where the atmosphere becomes optically thick at visible and infra-red wavelengths (here $\sim 10^{-3}$~bar):

\begin{equation}
    N_i(\theta,\phi) = \frac{1}{\mu m_H g}\int_{p_{top}}^{p_{max}}\frac{n_i(\theta,\phi,p_{gas})}{n_{tot}(\theta,\phi,p_{gas})}~dp_{gas}. 
\end{equation}

The second is a line-of-sight column density map, defined by integrating the number density along the slant path through the atmosphere during transit, as viewed from the host star:

\begin{equation}
    N_{i,los}(\Omega) = \int_{a}^{b}n_i(p_{gas},\theta,\phi)~ds.
\end{equation}

\noindent $N_{i,los}(\Omega)$ [molecules/cm$^2$] is the line-of-sight column density as a function of solid angle ($\Omega$). $a$ and $b$ are the intersection points of the star light path in the upper boundary of the atmosphere, and $ds$ is the length fraction in the direction of star light.

Together, these provide a link between the underlying three-dimensional chemical structure and the quantities probed by transmission spectroscopy.

We further identify the quench level for species affected by vertical mixing. In this work, the quench level of a species is defined as the first departure from chemical equilibrium in the deep atmosphere, corresponding to the pressure at which the absolute relative difference between the equilibrium and disequilibrium concentrations of the species first exceeds $10^{-3}$.

\subsection{Transmission Spectra}

Synthetic transmission spectra are calculated for the visible to mid-infrared wavelengths (0.4 – 12 $\mu$m) using \texttt{petitRADTRANS}, following the formalism as outlined in \citet[][]{molliere2019}. Here, the transit depth, $T_{Depth}$($\lambda$) ($=$ $R_{p}$($\lambda$)$^2$ / $R_{s}$($\lambda$)$^2$), is derived by calculating the optical depth of a ray of light grazing the atmosphere above the planetary center through a series of concentric spheres with $r_1$ to $r_N$, where $r_1$ is the radius of the outermost and $r_N$ the radius of the innermost sphere. The innermost sphere is assumed to be optically thick at all wavelengths \citep[see, e.g.,][for more details]{Carone2025}. For our models, this innermost sphere corresponds to a pressure $p_{bottom}$ = 700 bar and the outermost sphere corresponds to a pressure $p$ = $10^{-8}$ bar. We assume the planet is cloud-free and only consider opacities from the gas-phase species described above, including the collision-induced absorptions and Rayleigh scattering. This assumption allows us to isolate the influence of three-dimensional thermal structure and disequilibrium chemistry on the resulting transmission spectra without introducing additional degeneracies associated with cloud composition, particle size distributions, and spatially varying cloud coverage. Additionally, it is assumed that the planet passes through the center of the stellar disk during the transit, and we neglect any effects from stellar limb darkening and the so-called stellar light source effect \citep[see, e.g.,][]{rackham2018}. The planetary parameters used for our transmission spectra calculations can be found in Table \ref{Table:Input}.

The transmission spectra are calculated at the morning and evening terminators for individual planetary latitudes $\theta$ as well as for the full morning and evening limbs. In the latter case, the transmission spectra are averaged over all $\theta$ for each limb assuming the planet is mirror-symmetric about the equator. Synthetic transmission spectra are presented in Sections \ref{sec: Results} and \ref{sec: Asymmetry} and are all generated at a spectral resolution of $R \approx 100$.

\section{Sensitivity of Gas-Phase Chemistry to Metallicity, C/O, and Thermal Structure} \label{sec: Results}

In this section, we compare equilibrium and disequilibrium abundances for the dominant C-, N- , and S- bearing species in our base model ([M/H] = 1.0, C/O = 0.55), and then examine how metallicity ([M/H]), carbon-to-oxygen ratio (C/O), and longitudinal temperature structure influence the gas-phase chemistry of WASP-69~b. Because the underlying pressure–temperature (P–T) profiles vary between GCM models due to variations in [M/H] and C/O, changes in composition and thermal structure are intrinsically linked. Finally, we assess how these chemical differences manifest in synthetic transmission spectra across the 1-12$\mu$m wavelength range and identify the molecular features most sensitive to metallicity and C/O.

\begin{figure}
    \centering
    \includegraphics[width=0.6\linewidth]{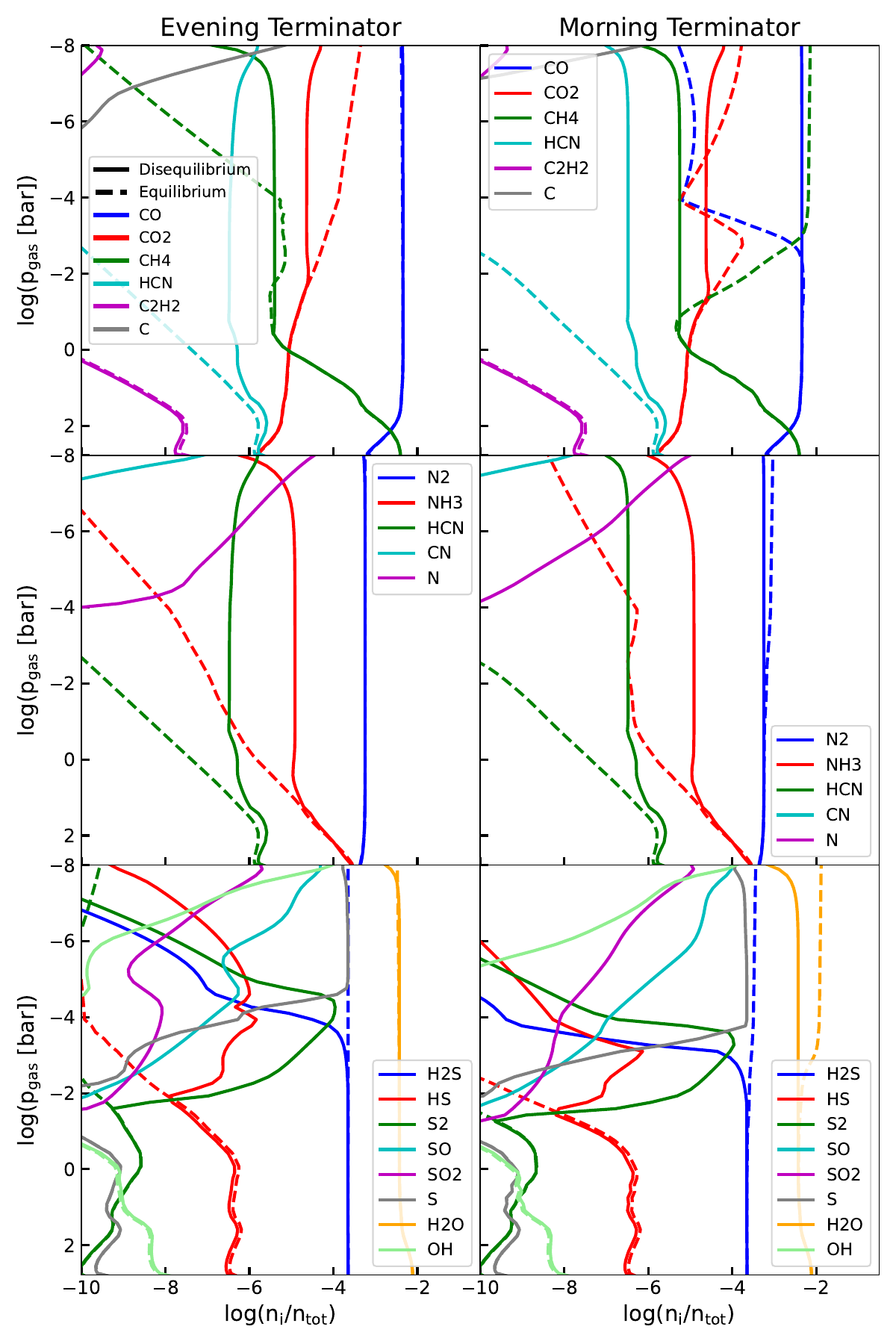}
    \caption{Concentrations of major carbon-bearing (top), nitrogen-bearing (middle), and sulfur- and oxygen-bearing (bottom) molecules at the evening (left) and morning (right) terminators of WASP-69~b, under [M/H] = 1.0, C/O = 0.55 assumptions. Dashed lines show thermochemical equilibrium concentrations, while solid lines show disequilibrium concentrations.}
    \label{fig:Eq vs diseq}
\end{figure}

Fig~\ref{fig:Eq vs diseq} compares the equilibrium vs disequilibrium chemical concentrations at the morning and evening terminators of WASP-69~b. The dominant carbon bearing species are \ce{CO}, \ce{CO2}, \ce{CH4} and HCN. Under chemical equilibrium, \ce{CH4} becomes the primary carbon carrier at low pressures on the cooler morning terminator. In disequilibrium, however, vertical quenching maintains similar \ce{CH4} concentrations at both terminators, leaving \ce{CO} as the dominant carbon reservoir throughout most of the atmosphere. 

Nitrogen chemistry is similarly affected by quenching. Although \ce{N2} remains the dominant nitrogen carrier, \ce{NH3} and \ce{HCN} are enhanced relative to their equilibrium predictions. In sulfur chemistry, \ce{H2S} dominates at depth, while photochemistry converts it into \ce{S}, \ce{S2}, \ce{SO} and \ce{SO2} in the upper atmosphere. We now examine how these disequilibrium species respond to variations in metallicity, C/O ratio, and thermal structure.  

\begin{figure}
    \centering
    \includegraphics[width=0.9\linewidth]{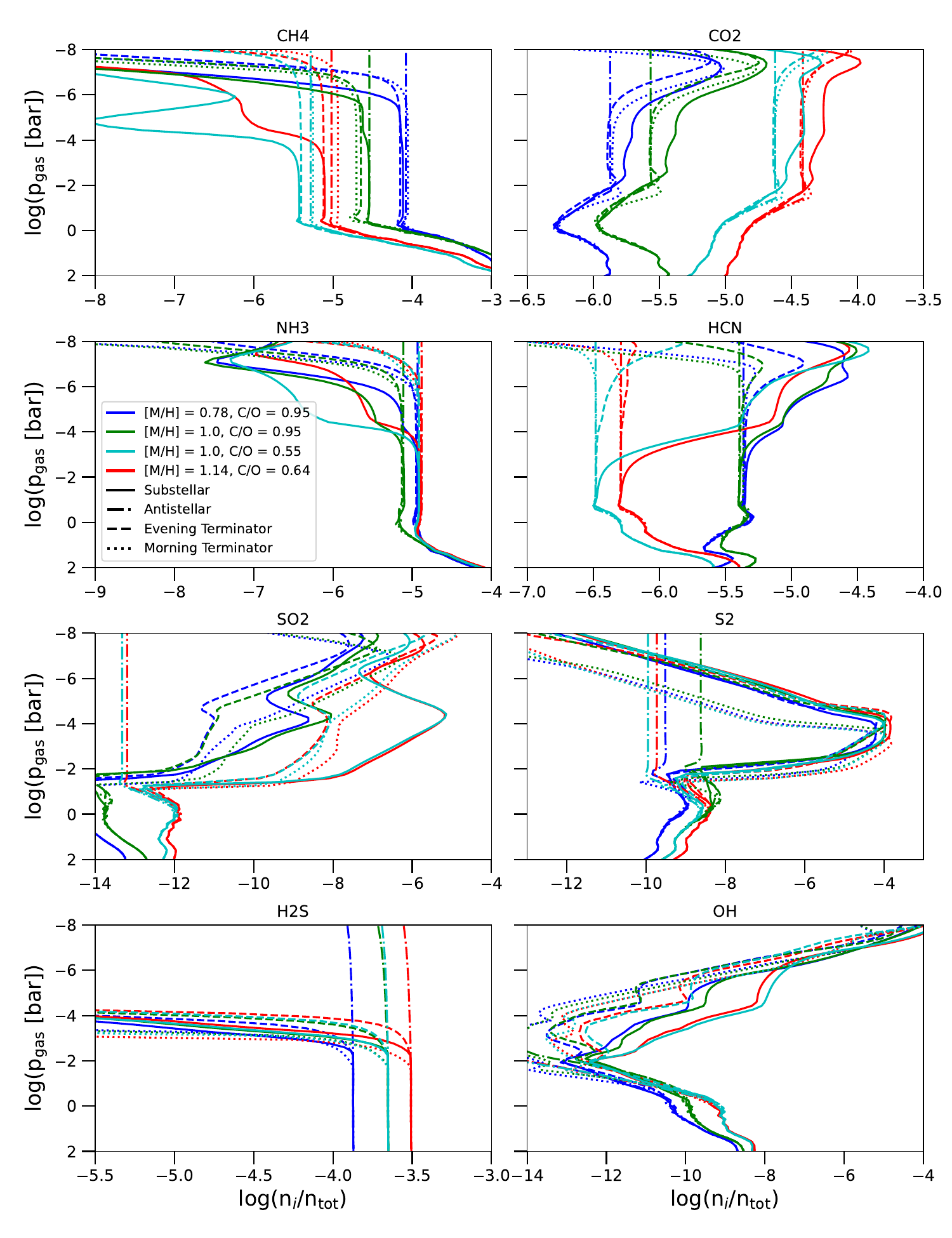}
    \caption{Disequilibrium concentrations of CH$_4$, CO$_2$, NH$_3$,  HCN, SO$_2$, S$_2$, \ce{H2S} and OH in the atmosphere of WASP-69~b for different [M/H] and C/O. Variations are most pronounced in the intermediate ($p_{\rm gas}\sim10^{-4}$~bar) and upper atmospheric regions ($p_{\rm gas}\sim10^{-8}$~bar), highlighting the influence of irradiation and vertical mixing on the atmospheric chemistry.}
    \label{fig:WASP-69~b}
\end{figure}

Figure~\ref{fig:WASP-69~b} shows the dependence of several disequilibrium species on metallicity, C/O ratio, and thermal structure.

Methane (\ce{CH4}) is primarily controlled by the C/O ratio and the local temperature at the quench point. The [M/H] = 0.78, C/O = 0.95 model produces the highest \ce{CH4} abundances, owing to both its enhanced carbon composition and relatively low quench temperature (see Fig.~\ref{GCM WASP69 b} near $p_{\rm gas}\sim0.1$~bar). In contrast, the [M/H] = 1.0, C/O = 0.95 model exhibits lower \ce{CH4} abundances despite having the same C/O ratio, due to its higher quench temperature. Further reductions in C/O lead to progressively lower methane abundances. Although \ce{CH4} undergoes photodissociation on the dayside, it displays only weak longitudinal variations because the quench temperatures remain similar within a given [M/H] and C/O scenario.

Carbon dioxide (\ce{CO2}) is sensitive to both metallicity and C/O ratio. While photochemistry enhances \ce{CO2} production on the dayside, its abundance remains relatively uniform with longitude compared to other species.

Ammonia (\ce{NH3}) is strongly controlled by the quench-point temperature. The [M/H] = 1.0, C/O = 0.95 model, which has the highest temperatures near $p_{\rm gas}\sim1$~bar, exhibits the lowest \ce{NH3} abundances, indicating that thermal structure dominates over composition in regulating ammonia chemistry.

Hydrogen cyanide (HCN) forms primarily through pathways involving \ce{CH4} and \ce{NH3} \citep{Moses2011}, and therefore increases with increasing C/O ratio. Its abundance is only weakly dependent on metallicity. HCN is moderately enhanced by photochemistry on the dayside but otherwise shows limited longitudinal variation.

Sulfur dioxide (\ce{SO2}) exhibits strong sensitivity to metallicity, C/O ratio, irradiation, and local temperature. Previous studies have shown that \ce{SO2} production peaks over a limited temperature range \citep{Hobbs2021, Mukherjee2025}. In our models, the morning terminator lies closer to this optimal temperature regime than the hotter evening terminator, resulting in locally enhanced \ce{SO2} concentrations. We emphasize, however, that this behavior applies to the local concentration only. As discussed in Sect. \ref{sec: Asymmetry}, the integrated line-of-sight column density of \ce{SO2} is instead larger at the evening terminator owing to the slant viewing geometry, and an eastward-shifted dayside. Since \ce{SO2} formation is favored in oxygen-rich environments, it is also sensitive to C/O. At pressures near $\sim0.01$~bar, the dominant production pathways are comparable to those identified by \citet{Hobbs2021} and \citet{Tsai2023}:

\begin{equation}
\begin{aligned}
        \ce{H2O &\xrightarrow{h\nu} OH + H}\\
        \ce{H2O + H &-> OH + H2}\\
        \ce{H2S + H &-> HS + H2}\\
        \ce{HS  + H &-> S + H2}\\
        \ce{S + OH  &-> SO + H}\\
        \ce{SO + OH  &-> SO2 + H}\\
        \hline
        {\rm Net:~} \ce{H2S + 2H2O &-> SO2 + 3H2}\\
        \hline
\end{aligned}
\end{equation}

We also find a second peak in \ce{SO2} concentrations at higher altitudes ($\sim$10$^{-7}$ bar) that is attributed to the higher concentrations of \ce{OH} at these altitudes.

Sulfur allotropes (\ce{S2} and higher) and \ce{H2S} show comparatively weak sensitivity to both [M/H] and C/O. Instead, elevated temperatures favour the photochemical conversion of \ce{H2S} into \ce{S2}, leading to higher \ce{S2} concentrations at the evening terminator as compared to the morning terminator.

Overall, across the modeled parameter space for WASP-69~b, \ce{SO2} emerges as the species most sensitive to metallicity, C/O ratio, and the underlying P–T structure. Meanwhile, \ce{CH4}, \ce{NH3}, and HCN are sensitive to the local gas temperature at the quench point and \ce{CO2} is primarily sensitive to [M/H] and C/O.

\begin{figure}
    \centering
    \includegraphics[width=0.7\linewidth]{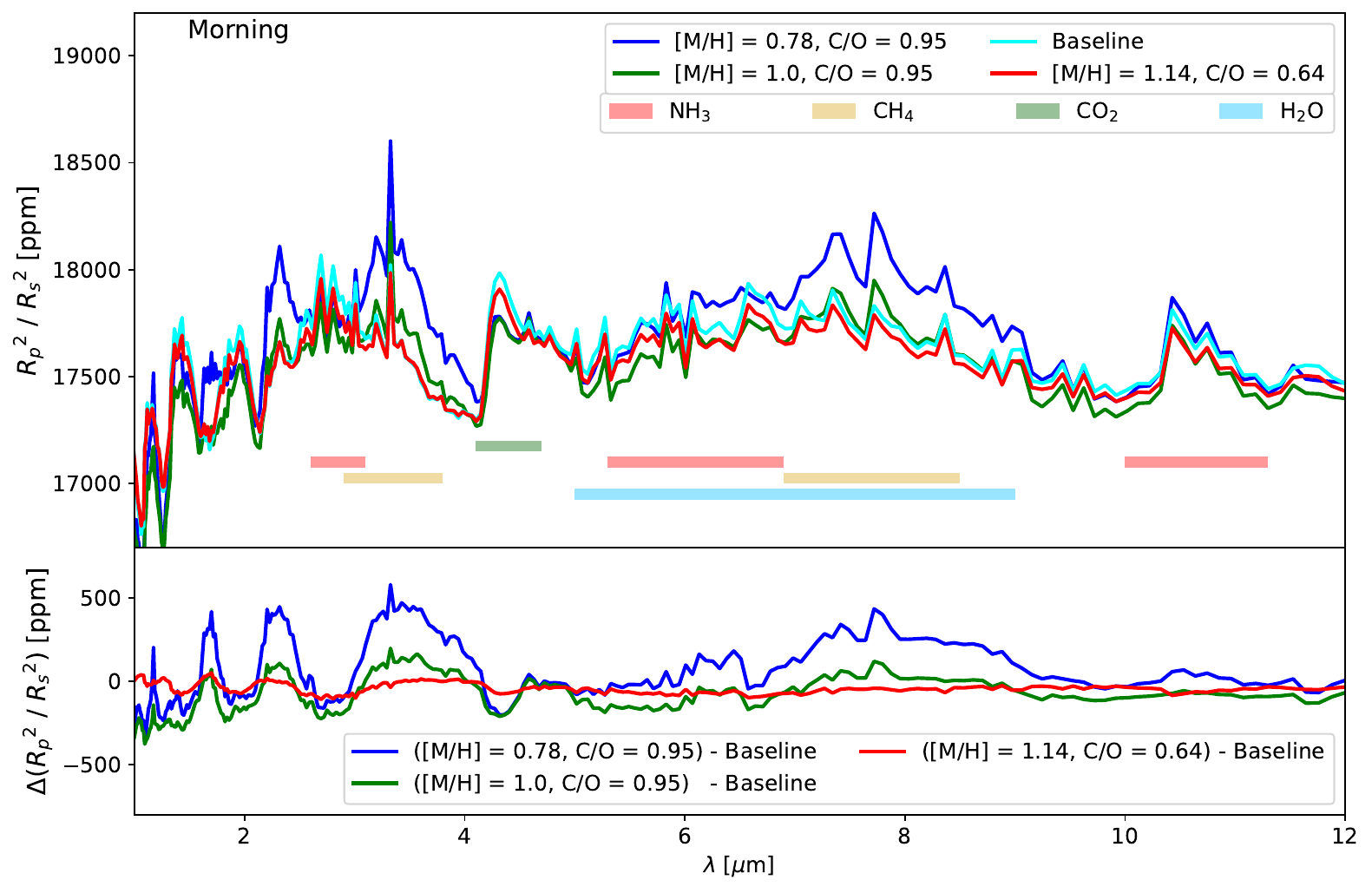}
    \includegraphics[width=0.7\linewidth]{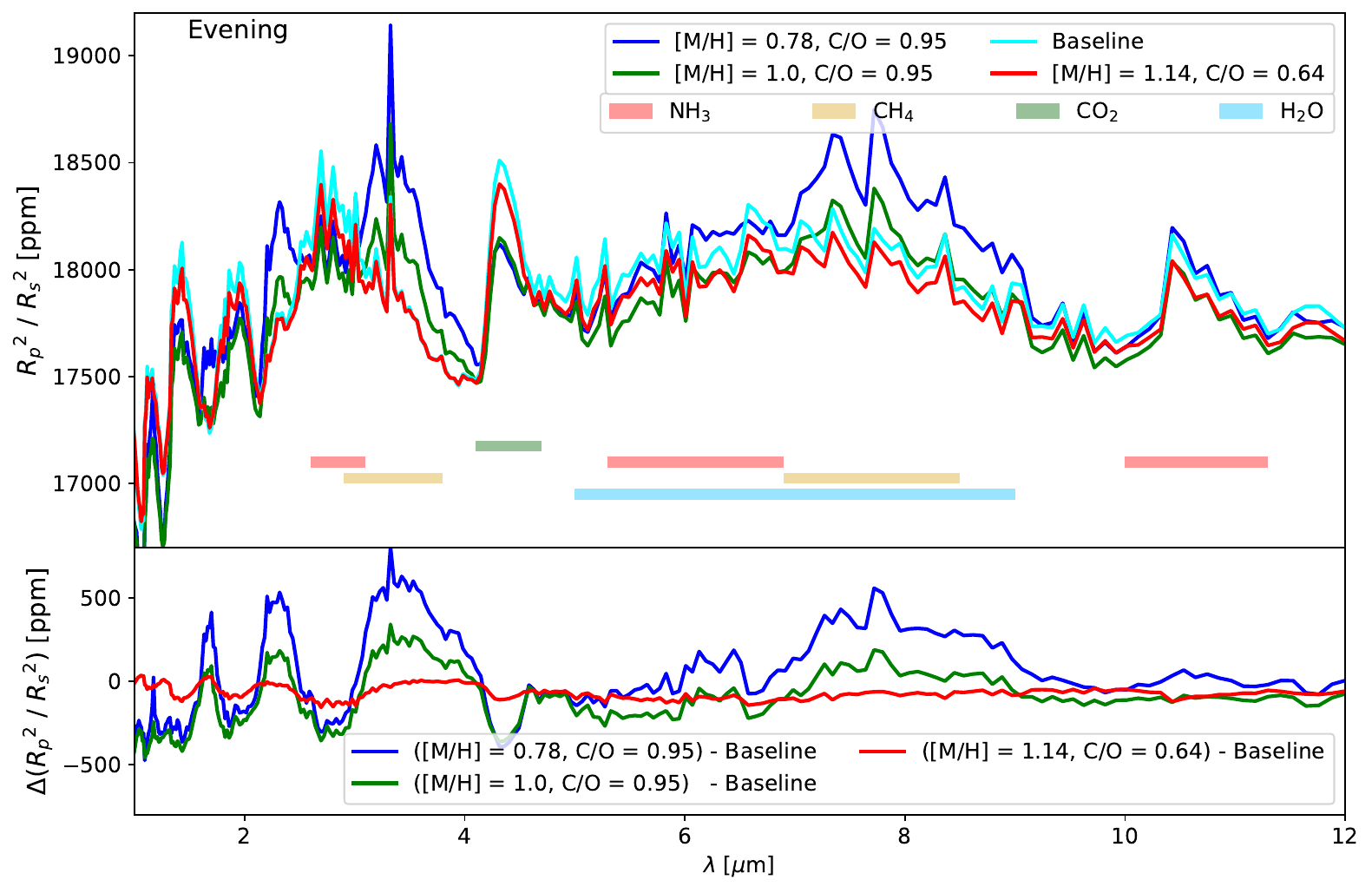}
    \caption{Synthetic transmission spectra for all four [M/H] and C/O combinations at the morning (top) and evening (bottom) terminators at $\theta$ = 0$\degree$. Here, the baseline model is taken as the [M/H] = 1.0 and C/O = 0.55 model. Residual spectra relative to this baseline model (individual combination spectra minus baseline spectrum) are shown below each panel. The molecular contributions at different wavelengths are shaded in different colors for various atmospheric species.}
    \label{fig:metallicityspectra}
\end{figure}

To assess observational implications, we generated synthetic transmission spectra at latitude $\theta$ = 0$\degree$ for all four [M/H] and C/O combinations across the wavelength ranges accessible to HST and JWST (1 - 12 $\mu$m; Fig. \ref{fig:metallicityspectra}). 
The different atmospheric scenarios are distinguished primarily by variations in the \ce{CH4} absorption bands near 3.3 and 7.7~$\mu$m, as well as the \ce{CO2} feature near 4.3~$\mu$m.
The largest variations occur in the 3.3 $\mu$m \ce{CH4} feature, where the most metal-poor model ([M/H] = 0.78, C/O = 0.95) differs by up to $\sim$850 ppm at the evening terminator, compared to $\sim$650 ppm at the morning. For the remaining metallicity cases, the variations are more modest, remaining near $\sim$200 ppm at both limbs. The \ce{CO2} feature near 4.3~$\mu$m also exhibits morning–evening differences, with metallicity-driven variations of $\sim$200 ppm at the morning terminator and up to $\sim$400 ppm at the evening terminator.
In both cases, the molecular features are stronger on the warmer evening limb, indicating that longitudinal thermal structure significantly affects the observable spectral signatures of these species.

These results suggest that the inferred sensitivity of molecular features to metallicity and C/O ratio depends strongly on the local thermal structure sampled during transit. Thus, in the next sections we address whether the spectral variations introduced by the metallicity uncertainties are larger or smaller than the variations caused by longitudinal/latitudinal thermal asymmetries.

\section{Spatial asymmetries in the chemistry on WASP-69~b} \label{sec: Asymmetry}

One of the major outcomes of recent observations are asymmetries in the atmospheres of gas giant exoplanets \citep[e.g.][]{Baeyens2024,Murphy2024,Murphy2025,Espinoza2024,Demangeon2024,Eherenreich2020}, that are observable with present (e.g., CHEOPS, JWST) and future (e.g., PLATO, Ariel) space missions. The role of chemical gas species in tracing asymmetries has been explored, for example, in \cite{Baeyens2021} and  \cite{2024ApJ...977...52S} with a specific focus on vertical mixing. Here, we focus on WASP-69~b to explore how kinetic chemistry may help with analyzing atmospheric asymmetries that result from global flows shaping the atmosphere's thermodynamic state. Species that are quenched at high pressures, such as \ce{CH4}, and species that are photochemically produced at low pressures, such as \ce{SO2}, may be particularly sensitive to these spatial variations. \ce{CH4} concentrations may reflect differences in the quench-point temperature across latitude and longitude \citep{2024ApJ...977...52S}, while \ce{SO2} concentrations are affected by local irradiation, temperature, and oxygen availability. Hence, these species may act as effective tracers of 3D atmospheric asymmetries, providing insight into the combined effects of chemistry and dynamics. 

Building on the terminator asymmetries identified in Section~\ref{sec: Results}, the spatial dependence of chemical concentrations is now considered in more detail. We focus on the [M/H] = 1.0 and C/O = 0.55 models in the following subsections.

\subsection{Thermodynamic latitudinal asymmetries.} \label{ss:latasym}

The climate of tidally locked irradiated exoplanets is characterized by an equatorial eastward wind jet and Rossby gyres on the nightside. The Rossby gyres manifest as part of the off-equatorial standing Rossby wave that is maintained by the planetary rotation. The shear between the Rossby and equatorial Kelvin wave is ultimately responsible for the formation and high wind speed ($\geq 1$~km/s) of the equatorial eastward jet \citep{Showman2011}. Both dynamical features shape the 3D gas temperature field and thus, potentially, also the chemistry of exoplanets. The eastward jet typically causes an eastward heat offset with respect to the substellar point, resulting in temperature asymmetries in the longitudinal (east-west) direction. The off-equatorial Rossby gyres encompass the coldest atmospheric regions in the planetary atmosphere, resulting in asymmetries in the latitudinal (north-south) direction.

The \texttt{ExoRad} simulations for WASP-69~b ([M/H] = 1.0, C/O = 0.55) exhibit Rossby gyres and eastward hot spot shifts from the top of our physical modeling domain ($p_{\rm gas}=10^{-4}$) down to $p_{\rm gas}=1$~bar (Fig. \ref{fig:gyres}). Thus, dynamics potentially impact the temperature in the pressure regions where \ce{CH4} and \ce{NH3} are quenched, as well as the atmospheric layers where \ce{SO2} is photochemically produced. Closer inspection reveals strong temperature gradients as a function of both latitude and longitude. At pressures greater than 0.1 bar, the temperature variations are up to $\sim$80~K between dayside and nightside for a given latitude, whereas in the upper atmosphere this variation can reach $\sim$400~K. These gradients reflect the reduced stellar insolation and weaker heat redistribution at high latitudes. In contrast, for a given longitude, latitudinal temperature differences are of the order $\sim$200~K both in the deep and upper atmosphere. 

The equatorial regions are therefore dominated by efficient day-to-night advection that smooths temperature variations at depth, while polar regions maintain cooler temperatures. Consequentially, species that quench at high pressures, like \ce{CH4}, may show significant latitudinal gradients in concentrations.

\begin{figure}
    \centering
    \includegraphics[width=0.65\linewidth]{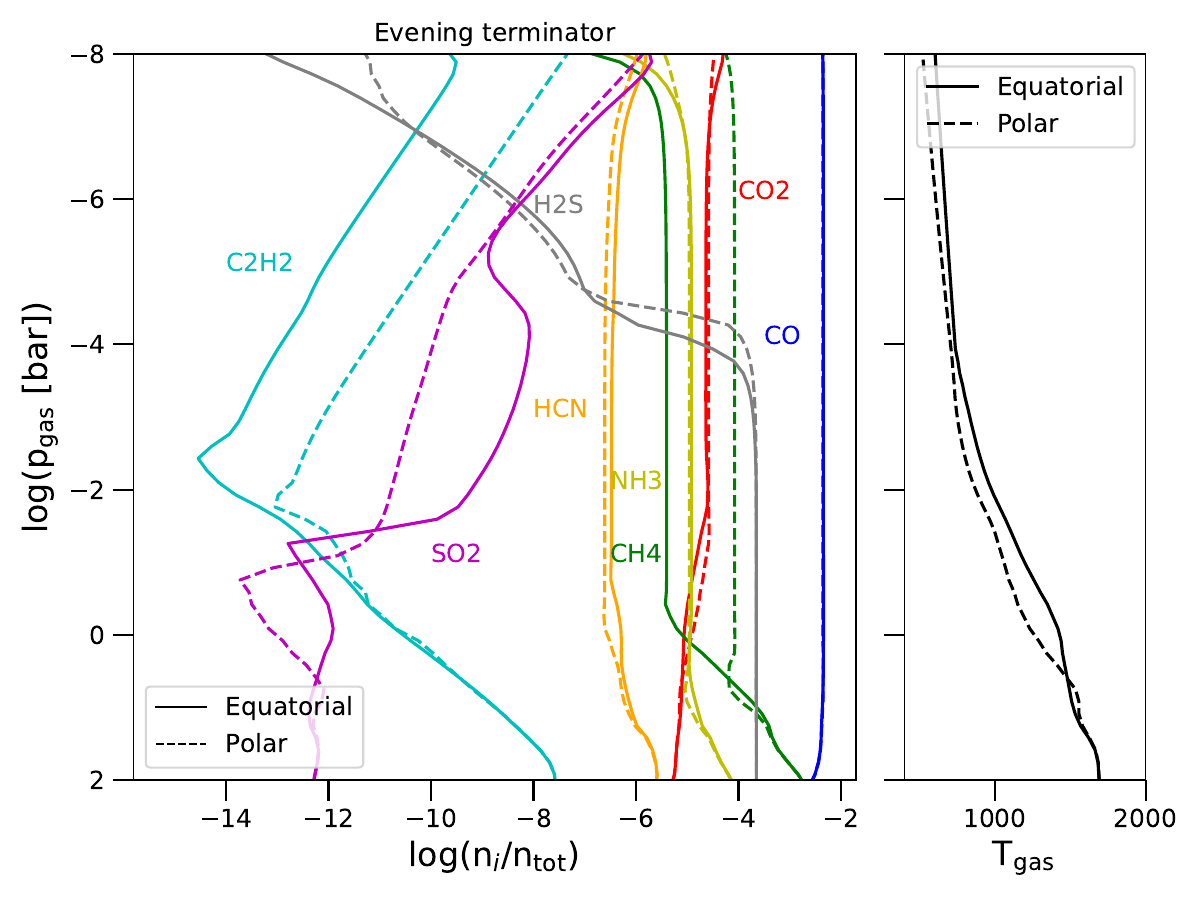}
    \includegraphics[width=0.65\linewidth]{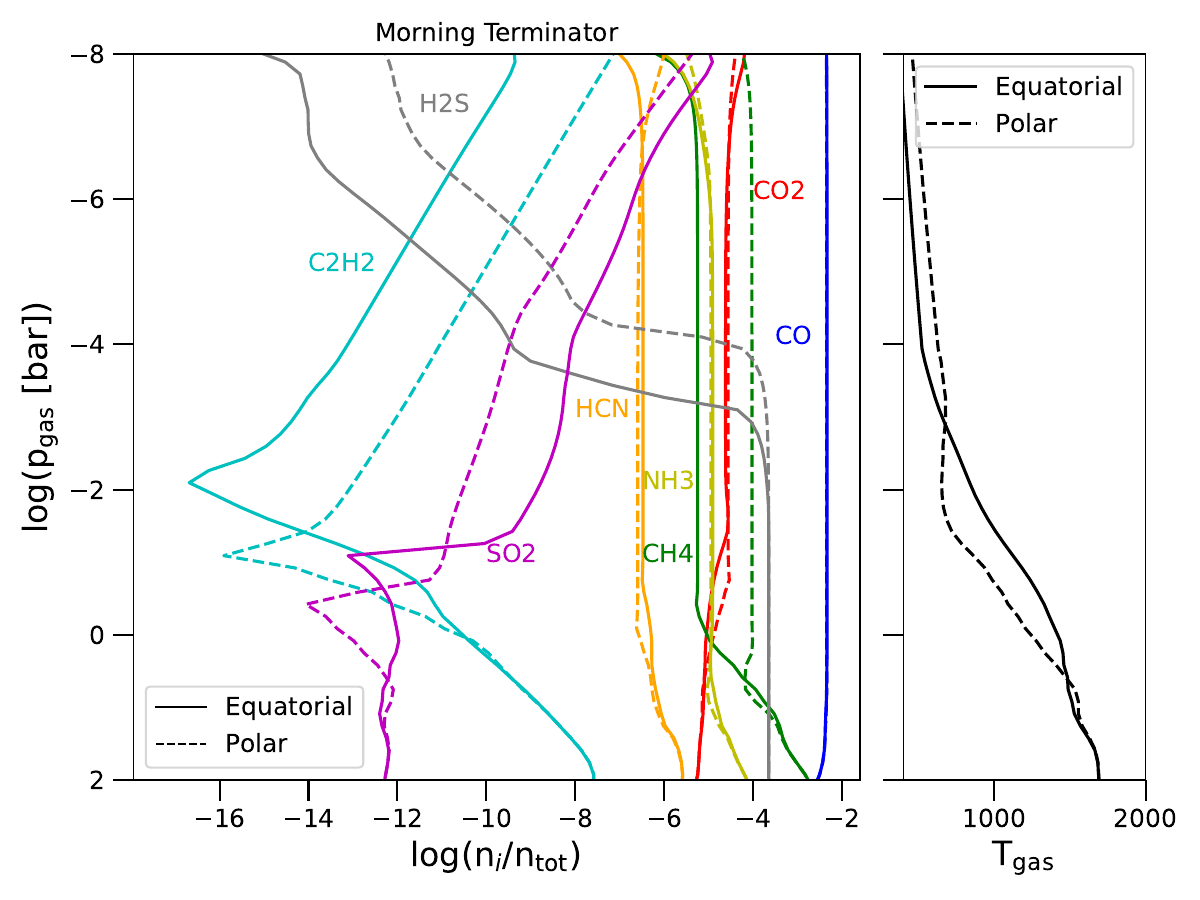}
    
    \caption{Concentrations of key molecules in the atmosphere of WASP-69~b at the (top) evening and (bottom) morning  terminator at latitudes 0$\degree$ and 76$\degree$, for [M/H] = 1.0 and C/O = 0.55, alongside their pressure-temperature profiles.}
    \label{fig:latitudinal assymetries}
\end{figure}

The strong latitudinal variations in temperature predicted by the GCM imply that the morning and evening limbs are not chemically homogeneous. This is consistent with the findings of \cite{Carone2025}, who demonstrated that latitudinal asymmetries can generate observable signatures in transmission spectra. Here, we extend this framework by examining how disequilibrium chemistry responds to these spatial gradients, and by identifying which molecular tracers are most sensitive to longitudinal versus latitudinal structure.

\subsection{Latitudinal asymmetries imprinted in chemistry}

The general trends for WASP-69~b have now been explored, including equatorial terminator asymmetries. Section~\ref{ss:latasym} demonstrated that Rossby gyres near the polar regions induce local variations in gas temperature. The question is therefore raised whether these temperature variations can be traced by specific gas-phase species.

\paragraph{Equator vs. pole in 1D:} Figure \ref{fig:latitudinal assymetries} compares the vertical profiles of key neutral species at the equator (0°) and high latitudes (76°) for the morning and evening terminators of WASP-69~b for [M/H] = 1.0 and C/O = 0.55. The results reveal strong latitudinal variations driven by temperature gradients. In the deep atmosphere, the equatorial profiles of both terminators have higher temperatures, due to stronger stellar radiation and efficient heat transport from the substellar point, while the polar regions remain cooler. These temperature variations lead to gradients in concentration, for example for quenched \ce{CH4}, consequentially also affecting the concentrations of photochemically produced \ce{C2H2}, of which \ce{CH4} is a parent molecule. 

\ce{SO2} concentrations are affected by the local gas temperature at p$_{\rm gas}\sim 10^{-5} - 10^{-1}$ bar. At the morning terminator, the latitudinal temperature differences are $\sim$200~K, resulting in \ce{SO2} concentrations that differ by up to two orders of magnitude. The equatorial-polar contrast is similarly pronounced for \ce{H2S} at the morning terminator, although it remains the major sulfur reservoir for p$_{\rm gas} > 10^{-4}$ at both terminators.

These compositional asymmetries may have measurable consequences for transmission spectra. In particular, \ce{CH4}, being highly sensitive to temperature, shows strong latitudinal dependence. Because transmission spectroscopy integrates over the entire limb, and the cooler polar regions contribute disproportionately to the observed signal, neglecting these temperature contrasts could lead to biased estimates of atmospheric C/H or C/O ratios. Furthermore, for [M/H] = 1.0 and C/O = 0.55, both photochemically produced \ce{C2H2} and \ce{SO2} may coexist at comparable concentrations at the polar regions (see Fig. \ref{fig:latitudinal assymetries}). High concentrations of \ce{CH4} and \ce{C2H2} are produced even in this oxygen-rich environment. This could partly explain the contrasting C/O ratios inferred for WASP-69~b from high-resolution transmission spectra compared to emission spectra \citep{Guilluy2022, Schlawin2024}, although the concentrations of these species remain low enough that the resulting spectral features are expected to be weak.

\paragraph{Column density 2D maps:} To assess whether chemical species can trace the latitudinal and longitudinal temperature gradients on WASP-69~b, we model the chemistry at each sampled latitude-longitude column using ARGO. Since our approach neglects horizontal transport, our results isolate the effect of local temperature structure and photochemistry on the resulting concentrations. This contrasts with the 3D-GCM framework of \cite{Zamyatina2024}, where horizontal mixing substantially dampens the chemical contrast on the hotter planet WASP-96~b. They show that \ce{CH4} can trace the gyre pattern under pure thermochemical equilibrium conditions, but that this signature disappears when horizontal quenching is included. While our models do not include horizontal quenching, they do account for vertical quenching; consistent with this, we find that any dependency on the gyre structure is not imprinted on \ce{CH4} concentrations, as \ce{CH4} quenches at deeper pressures (p$_{\rm gas}\approx$ 0.1-0.4 bar) and is insensitive to the temperatures at the lower pressure layers (p$_{\rm gas}\lesssim$ 0.1 bar) where the gyres reside. 

\begin{figure}
    \centering

    \includegraphics[width=0.48\linewidth]{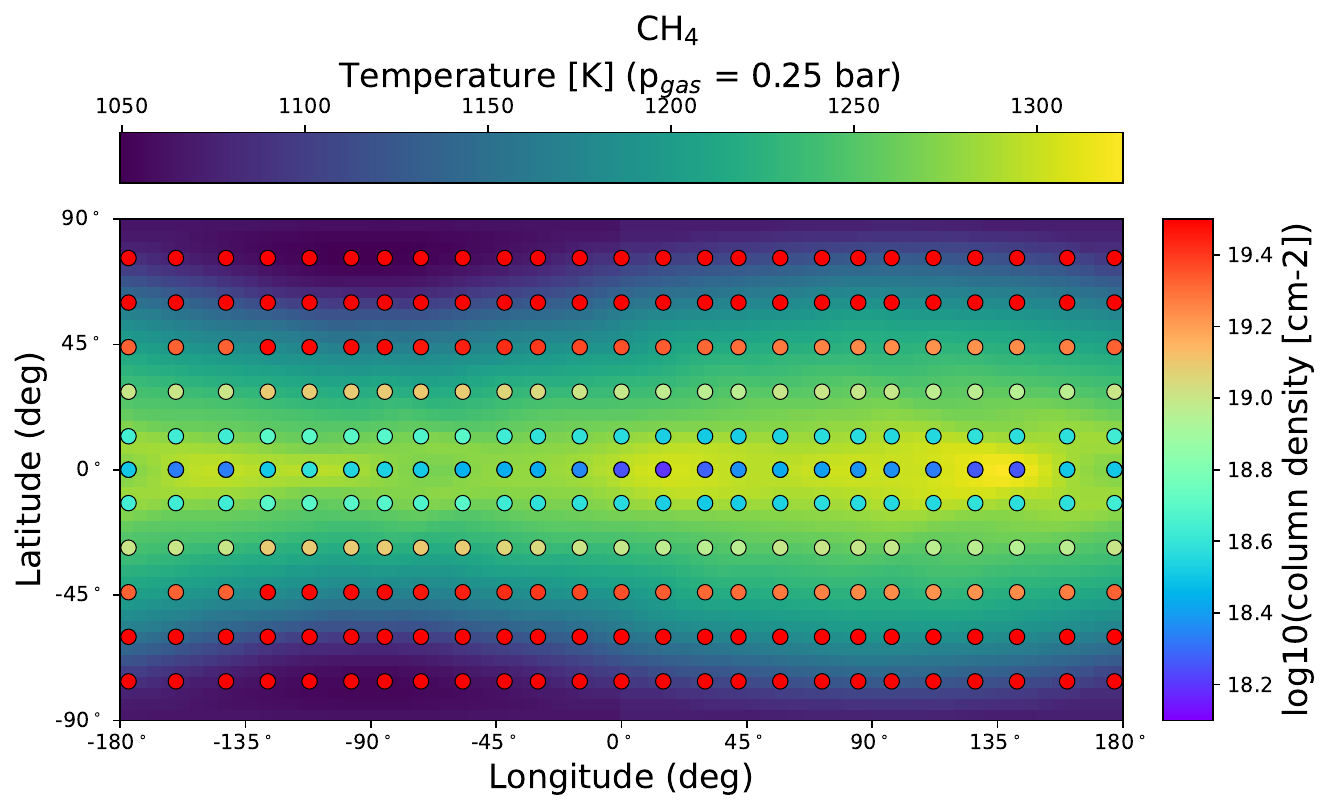}
    \includegraphics[width=0.48\linewidth]{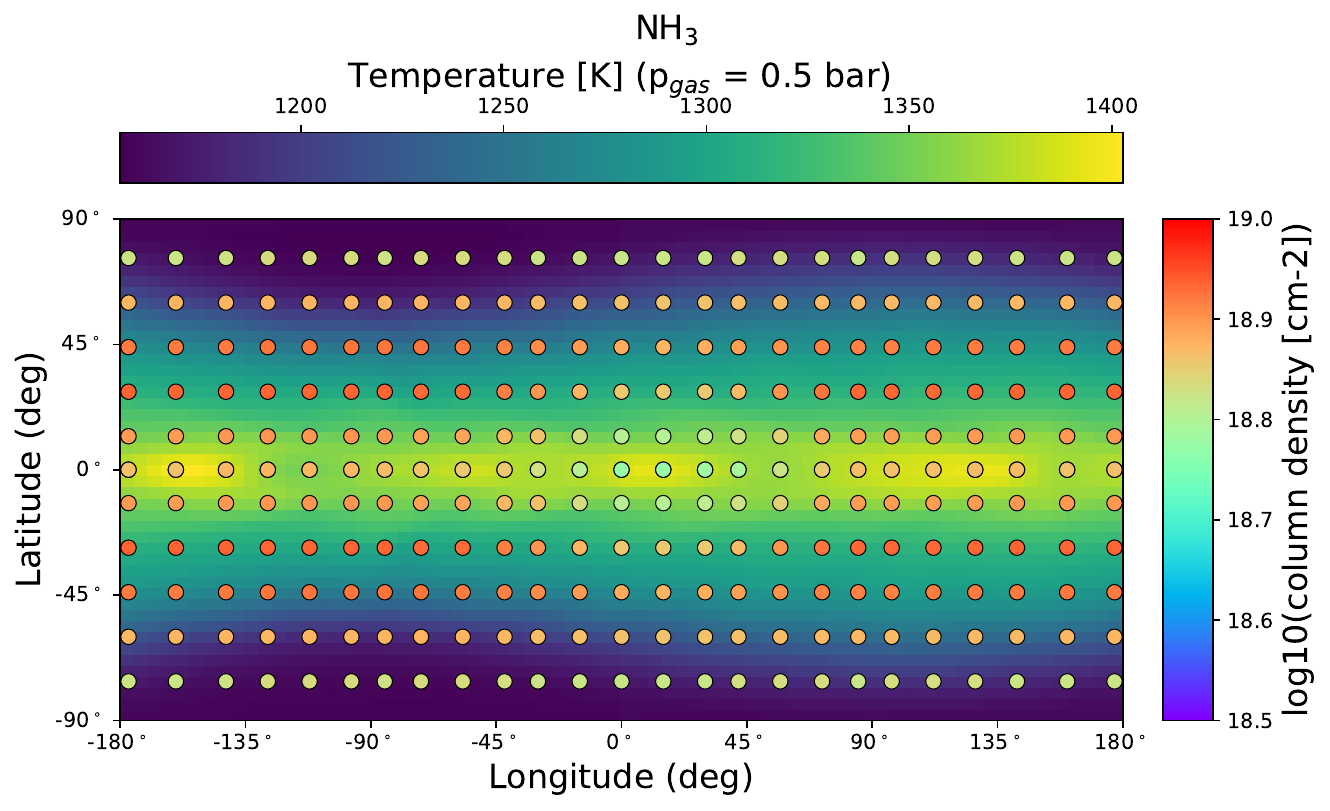}
    \includegraphics[width=0.48\linewidth]{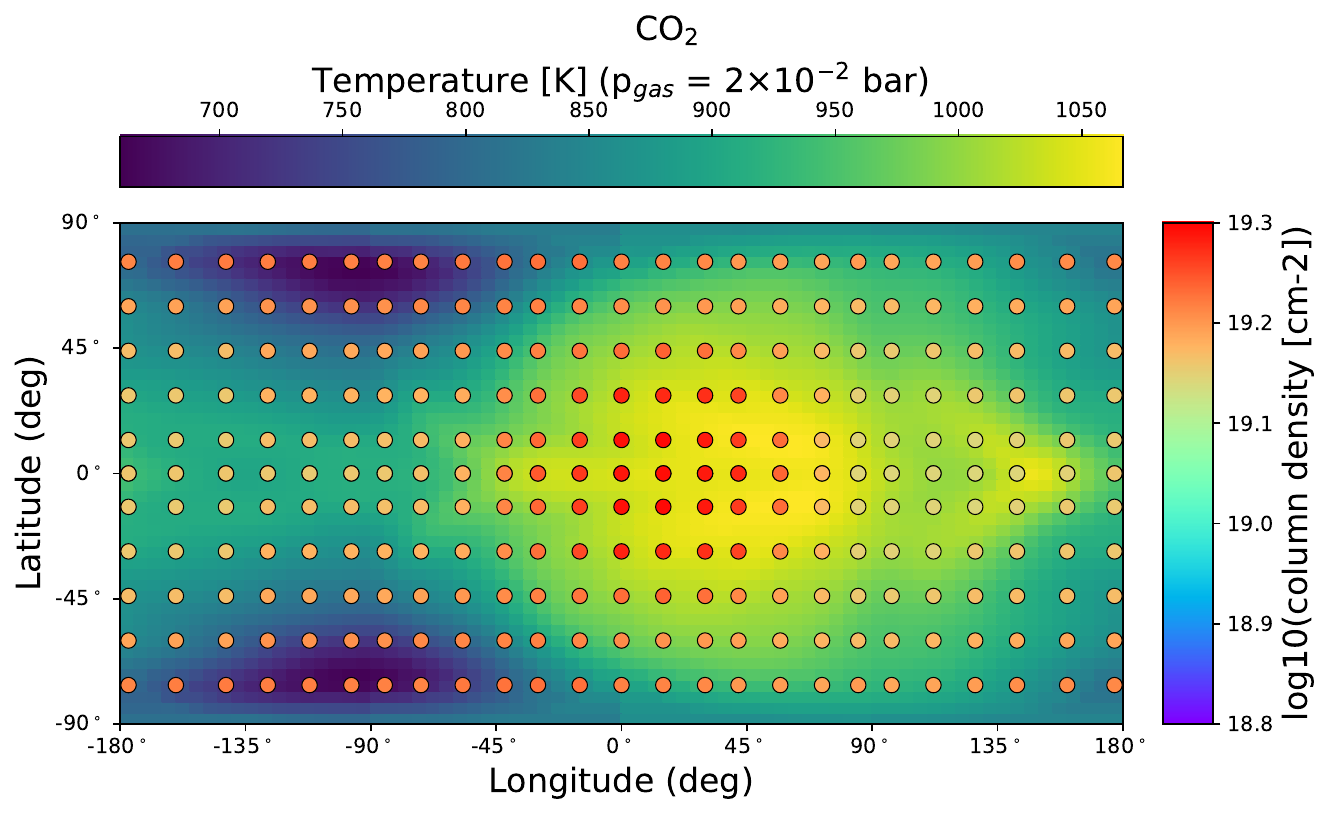}
    \includegraphics[width=0.48\linewidth]{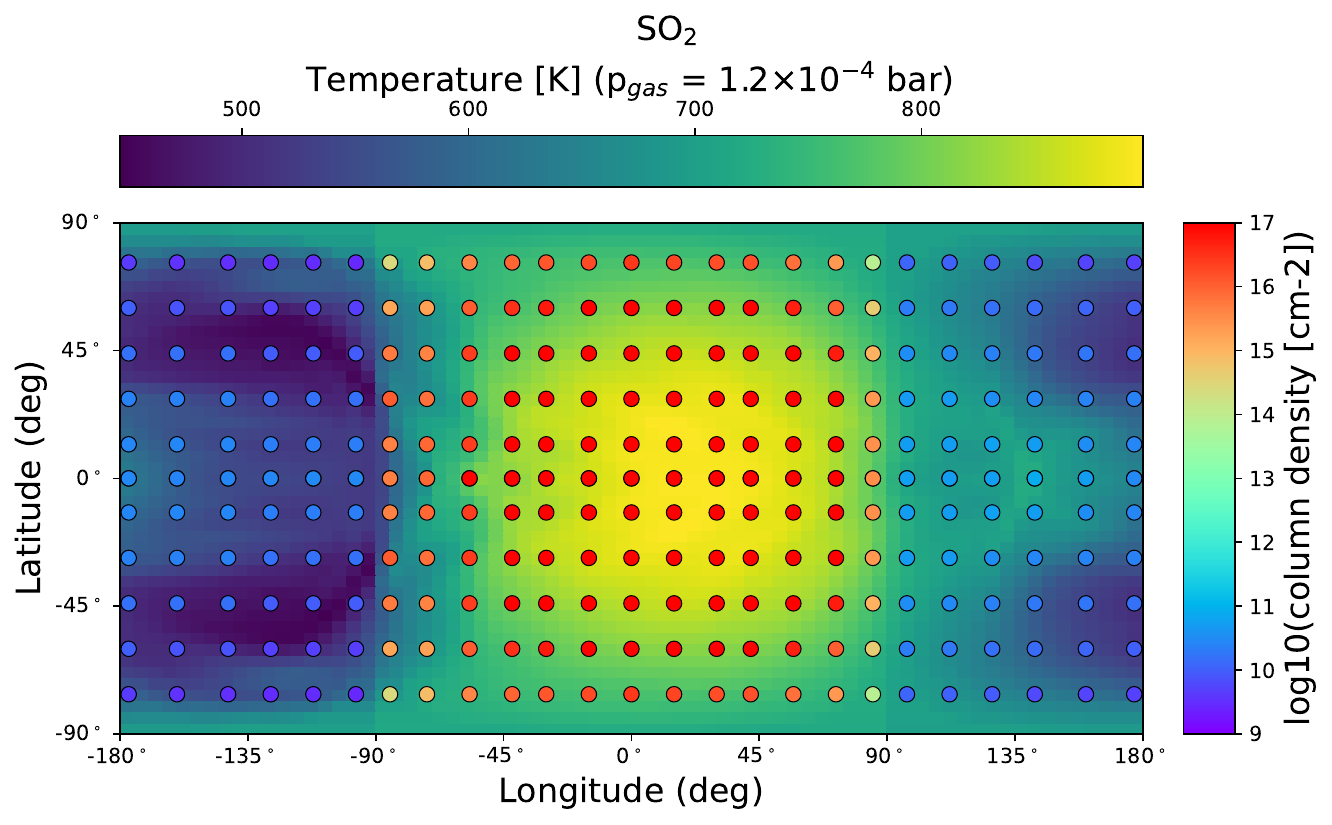}
    \includegraphics[width=0.48\linewidth]{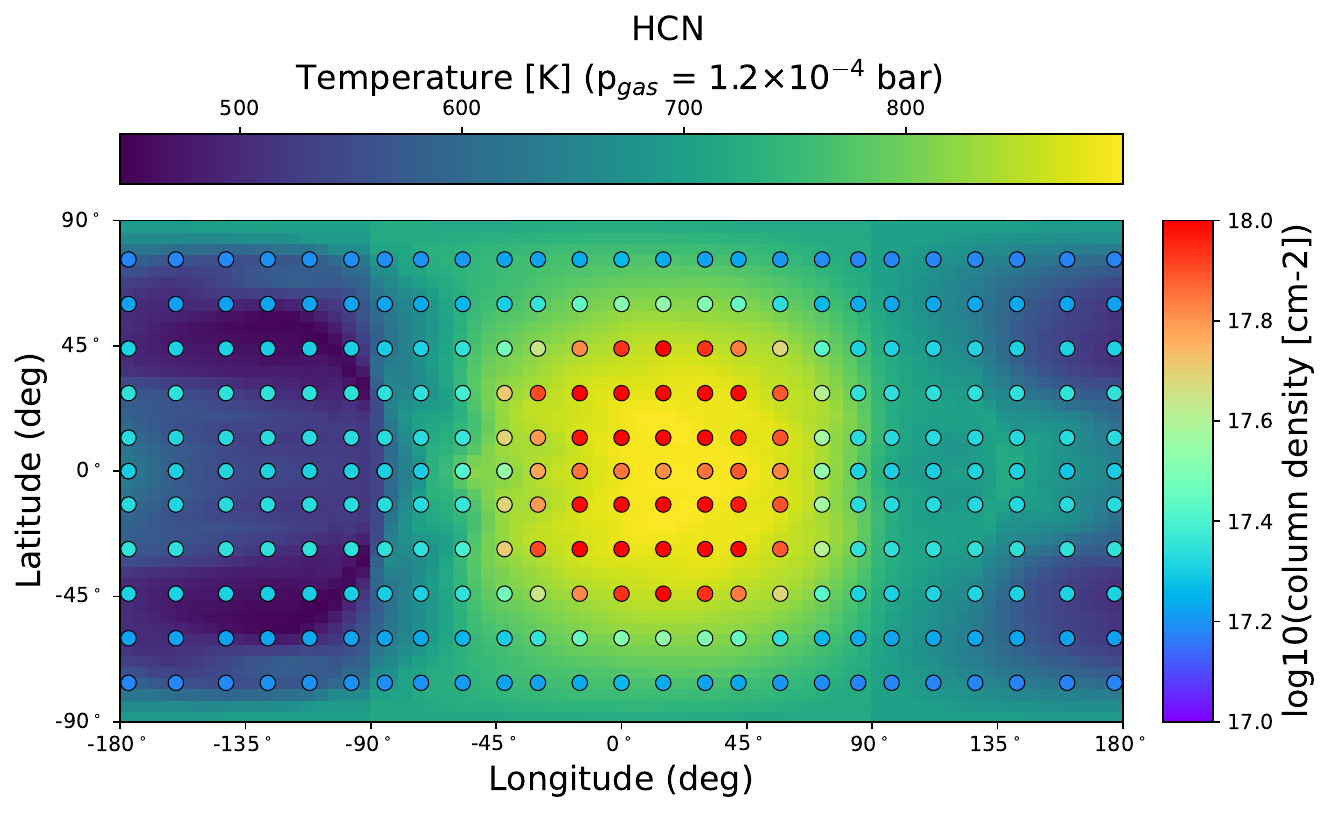}
    \includegraphics[width=0.48\linewidth]{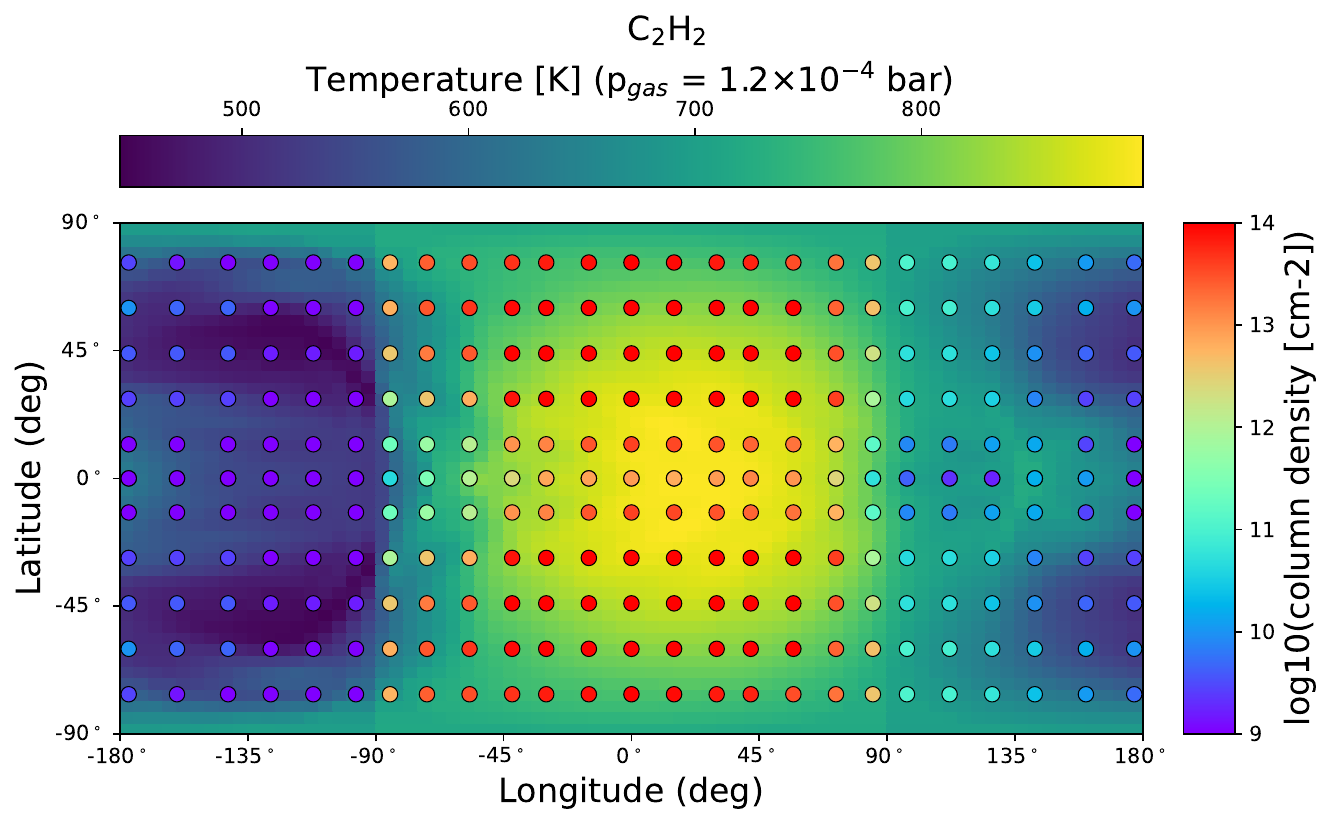}
    
    \caption{Integrated column densities for WASP-69~b overplotted on the local gas temperature map at a pressure level representative of the bulk production of each species. For CH$_4$ at 0.25~bar (top left), NH$_3$ at 0.5~bar (top right) and CO$_2$ at 0.02~bar (center left), the pressures corresponds to a representative pressure within the quench region, whereas SO$_2$ (center right), HCN (bottom left) and \ce{C2H2} (bottom right) are mostly sensitive to temperatures at $1.2\times 10^{-4}$~bar, which corresponds to the photochemically active layer where production is strongest. The column densities are shaped by the vertically varying gas temperature (Fig.~\ref{GCM WASP69 b}) and the velocity field (Fig.~\ref{fig:gyres}). The center of each panel indicates the substellar point (dayside), while the left rim shows the antistellar point (nightside). 
    Note that the color scale for the integrated column densities differs between species.}
    \label{fig:Gyres_plus_Chemistry}
\end{figure}

Figure~\ref{fig:Gyres_plus_Chemistry} presents vertically integrated column density maps for key species. The background color scale shows the local temperature at a representative pressure level relevant for each molecule: the quench level for species such as \ce{CH4},\ce{NH3}, and \ce{CO2}, and the peak photochemical production region for species such as \ce{SO2}, \ce{C2H2} and \ce{HCN}. Overlaid contours indicate the integrated column densities, calculated from the top of the modeled atmosphere down to the $\tau = 1$ level.

For \ce{CH4} (Fig.~\ref{fig:Gyres_plus_Chemistry}, top left), the deep quench point suppresses sensitivity to the gyre structure itself, and the latitudinal asymmetries are reflective of temperatures at the quench layers of p$_{\rm gas}\approx$ 0.1 - 0.4 bar. Differences of $\sim$0.2 dex are evident between the cooler morning terminator and the warmer evening terminator (e.g., at latitude 45$^\circ$ in Fig.~\ref{fig:Gyres_plus_Chemistry}). At equatorial latitudes, \ce{CH4} is further depleted on the evening terminator due to higher temperatures associated with the eastward-shifted hotspot. Latitudinal temperature variations of $\sim$200~K produce over an order-of-magnitude variation in column density, while dayside photodissociation further reduces abundances by up to $\sim$0.5 dex at a given latitude. Overall, \ce{CH4} exhibits clear temperature-driven gradients in latitude, but only weak longitudinal variation.

\ce{NH3} and \ce{CO2} (top right and center left panels of Fig.~\ref{fig:Gyres_plus_Chemistry}, respectively) similarly reflect temperatures at their respective quench pressures. As the high temperatures favor \ce{N2} production, \ce{NH3} exhibits only modest variations in column density even with global differences in temperature of $\sim$400~K. \ce{CO2} is additionally influenced by dayside photochemical production, resulting in enhanced column densities on the dayside. However, since its formation is primarily dependent on the \ce{H2O} and \ce{CO} concentrations, both of which are present abundantly, its overall spatial variation remains within a factor of two. As a result, \ce{CO2} remains a robust tracer of atmospheric metallicity and C/O, comparatively insensitive to local temperature variations and irradiation geometry.

In contrast, \ce{SO2} (center right panels Fig.~\ref{fig:Gyres_plus_Chemistry}) shows strong spatial asymmetries. Photochemically formed on the dayside, its production peaks around 10$^{-4}$ bar, where the longitudinal temperature contrast is $\sim$ 400~K (center right panel Fig. \ref{fig:Gyres_plus_Chemistry}). Here, its column densities trace the local temperature structure, reproducing the eastward-shifted hotspot pattern evident in the GCM. The column density varies by eight orders of magnitude across the planet, dropping to negligible values at longitudes $< -90\degree$ and $> 90\degree$, where the planet receives no stellar irradiation. As a result, any direct correspondence of SO$_2$ concentrations with the gyre structure on the night side is inhibited. 

Both \ce{HCN} and \ce{C2H2} (Fig.~\ref{fig:Gyres_plus_Chemistry}, bottom panels) display dayside-dominated distributions, with depletion at equatorial latitudes. As both species are primarily produced from \ce{CH4} \citep[e.g.,][]{Moses2011,Bangera2025}, they inherit its spatial variability. On the dayside, production is reduced at equatorial latitudes where higher temperatures suppress \ce{CH4}. \ce{HCN}, which forms through both thermochemical and photochemical pathways, exhibits moderate spatial variability of order one dex while remaining relatively abundant throughout the observable atmosphere. In contrast, \ce{C2H2}, formed predominantly via photochemistry, shows much stronger day–night and latitudinal contrasts, with variations of up to five orders of magnitude, although its overall abundance remains low.

\begin{figure}
    \centering
    \includegraphics[width=0.45\linewidth]{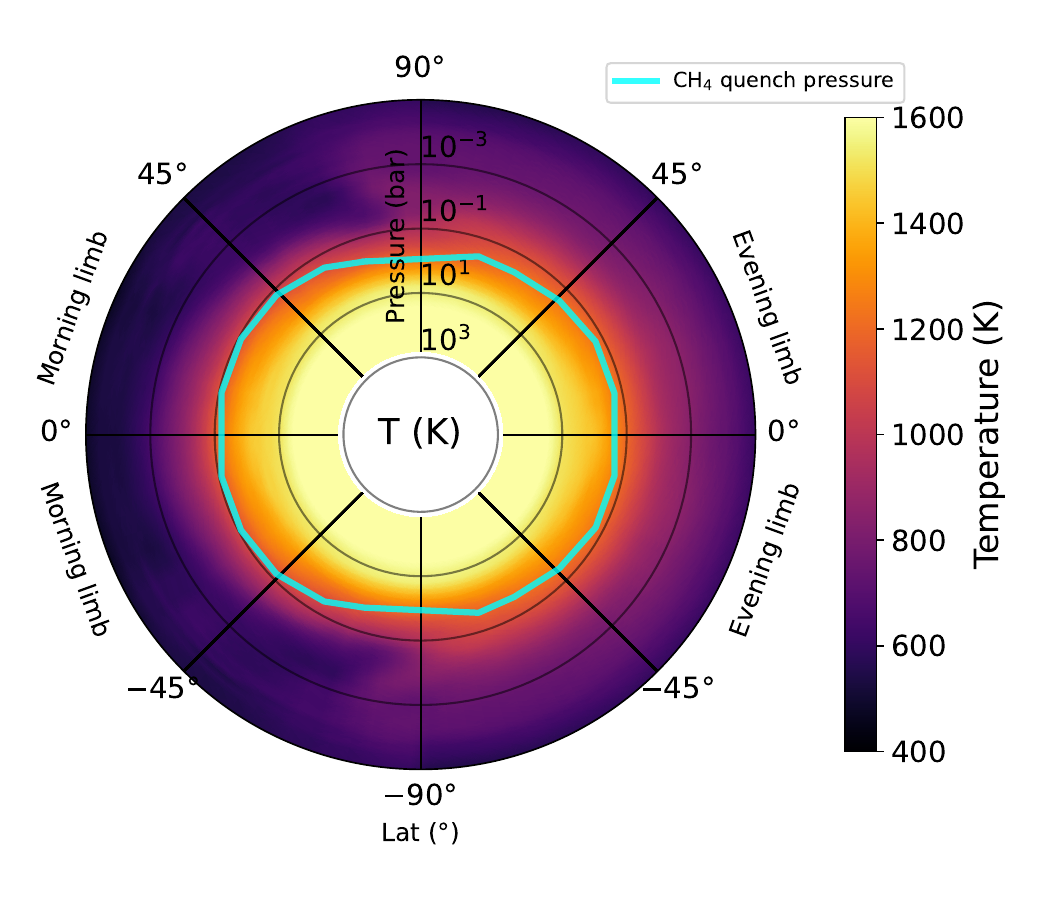}
    \includegraphics[width=0.45\linewidth]{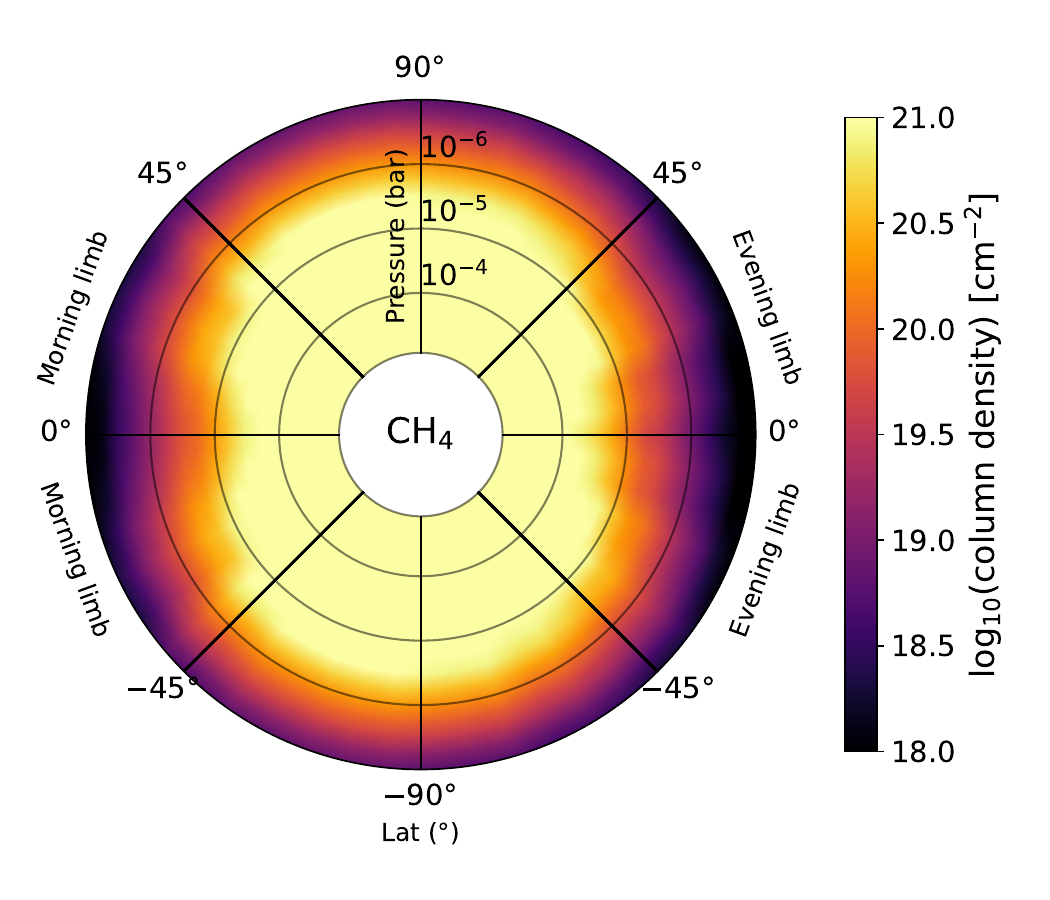}
    \includegraphics[width=0.45\linewidth]{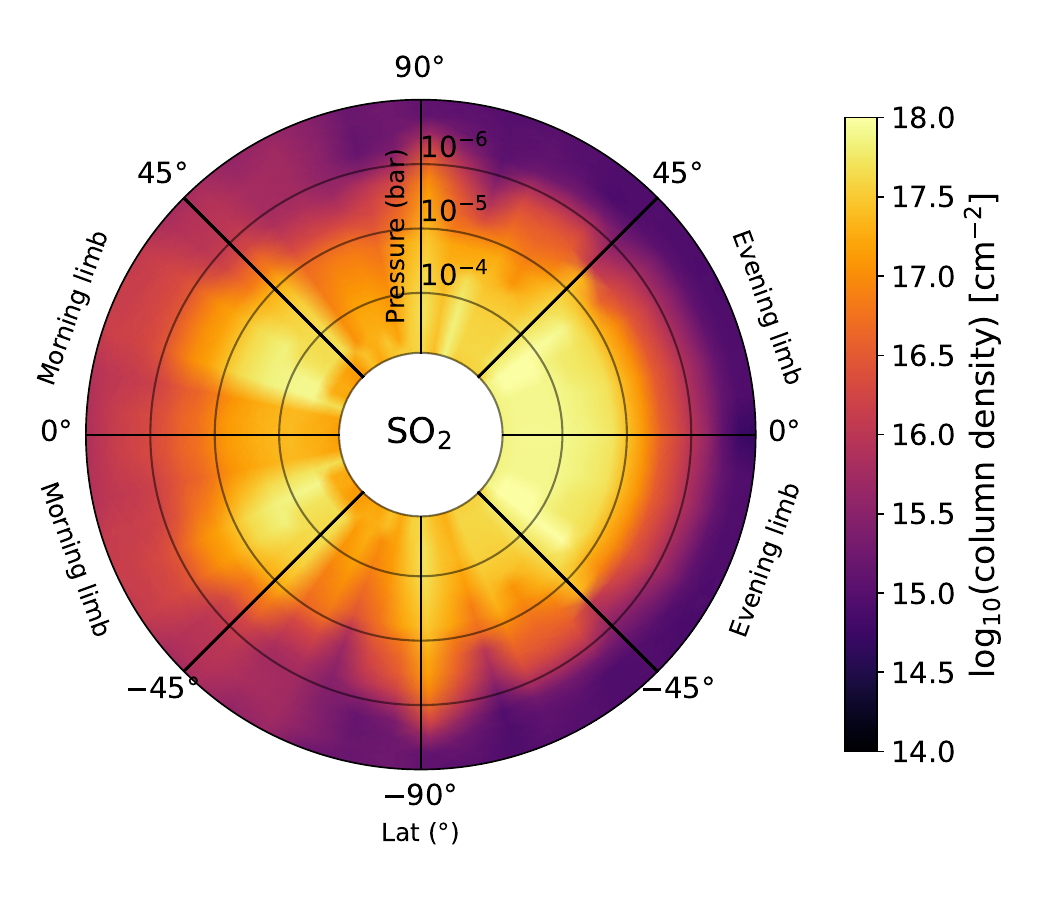}
    \includegraphics[width=0.45\linewidth]{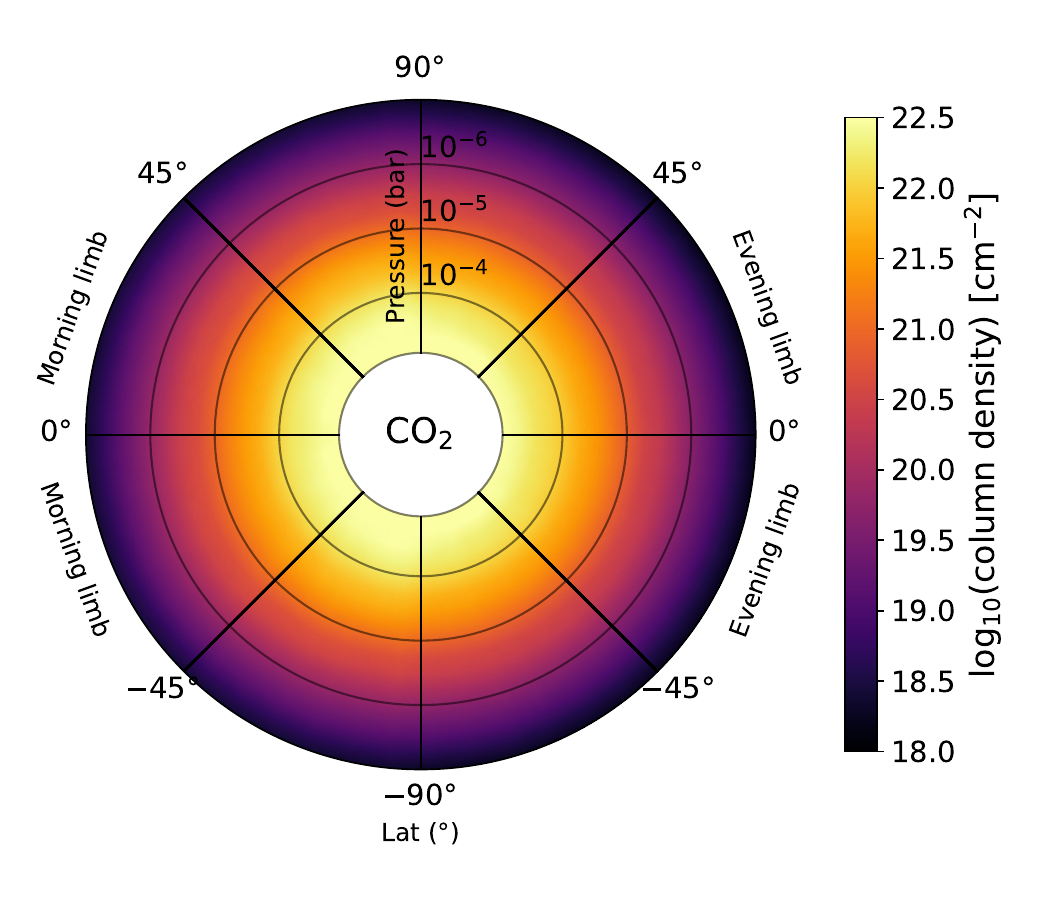}
    
    \caption{{\bf Top left:} (p$_{\rm gas}$, T$_{\rm gas}$) cross-section through the terminators of WASP-69~b, with the location of the CH$_4$ quench point overlaid (light blue line). We define the \ce{CH4} quench level as the pressure where the absolute relative difference between the equilibrium and disequilibrium concentrations first exceeds $10^{-3}$. {\bf Top right:} Column densities in the line-of-sight direction in the atmosphere of WASP-69~b for [M/H] = 1.0, and C/O = 0.55 for CH$_4$.  {\bf Bottom:} The same as for CH$_4$ but now for SO$_2$ (left) and CO$_2$ (right). The terminator pressure scale is used as a radius axis. The hotspot offset  at 10$^{-4}$ bars is reflected in the column density of \ce{SO2}, with larger concentrations being formed on the evening terminator.}
    \label{fig:l-o-s cd}
\end{figure}

Figure~\ref{fig:l-o-s cd} shows line-of-sight integrated column densities, which approximate what transmission spectroscopy would sample, for the quenched \ce{CH4}, photochemically produced \ce{SO2}, and thermochemically stable species \ce{CO2}. The quench pressure illustrated are determined using the definition described in Section \ref{sec:ARGO}. Using this definition, the \ce{CH4} quench pressure varies between approximately 0.1 and 0.4 bar across the different latitude-longitude columns owing to spatial variations in the pressure-temperature structure. \ce{CH4} appears nearly symmetric around the limb, with slight depletion at the equator driven by a combination of warmer temperatures and stronger photodissociation. \ce{SO2} shows stronger longitudinal asymmetry, with column densities $\sim$2 orders of magnitude higher on the evening terminator than on the morning terminator. \ce{CH4} column densities peak at high latitudes, whereas that of \ce{SO2} peaks at the equator. These two species therefore sample different regions in latitude and altitude, and their combined spectral signatures would reflect information about WASP-69~b's 3D structure. In contrast, \ce{CO2} shows an almost symmetric distribution around the limb. 

\subsection{Observational Consequences}

In this section, we examine how the spatial chemical asymmetries identified above translate into transmission spectra. The full PLATO-HST–JWST wavelength range (0.4 – 12~$\mu$m) is shown in the Appendix. Here, we focus on the \ce{CH4} absorption bands near 3.3 and 7.7~$\mu$m and the \ce{CO2} feature near 4.3~$\mu$m, as these exhibit both the strongest latitudinal dependence and the largest compositional variations in Fig.~\ref{fig:metallicityspectra}.

\begin{figure}
    \centering
    \includegraphics[width=0.85\linewidth]{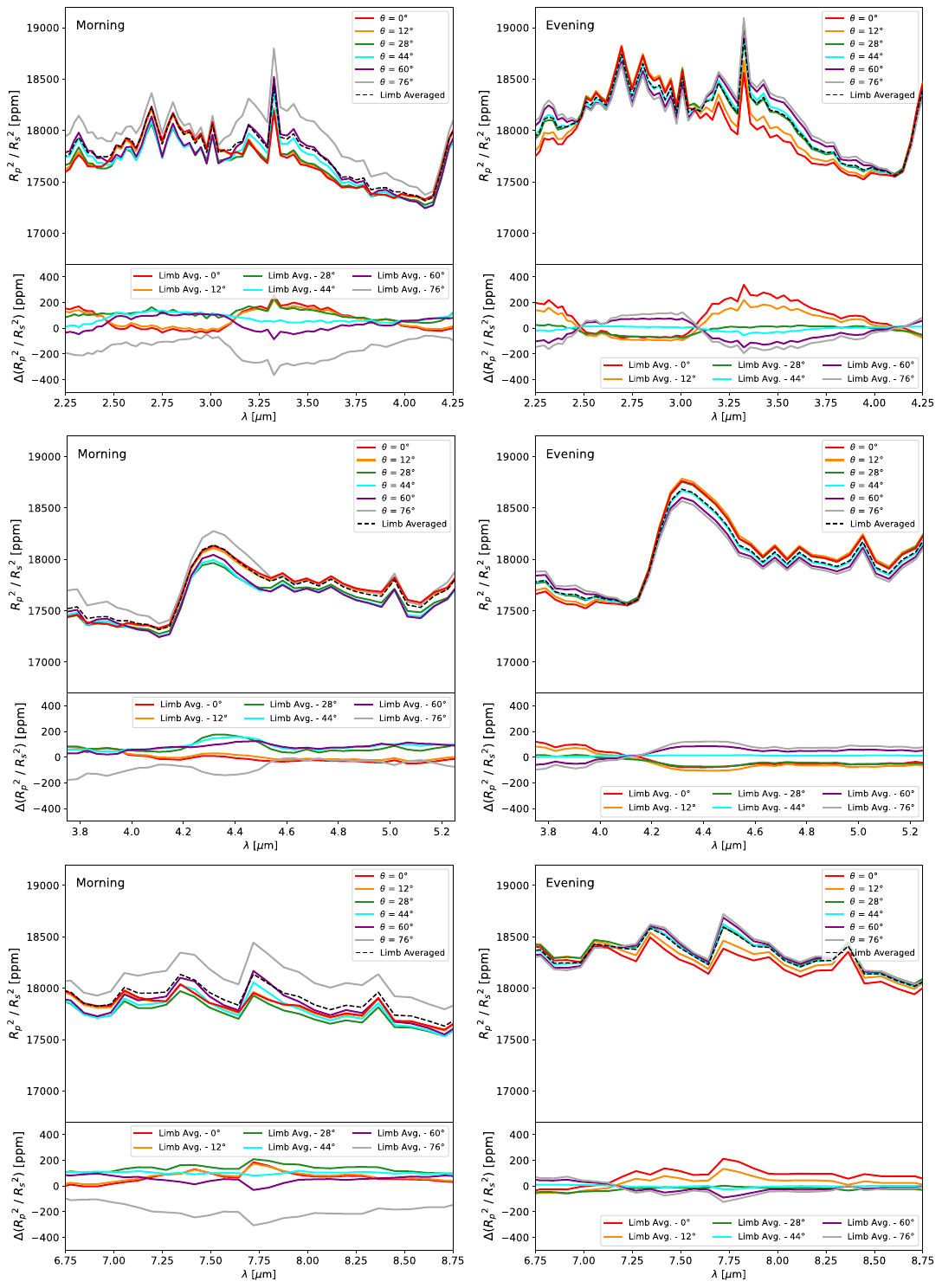}
    \caption{Synthetic transmission spectra of the morning (left panels) and evening (right panels) terminators of the baseline WASP-69~b model across multiple latitudes, with the corresponding limb-averaged spectra overlaid as dashed black lines. The top panels highlight the \ce{CH4} absorption feature near 3.3~$\mu$m, the middle panels the \ce{CO2} feature near 4.3~$\mu$m, and the bottom panels the \ce{CH4} feature near 7.7~$\mu$m. Residual spectra relative to the limb-averaged model (limb-averaged minus latitude-dependent spectrum) are shown below each panel. The strongest latitudinal variations occur in the CH$_4$ features, reflecting enhanced methane abundances in the cooler high-latitude regions.}
    \label{fig:nominalspectra}
\end{figure}

Figure~\ref{fig:nominalspectra} shows synthetic transmission spectra for the baseline composition case ([M/H] = 1.0, C/O = 0.55) at multiple latitudes for the morning (left panels) and evening (right panels) terminators. The corresponding limb-averaged spectra are overplotted as dashed black curves, while the lower panels show residuals relative to the limb average. The strongest latitudinal differences occur between $\sim$2–4~$\mu$m and $\sim$7–9~$\mu$m, where the spectra are dominated by \ce{CH4} absorption.

At both terminators, the high-latitude regions ($\theta \sim 60^\circ$) contribute more strongly to the limb-averaged \ce{CH4} feature than the equatorial latitude. This behavior directly reflects the enhanced \ce{CH4} line-of-sight column densities at high latitudes (Fig.~\ref{fig:l-o-s cd}), demonstrating that the equatorial atmosphere alone is not representative of the integrated transmission spectrum. The resulting variations in the 3.3~$\mu$m \ce{CH4} feature reach amplitudes of $\sim$400 ppm at the morning terminator and $\sim$350 ppm at the evening terminator. Although these variations are smaller than the maximum metallicity-driven differences of $\sim$650 ppm and $\sim$850 ppm predicted at the morning and evening terminators respectively for the most metal-poor, carbon-rich model ([M/H] = 0.78, C/O=0.95), they exceed the $\sim$200 ppm cases produced by the other metallicity cases explored in Section~\ref{sec: Results}. These results indicate that three-dimensional thermal structure can modify \ce{CH4} spectral signatures at a level comparable to plausible uncertainties in the atmospheric metallicity of WASP-69~b.

In contrast, the \ce{CO2} feature near 4.3~$\mu$m exhibits weaker latitudinal variations, reaching only $\sim$200 ppm at the morning terminator and $\sim$120 ppm at the evening terminator. These variations are comparable to the $\sim$200 ppm metallicity-driven changes at the morning terminator, but smaller than the $\sim$400 ppm changes at the evening terminator. This indicates that \ce{CO2} remains primarily sensitive to atmospheric composition rather than to local thermal asymmetries. However, the relative contributions to the \ce{CO2} feature differ between the two limbs. On the morning terminator, the feature is dominated by equatorial regions, whereas on the evening terminator the dominant contribution shifts toward higher latitudes ($\theta > 44^\circ$). We attribute this behavior to the eastward hotspot offset in the GCM, which reduces equatorial \ce{CO2} abundances on the evening limb.

Overall, the predicted latitudinal structure imprints observable signatures on the transmission spectra at a level of several hundred ppm. These results demonstrate that spatial thermal and chemical asymmetries can strongly influence spectral interpretations and should therefore be considered when analyzing spectral observations of warm giant exoplanets.

\section{Discussion} \label{sec: Discussion}

The chemistry of sulfur and carbon-bearing species provides a powerful probe of both the physical structure and the formation history of exoplanetary atmospheres \citep{Turrini2021,Pacetti2022,SommervilleThomas2026,Nakazawa2026}. In this section we evaluate the sensitivity of sulfur and carbon chemistry to latitudinal and longitudinal asymmetries, assess the assumptions made in our modeling, and examine the consequences for interpreting current and future spectroscopic observations. 

\subsection{Chemical asymmetries as tracers for different atmospheric layers}

The 3D-GCM simulations of WASP-69~b reveal latitudinal and longitudinal temperature gradients, with day-night differences of 200 - 400~K and cooler poles relative to the equator. In \texttt{ExoRad} these differences are shaped to varying degrees by Rossby gyres and the equatorial eastward jet, and imprint distinct signatures on the atmospheric chemistry depending on where individual species are formed. 

A key result of this work is that different molecules trace different atmospheric layers and therefore probe different dynamical regimes. Species controlled by deep atmosphere quenching, such as \ce{CH4} and \ce{NH3}, primarily reflect the thermal structure at pressures of $\sim$0.1-1.0 bar. \ce{CH4} is also enhanced in the cooler high latitude regions. In contrast, photochemically produced species such as \ce{SO2} trace the upper atmospheric pressures (10$^{-5}$ - 10$^{-1}$~bar), where the longitudinal temperature contrasts are largest. Along the limb, the lowest temperatures occur at the morning terminator owing to the presence of the Rossby gyres, which favors local SO$_2$ production. However, because of the eastward hotspot offset, the integrated line-of-sight column densities of SO$_2$ are largest at the evening terminator. 

\ce{HCN} and \ce{C2H2} occupy an intermediate regime. Both species inherit part of the spatial variability of \ce{CH4} through their chemical formation pathways, while also being affected by upper atmosphere photochemistry. HCN therefore exhibits moderate spatial structure across the observable atmosphere, whereas \ce{C2H2} shows stronger day–night contrasts but remains relatively low in abundance.

In contrast to \ce{CH4} and \ce{SO2}, \ce{CO2} exhibits comparatively weak spatial variability across WASP-69~b and therefore remains a robust tracer of atmospheric composition. Although \ce{CO2} experiences modest photochemical enhancement on the dayside, its abundances vary by less than a factor of two across the observable atmosphere, and the resulting variations in the transmission spectra remain smaller than those driven by changes in metallicity and C/O ratio.

The predicted spatial gradients in concentrations from this work may have direct observational implications. While observational efforts are beginning to probe atmospheric asymmetries \citep[e.g.,][]{Eherenreich2020,Prinoth2022,Espinoza2024}, most analyses still rely on limb-averaged spectra that combine contributions from both terminators and a range of latitudes \citep{barstow2020}. As a result, temperature-driven chemical inhomogeneities can bias retrieved abundances. In particular, cooler, \ce{CH4}-rich high-latitude regions can contribute to the observed signal, potentially leading to overestimated \ce{CH4} abundances and inferred C/O ratios. Likewise, strong longitudinal asymmetries in \ce{SO2}, such as the contrast between the morning and evening terminators, may bias constraints on sulfur abundance or metallicity if interpreted with one-dimensional models. As highlighted by \citet{Carone2025}, contributions from high latitudes are non-negligible and must be accounted for when interpreting limb-averaged transmission spectra.

There is growing observational evidence that such three-dimensional effects are measurable. Longitudinal variations have already been inferred for WASP-107~b, which is even colder than WASP-69~b \citep{Murphy2024,Murphy2025}. Even the interpretation of dayside planetary flux measurements can be biased by assuming a 1D temperature column \citep{Taylor2020}. JWST has demonstrated recently that it is indeed possible to constrain longitudinal and latitudinal temperature variations for ultra-hot Jupiters like WASP-18~b \citep{Challener2025}. Together, these developments highlight the importance of understanding the 3D temperature-structure and chemistry when interpreting current and future observations. 

\subsection{Role of horizontal transport}

A key limitation of the present study is the neglect of horizontal transport. The column-by-column approach of this work should be interpreted as an upper limit on the strength of chemical contrasts that can arise from local temperature structure and photochemistry alone. Three-dimensional circulation models show that advection by the equatorial jet can substantially modify these patterns. In particular, horizontal transport can amplify the longitudinal asymmetry of photochemically produced species. For example, \citet{Baeyens2024} predicted strong enhancements of HCN at the cold morning terminator of the ultra-hot Jupiter WASP-76~b. Similarly, for WASP-39~b, the inclusion of advection in the \texttt{ExoRad} framework leads to several orders of magnitude enhancement of SO$_2$ at the morning terminator, making it up to an order of magnitude more abundant there than at the evening terminator \citep[][Fig.~7]{Steinrueck2025}. In contrast, kinetically quenched species such as CH$_4$ tend to be smoothed in longitude once eastward advection is included, effectively erasing the longitudinal gradients seen in column-by-column models \citep{Drummond,Zamyatina2024,Steinrueck2025}. 

Latitudinal variations in CH$_4$, however, are generally preserved in full 3D kinetic models, although with reduced amplitude \citep{Zamyatina2024,Steinrueck2025}. For WASP-39~b, variations of up to one order of magnitude are still predicted \citep[Fig.~5 of][]{Steinrueck2025}. As also found in this work, these latitudinal contrasts are primarily driven by the cold Rossby gyres at the quench level. The gyres extend throughout the observable atmosphere, are located north and south of the equatorial jet, and maintain localized temperature minima at each pressure level due to their closed circulation patterns (Figs.~\ref{fig:Gyres_plus_Chemistry} and \ref{fig:gyres}). For WASP-69~b, we therefore expect the strong \ce{CH4} latitudinal asymmetries identified here to remain qualitatively robust even in fully coupled 3D chemical simulations, although likely with reduced amplitude.

Photochemically produced \ce{SO2} is expected to respond differently. In our models, \ce{SO2} is produced most efficiently near the substellar region, where concentrations exceed those at the terminators by a factor $\sim$500 or more, due to the intense radiation. Although the terminator abundances in our nominal models remain too low ($n_\mathrm{SO2}/n_\mathrm{tot}\sim$10$^{-6}$) to generate detectable spectral features, horizontal advection could transport \ce{SO2}-rich gas from the dayside towards the limbs. Motivated by this possibility, we ran additional test cases where we artificially increased the \ce{SO2} concentrations by a factor of 500 at the limbs, and found that under these conditions, a large \ce{SO2} feature emerged near 7.5~$\mu$m as well as a smaller feature near 4~$\mu$m. Consistent with the column density maps (Fig.~\ref{fig:Gyres_plus_Chemistry}), these features were contributed by the lower latitudes, where \ce{SO2} is most abundant due to stronger irradiation and higher local temperatures. This demonstrates that redistribution of dayside produced \ce{SO2} could produce observable signatures of its latitudinal asymmetries. \\

\noindent The STAND2025 chemical network used in this work includes ion-neutral chemistry and thermal ionization processes. At the temperatures relevant for WASP-69~b (T$_{\rm{eq}}$ $\sim$ 900~K), ion chemistry does not substantially modify the dominant neutral disequilibrium species. However, ion-neutral chemistry may become increasingly important for hotter or more strongly irradiated atmospheres, motivating future studies coupling ion chemistry to fully three-dimensional chemical analysis (Bangera et al., in prep).

\section{Conclusions} \label{sec: Conclusions}

This study investigates how the atmospheric chemistry on WASP-69~b is shaped by the interplay between atmospheric composition, photochemistry, and three-dimensional atmospheric structure, using GCM-derived temperature fields as inputs for 1D photochemical-kinetics models. With an intermediate equilibrium temperature ($T_{\mathrm{eq}}\approx 900$~K), WASP-69~b resides in a regime where disequilibrium processes are expected to dominate, making it an ideal test case for assessing how modest variations in composition and temperature structure translate into observable spectroscopic variations.

The key results of this work are as follows: 
\begin{itemize}
    \item \textbf{Dependence on atmospheric metallicity and C/O ratio:} Across the explored parameter space ($[{\rm M/H}] \simeq 0.78$--1.14 and C/O = 0.55--0.95), \ce{CO2} is the most sensitive tracer of atmospheric composition, exhibiting variations corresponding to $\sim$200 ppm at the morning terminator and $\sim$400 ppm at the evening terminator. \ce{CH4} shows even larger spectral variability, reaching $\sim$650 ppm at the morning terminator and $\sim$850 ppm at the evening terminator; however \ce{CH4} is more sensitive to the deep atmosphere quench temperatures, which differ between the GCM models for the different metallicities. \ce{SO2} and \ce{HCN} concentrations are more sensitive to the C/O ratio than to metallicity, but both species do not show strong spectral signatures in our models.
    \item \textbf{Thermal structure as a driver of chemical asymmetry:} The 3D ExoRad simulations predict large longitudinal and latitudinal temperature gradients (200–400~K), shaped by Rossby gyres and an eastward-shifted hotspot. These thermal contrasts translate directly into chemical inhomogeneities. Quenched species such as \ce{CH4} and \ce{NH3} primarily reflect deep atmospheric conditions and exhibit latitudinal variations. In particular, \ce{CH4} is quenched at pressures of $\sim$0.1–0.4 bar and shows up to $\sim$1 dex latitudinal variation driven by differences in quench-level temperatures, while remaining largely insensitive to upper-atmospheric thermal structure. In contrast, photochemically produced species such as \ce{SO2}, \ce{HCN}, and \ce{C2H2} trace upper-atmospheric temperature profiles and irradiation patterns. \ce{SO2}, forming at pressures of $10^{-5}$–$10^{-1}$ bar, exhibits day–night contrasts spanning up to eight orders of magnitude and closely follows the dayside hotspot offset, however, the role of horizontal mixing has not been taken into account in this work and may impact \ce{SO2} asymmetries. We do not identify any species in our models that robustly trace the nightside gyre-driven circulation patterns.
    \item \textbf{Spectral implications:} Latitudinal chemical asymmetries produce observable variations in transmission spectra, with \ce{CH4} showing variations of up to $\sim$400 ppm. \ce{SO2} features remain too weak to have detectable spectral signatures in our models. \ce{CO2} exhibits moderate spatial variability ($\leq$200 ppm), in comparison to its metallicity-driven variations (200-400 ppm), reinforcing its robustness as a tracer of metallicity. In contrast, \ce{CH4} is sensitive to both thermal asymmetries and compositional variations.       
\end{itemize}

\noindent  This work demonstrates that disequilibrium species do not simply trace atmospheric composition, but also the three-dimensional thermal and dynamical structure of irradiated exoplanet atmospheres. Incorporating such three-dimensional effects will be essential for robustly constraining atmospheric properties from current and future JWST observations of warm, hydrogen-dominated exoplanets.

\begin{acknowledgments}
The GCM simulations (project id 72245) were performed on the Austrian Scientific Computing (ASC) infrastructure, in particular the Vienna Science Cluster (VSC). LC, ChH, PW and H.L.M acknowledge support from the MC ITN EJD CHAMELEON. PW thanks the Austrian Science Fund (FWF) for the support of the VeReDo research project, grant I6857-N. H.L.M. acknowledges the support from the Swiss National Science Foundation under the grant 200021-231596. NB thanks Tereza Constantinou for helpful conversations regarding STAND and ARGO.
\end{acknowledgments}

\begin{contribution}
NB was responsible for the modeling and analysis of atmospheric disequilibrium chemistry, and writing of the manuscript. LC performed the GCM simulations, lead their application and evaluation, and guided and contributed to the paper writing. VS generated the line-of-sight integrated column density maps and contributed to the manuscript text. KGG performed the simulations of the synthetic transmission spectra and contributed to the manuscript text. LF supported the interpretation of the results and commented on the manuscript. H.L.M. and PW reviewed the chemical data used in the models and provided comments on the manuscript. P.B.R. developed and maintains the ARGO code. ChH conceptualized the project, individual project steps and the result presentation, structured and edited the paper, and provided all funding.

\end{contribution}

\appendix

This appendix summarizes the chemical network, transport treatment, and supplementary figures supporting the disequilibrium chemistry analysis presented in the main text.

\section{Disequilibrium chemistry} \label{appendix: argo}
To perform the atmospheric chemistry modeling of WASP-69~b, we employ the 1D photochemical-diffusion code \textsc{ARGO} \citep{Rimmer&Helling,ErratumRimmer_2019}, which was recently coupled with the radiative transfer code CDISORT to calculate the attenuated stellar flux \citep{Bangera2025}. 

The \textsc{ARGO} model solves the 1D continuity equation for each gas species $i$,
\begin{equation}
    \frac{\partial n_i}{\partial t} = P_i - L_i - \frac{\partial \phi_i}{\partial z},
\end{equation}
\noindent
where $n_i$, in cm$^{-3}$, is the number density of species $i$, $P_i$ and $L_i$ are the production and loss rates of species $i$, respectively [cm$^{-3}$s$^{-1}$], and $\frac{\partial \phi_i}{\partial z}$ is the vertical change in flux $\phi_i$ [cm$^{-2}$s$^{-1}$]. The production and loss terms $P_i$ and $L_i$ describe two-body neutral-neutral and ion-neutral reactions, three-body neutral reactions, dissociation reactions, radiative-association reactions, thermal ionization, recombination reactions, and photochemical reactions described in the chemical network STAND2025 which is an extension of STAND2020 \citep{Hobbs2021}. STAND2020 included all H/C/N/O/S reactions for species containing up to six H, two C, two N, three O atoms, and 30 species containing sulfur. Bangera et al. (in prep) describes the extensions made to STAND2020 to include ionic sulfur species and create STAND2025. 

In \textsc{ARGO}, vertical transport is assumed to occur via both eddy ($K_{\rm{zz}}$ [cm$^{2}$s$^{-1}$]) and molecular ($D_{\rm{zz}}$ [cm$^{2}$s$^{-1}$]) diffusion,
\begin{equation}
\begin{aligned}
    \phi_i & = - K_{\rm{zz}} \left[ \frac{\partial n_i}{\partial z} + n_i \left( \frac{1}{H_0} + \frac{1}{T_{gas}} \frac{dT_{gas}}{dz} \right) \right] \\ 
    & - D_{\rm{zz}}\left[ \frac{\partial n_i}{\partial z} + n_i \left( \frac{1}{H_i} + \frac{1 + \alpha_T}{T_{gas}}\frac{dT_{gas}}{dz} \right) \right],
\end{aligned}
\end{equation}
\noindent
where $H_0$ is the atmospheric pressure scale height at vertical height $z$, $H_i$ is the atmospheric scale height for species $i$, $T$ is the temperature, and $\alpha_T$ the thermal diffusion coefficient \citep{Banks1973,Zahnle2006}. The molecular diffusion coefficients in ARGO are adopted from Chapman-Enskog theory \citep{Enskog1917,Chapman1991}. 

The network is then applied to WASP-69~b—an important target lying near the proposed SO$_2$ “shoreline”. Here the parametrization for eddy diffusion coefficient of \cite{Moses2022} for exo-Neptune exoplanets is employed, as was done in our previous work \citep{Bangera2025}:

\begin{equation}
    K_{zz}~[\mathrm{cm^2}\mathrm{s}^{-1} ] = 5 \times 10^8 \; [p_{gas}(\mathrm{bar})]^{-0.5} \; \left( \frac{H_{1\mathrm{mbar}}}{620\mathrm{km}}\right)\left(\frac{T_\mathrm{eff}}{1450 K}\right)
\end{equation}
where  K$_{\mathrm{zz}}$ is the eddy diffusion coefficient, $p_{\rm{gas}}$ is the atmospheric pressure, and H$_{\mathrm{1 mbar}}$ is
the atmospheric pressure scale height at 1 mbar. This formulation reproduces the GCM-derived vertical profile of $K_{\mathrm{zz}}$ for HD 209458~b inferred from tracer transport studies \citep{Parmentier2013}. For the adopted atmospheric parameters of WASP-69~b ($T_{\rm{eff}} = 915$~K and $H_{\rm{1mbar}} = 573$~km), the resulting scaling factor is 0.57, yielding  $K_{zz} [\mathrm{cm^2}\mathrm{s}^{-1} ] = 2.8 \times 10^8 \; [P(\mathrm{bar})]^{-0.5}$. Following \cite{Moses2022}, we restrict K$_{\mathrm{zz}}$ from exceeding 10$^{11}$ cm$^2$s$^{-1}$ in the upper atmosphere.

The incident stellar flux for WASP-69~b is adopted from \citet{Bangera2025}, which combines a PHOENIX model in the UV, optical, and IR \citep{Husser2013} with a scaled solar flux in the UV regime \citep{Claire2012}. The effect of cosmic rays is not considered.

\section{Supplementary chemistry and spectral diagnostics} \label{appendix: molecules}

\begin{figure}
    \centering
    \includegraphics[width=0.65\linewidth]{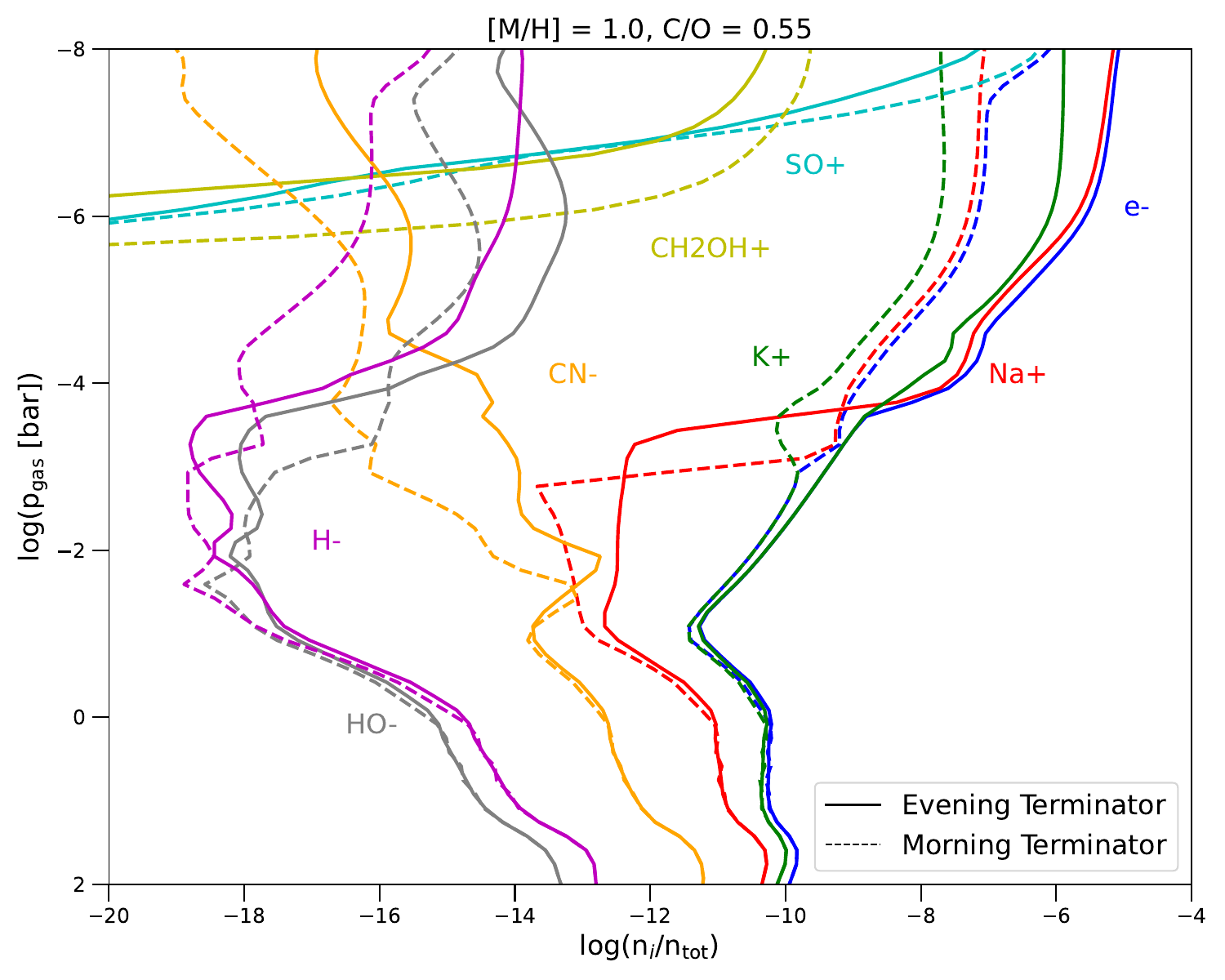}
    \caption{Concentrations of the most abundant ions in the atmosphere of WASP-69~b at both terminators for the base model.}
    \label{fig:molecules_and_ions}
\end{figure}

While ion-neutral reactions are included in STAND2025, ion abundances remain too low to significantly modify the neutral chemistry for the WASP-69 b parameter space explored in this work. We therefore do not discuss ion chemistry in the main text. Figure~\ref{fig:molecules_and_ions} is included to document the predicted ion distributions and to facilitate comparison with future studies.

The ionization fraction remains low throughout most of the atmosphere, increasing from $\sim10^{-10}$ in the lower atmosphere to $\sim$ 10$^{-5}$ in the upper atmosphere due to photochemical ionization. The dominant cations are \ce{Na+} and \ce{K+}, with smaller contributions from \ce{SO+} and \ce{CH2OH+} at low pressures. The dominant anions, including \ce{CN-}, \ce{OH-}, and \ce{H-}, remain below $\sim$ 10$^{-10}$ throughout the modeled atmosphere.

\begin{figure}
    \centering
    \includegraphics[width=0.49\linewidth]{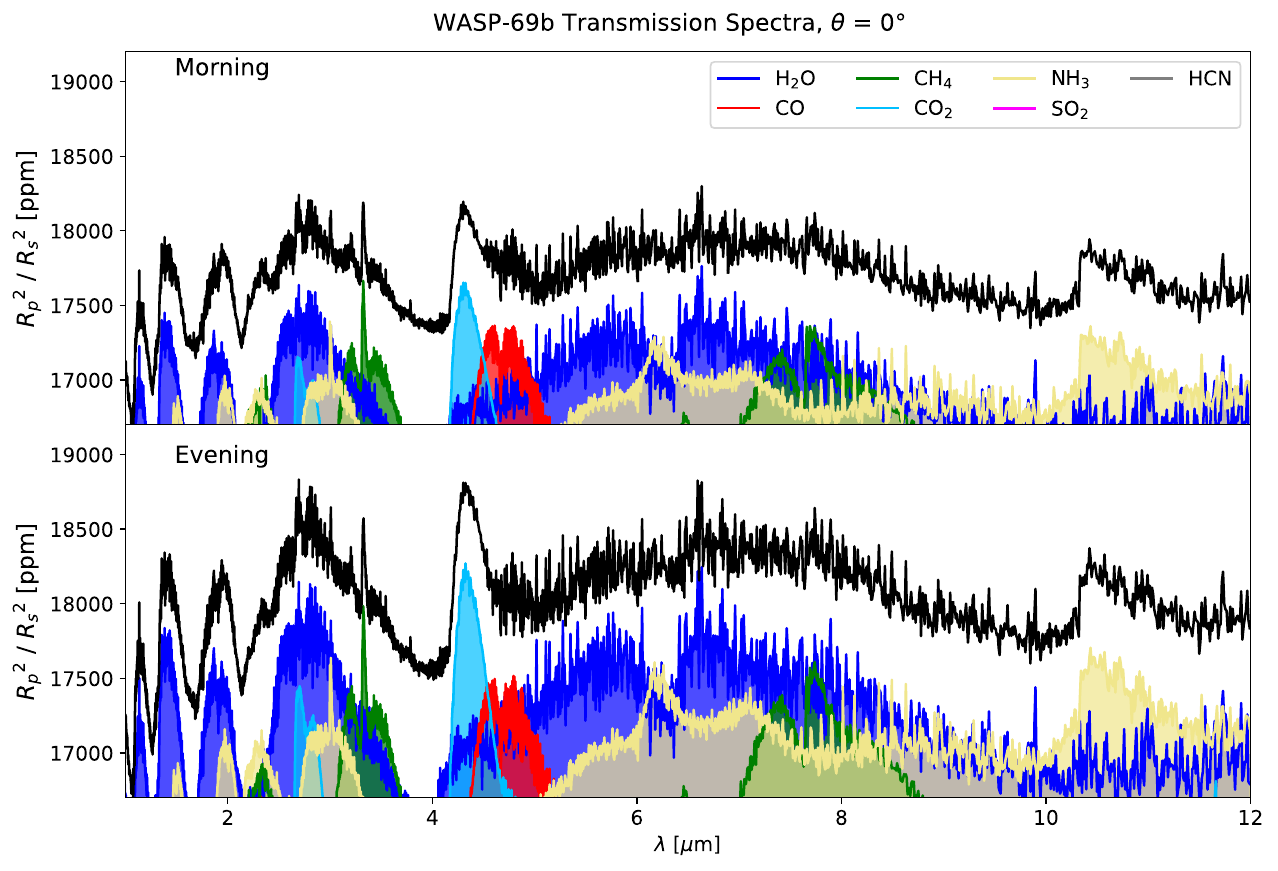}
    \includegraphics[width=0.42\linewidth]{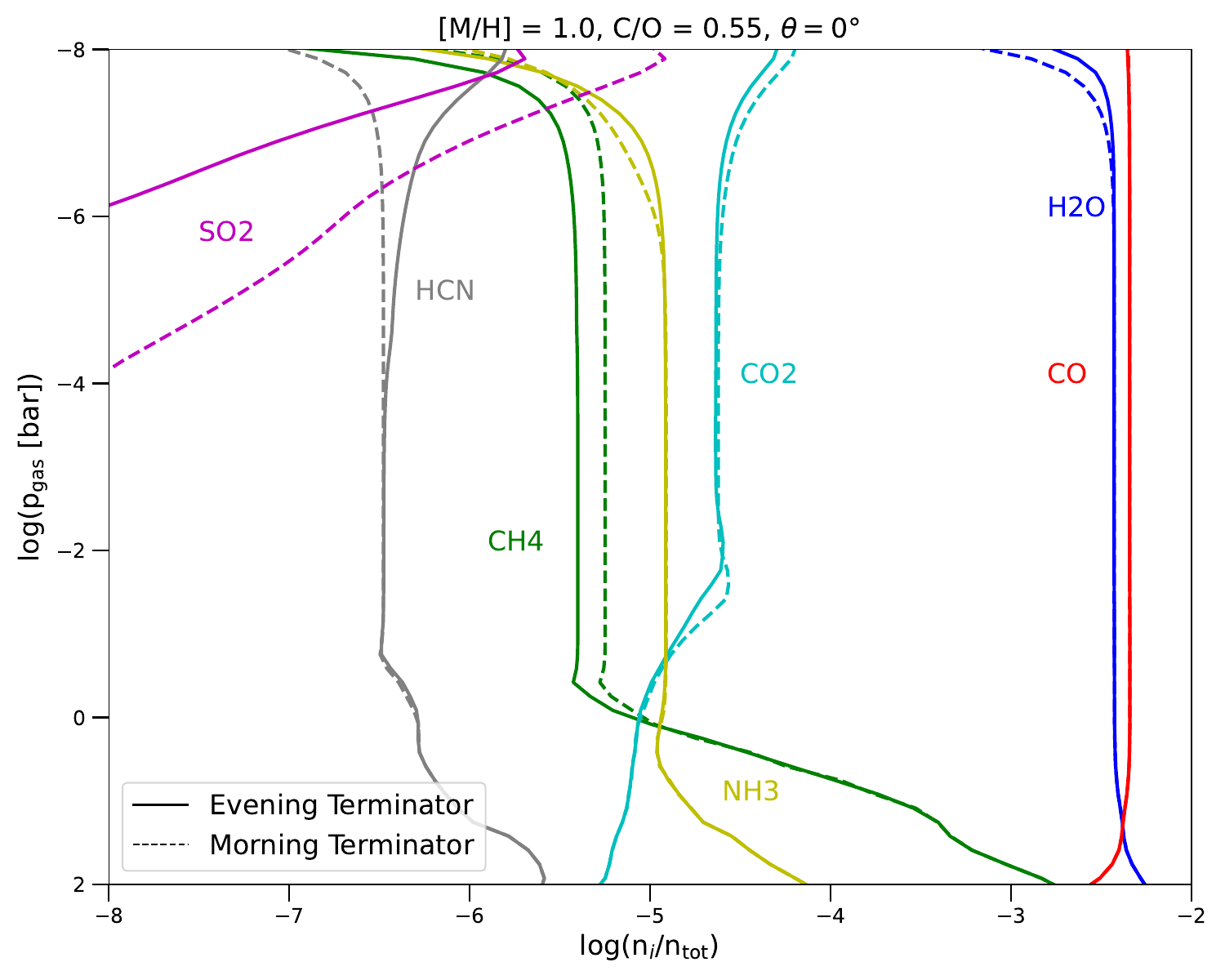}
    \includegraphics[width=0.49\linewidth]{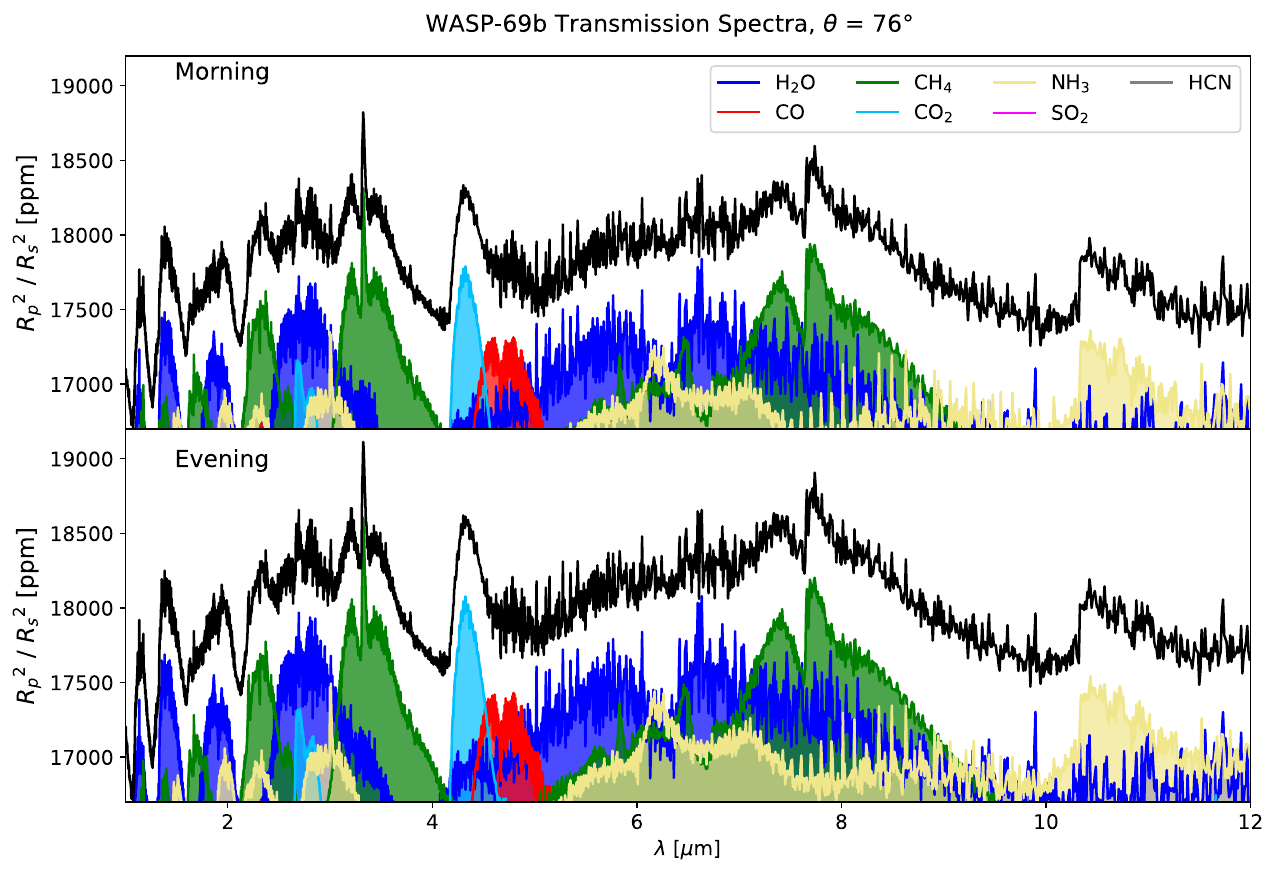}
    \includegraphics[width=0.42\linewidth]{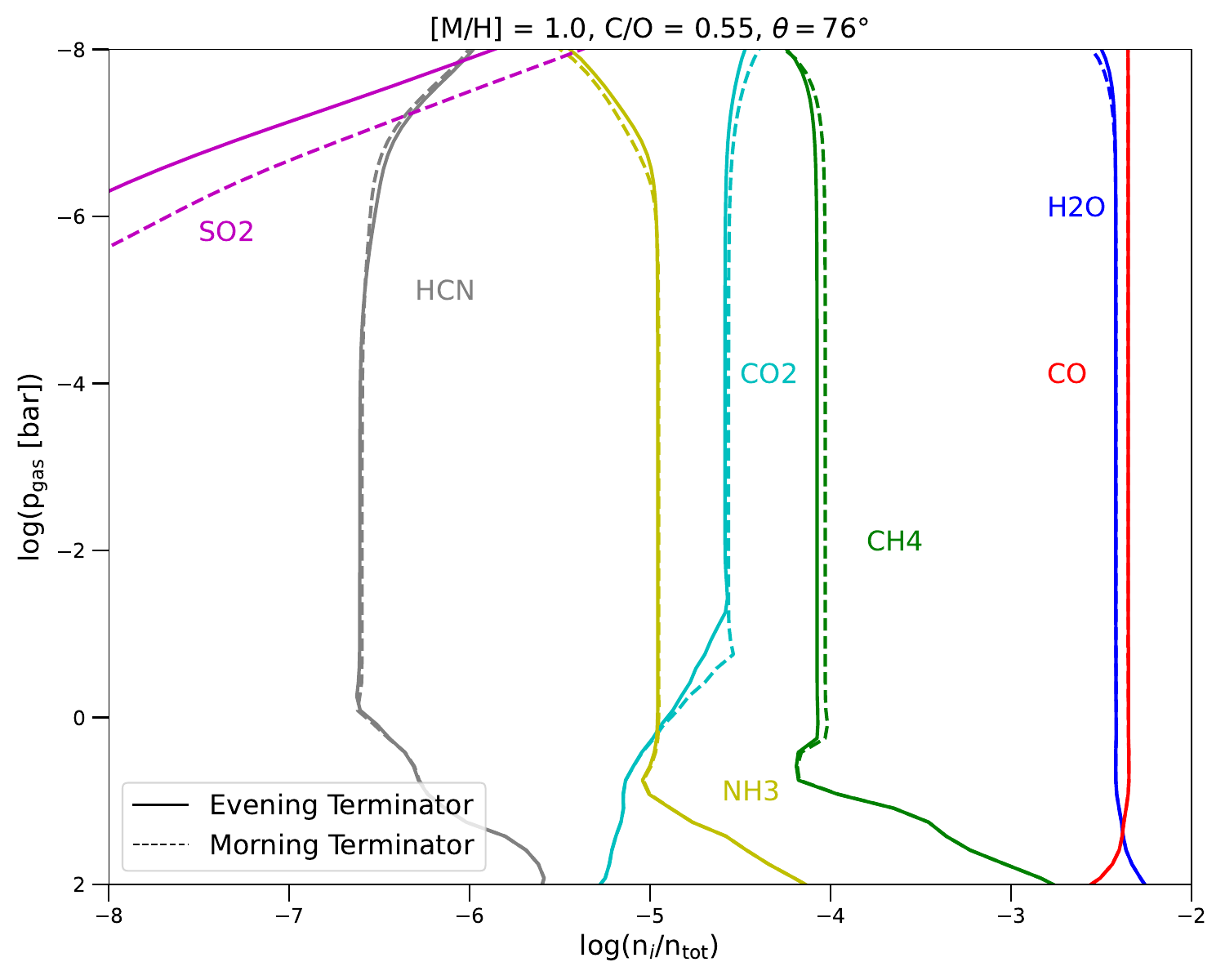}
    \caption{Synthetic transmission spectra of the WASP-69~b baseline model at equatorial ($\theta = 0^\circ$; top) and polar ($\theta = 76^\circ$; bottom) latitudes for the morning and evening terminators, shown together with the concentrations of the dominant spectroscopically active molecules. Here, the transmission spectra are generated at $R \approx 1000$ to highlight the contributions to the spectral features.}
    \label{fig:spectroscopically_active}
\end{figure}

Figure~\ref{fig:spectroscopically_active} compares the dominant spectroscopically active molecules at equatorial ($\theta = 0^\circ$) and polar ($\theta = 76^\circ$) latitudes for both terminators, together with the resulting transmission spectra. Here, the transmission spectra are generated at a resolution of $R \approx 1000$ in order to more closely match the expected resolution of JWST NIRSpec observations. At equatorial latitudes, the morning and evening spectra are broadly similar because the molecular abundances remain comparable between the two limbs. In contrast, stronger spatial asymmetries emerge at polar latitudes, where cooler temperatures enhance the abundances of species such as \ce{CH4}. As a result, the CH$_4$ absorption features become more prominent at high latitudes, demonstrating that the polar regions can contribute significantly to the limb-averaged transmission spectrum.

\begin{figure}
   \centering
    \includegraphics[width=0.65\linewidth]{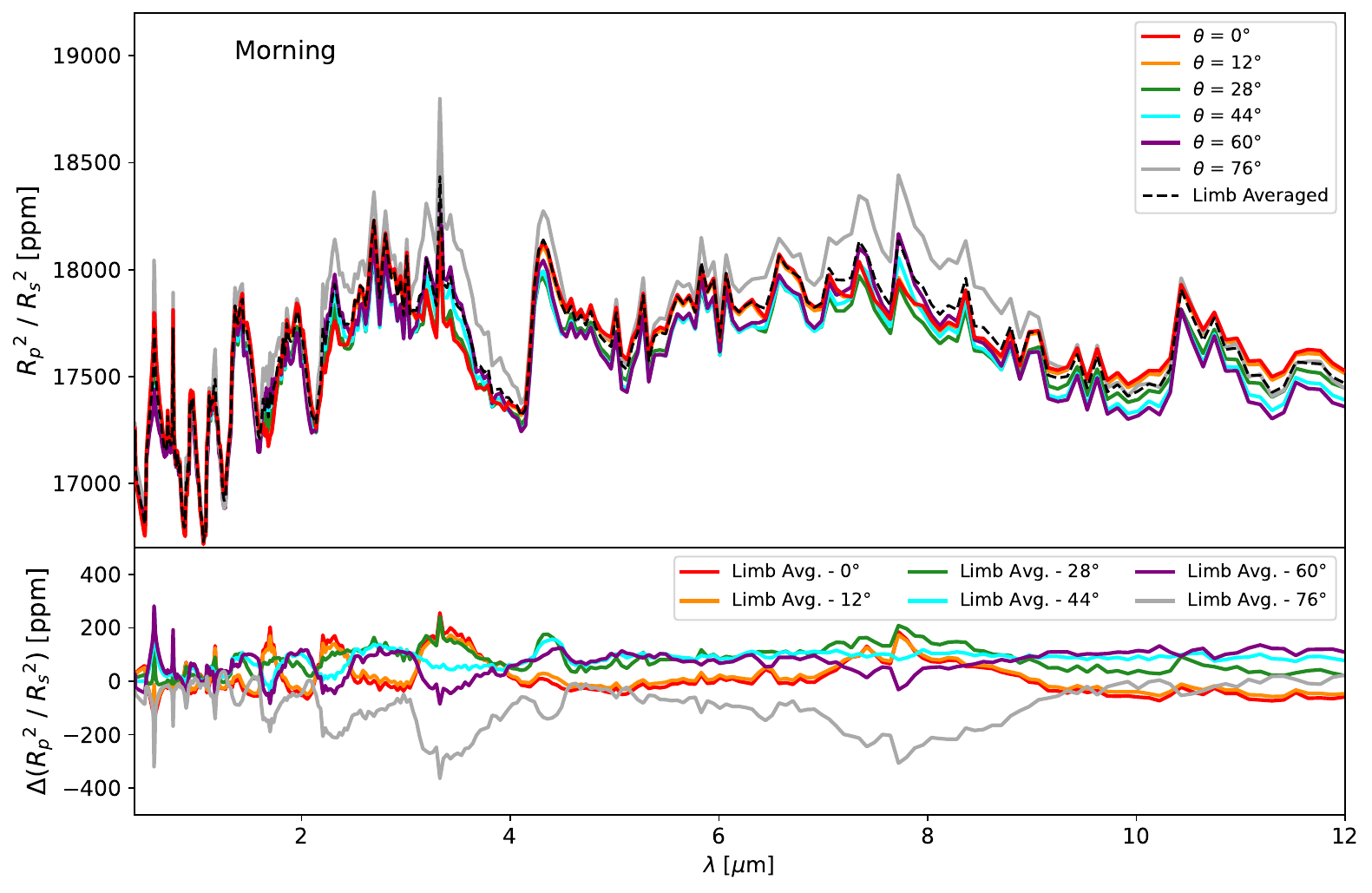}
    \includegraphics[width=0.65\linewidth]{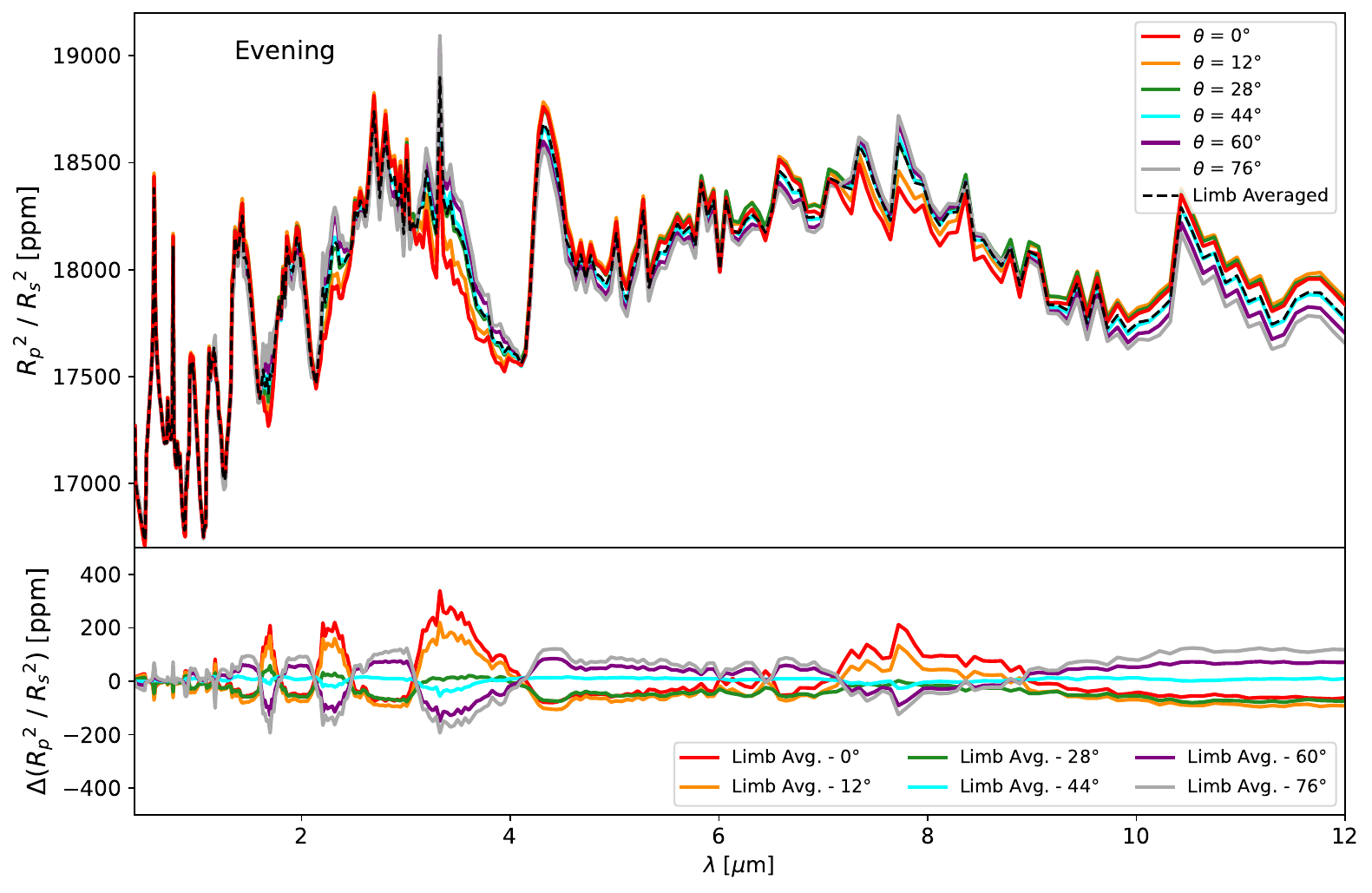}
    \caption{Synthetic transmission spectra of the WASP-69 b baseline model for the full PLATO-HST-JWST wavelength range at the morning (\textbf{top}) and evening (\textbf{bottom}) terminators across multiple latitudes, with the limb-averaged spectra overlaid as black dashed lines. Residual spectra of each latitude relative to the limb-averaged model are also provided in the bottom panels for both terminators.}
    \label{fig:fullHST-JWSTspectra}
\end{figure}

Figure~\ref{fig:fullHST-JWSTspectra} shows the synthetic transmission spectra over the full PLATO-HST–JWST wavelength range (0.4 – 12~$\mu$m) at the morning (left) and evening (right) terminators for the baseline WASP-69~b model across multiple latitudes. The largest latitudinal variations occur between $\sim$2–4~$\mu$m and $\sim$7–9~$\mu$m, where the spectra are dominated by \ce{CH4} absorption.

The resulting latitudinal variations in the \ce{CH4} feature near 3.3~$\mu$m reach amplitudes of $\sim$400 ppm at the morning terminator and $\sim$350 ppm at the evening terminator. In contrast, features from \ce{SO2}, \ce{HCN}, and \ce{C2H2} remain weak and undetectable throughout the spectra because of the relatively low terminator abundances for these species ($n_{\ce{X}}/n_{\mathrm{tot}} \sim 10^{-8}$, $\sim 10^{-7}$, and $\sim 10^{-10}$, respectively; see Figs.~\ref{fig:latitudinal assymetries} and \ref{fig:Gyres_plus_Chemistry}). The latitudinal variations in the \ce{CO2} feature near 4.3~$\mu$m remain $\lesssim$~200 ppm at both terminators. Finally, \ce{NH3} contributes strongly to the spectra at longer wavelengths ($\lambda > 10$ $\mu$m) but with latitudinal variations that do not exceed $\sim$150 ppm at either terminator. This comparatively weak spatial dependence reflects the similar \ce{NH3} abundances at the deep pressure levels where the species is chemically quenched.

\section{Supplementary atmospheric structure figures}

\begin{figure}
   \centering
    \includegraphics[width=0.45\linewidth]{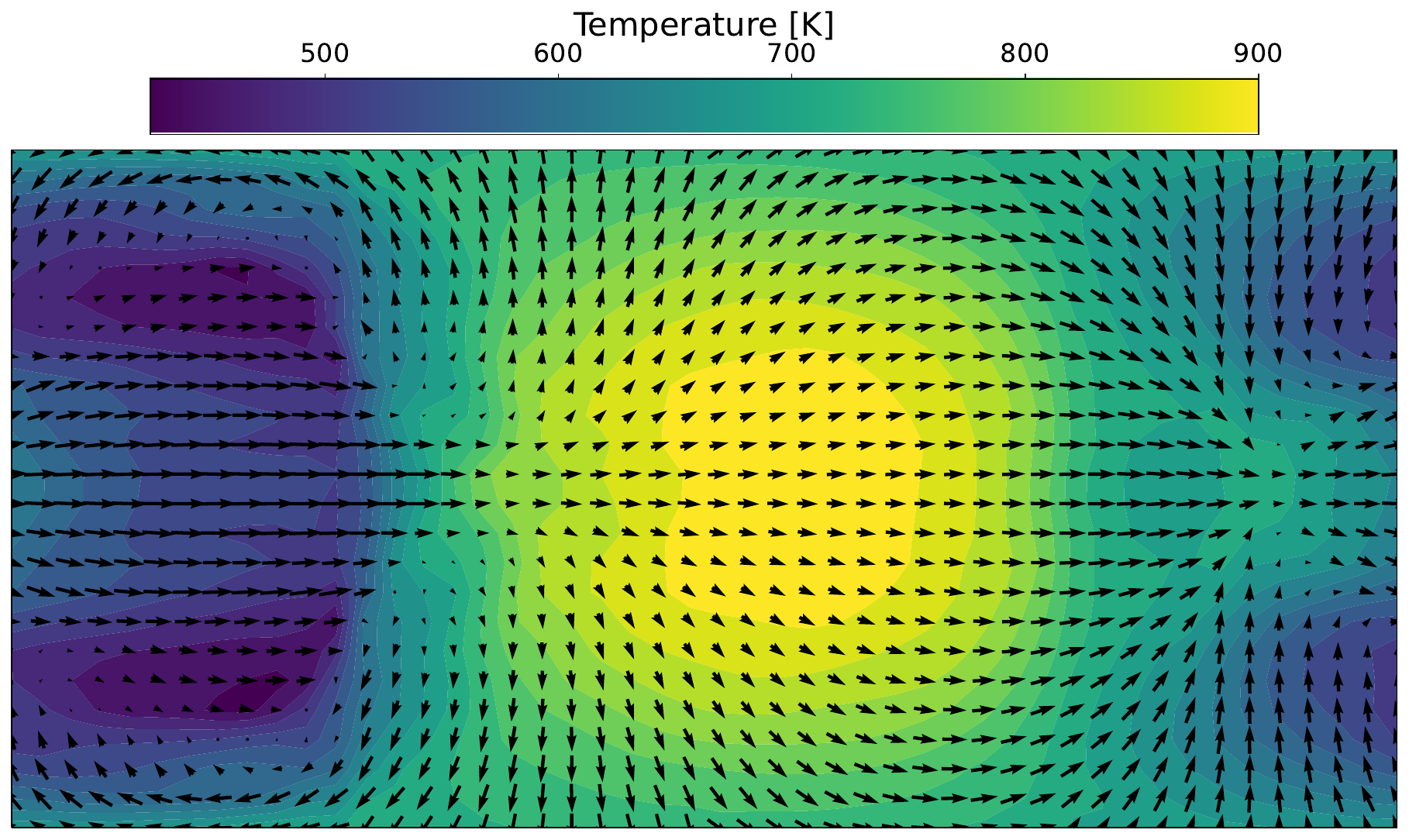}
    \includegraphics[width=0.45\linewidth]{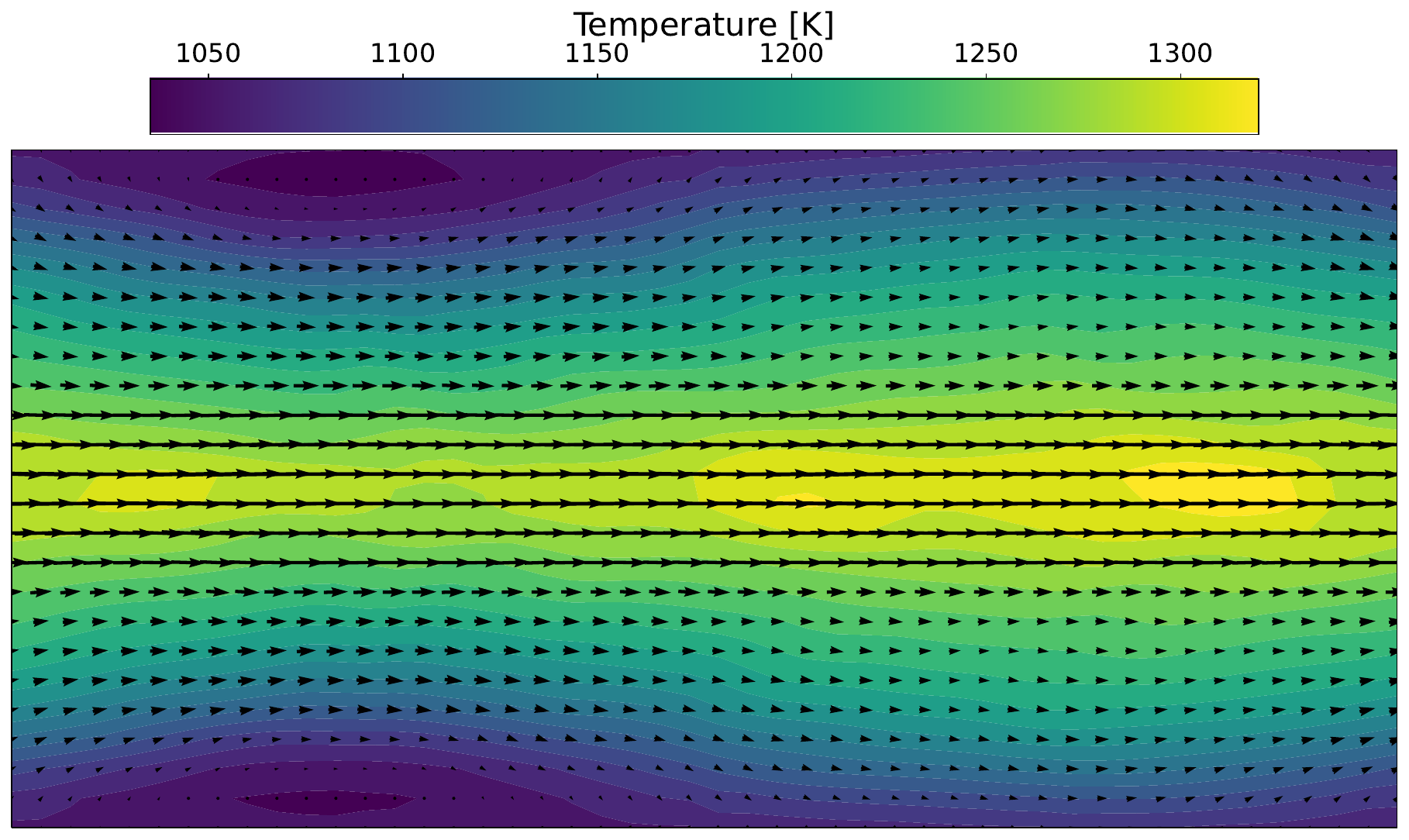}
    \caption{The spatial temperature variations responsible for the WASP-69~b latitudinal and longitudinal asymmetries and the direction of wind flow. Left panel: Wind flow and 3D temperature distribution are shown at $p=1.2\times 10^{-4}$~bar that shape the photochemically produced \ce{SO2} concentrations. Right panel: Wind flow and 3D temperature distribution are shown at $p=2.5\times 10^{-1}$~bar that shape \ce{CH4} concentrations via kinetic chemistry. } 
    \label{fig:gyres}
\end{figure}

Figure \ref{fig:gyres} shows the horizontal wind vectors at two representative pressure levels corresponding to (i) a representative point within the CH$_4$ quench region ($\sim$0.25 bar) and (ii) the SO$_2$ production region ($\sim10^{-4}$ bar). At the CH$_4$ quench level, the flow is dominated by equatorial superrotation. At the SO$_2$ formation level, the circulation becomes more strongly day–night driven, with large‐scale Rossby gyres and a hotspot offset. The wind maps therefore provide physical context for the longitudinal and latitudinal chemical asymmetries shown in Section \ref{sec: Asymmetry} and for the predicted differences in CH$_4$ and SO$_2$ column densities.

\bibliography{biblio}{}

@ARTICLE{Rimmer&Helling,
       author = {{Rimmer}, P.~B. and {Helling}, Ch},
        title = "{A Chemical Kinetics Network for Lightning and Life in Planetary Atmospheres}",
      journal = {\apjs},
         year = 2016,
        month = may,
       volume = {224},
       number = {1},
          eid = {9},
        pages = {9},
          doi = {10.3847/0067-0049/224/1/9},
archivePrefix = {arXiv},
       eprint = {1510.07052},
 primaryClass = {astro-ph.EP},
       adsurl = {https://ui.adsabs.harvard.edu/abs/2016ApJS..224....9R}
}

@article{ErratumRimmer_2019,
doi = {10.3847/1538-4365/ab5192},
url = {https://dx.doi.org/10.3847/1538-4365/ab5192},
year = {2019},
month = {nov},
publisher = {The American Astronomical Society},
volume = {245},
number = {1},
pages = {20},
author = {Paul B. Rimmer and Christiane Helling},
title = {Erratum: “A Chemical Kinetics Network for Lightning and Life in Planetary Atmospheres” (2019, ApJS, 224, 9)},
journal = {The Astrophysical Journal Supplement Series}
}

@BOOK{Banks1973,
       author = {{Banks}, P.~M. and {Kockarts}, G.},
        title = "{Aeronomy.}",
         year = 1973,
       adsurl = {https://ui.adsabs.harvard.edu/abs/1973aero.book.....B}
}

@ARTICLE{Zahnle2006,
       author = {{Zahnle}, K. and {Claire}, M. and {Catling}, D.},
        title = "{The loss of mass‑independent fractionation in sulfur due to a Palaeoproterozoic collapse of atmospheric methane}",
      journal = {Geobiology},
         year = 2006,
        month = dec,
       volume = {4},
       number = {4},
        pages = {271-283},
          doi = {10.1111/j.1472-4669.2006.00085.x},
       adsurl = {https://ui.adsabs.harvard.edu/abs/2006Gbio....4..271Z}
}

@BOOK{Enskog1917,
       author = {{Enskog}, David},
        title = "{Kinetische Theorie der Vorgaenge in maessig verduennten Gasen. I. Allgemeiner Teil}",
         year = 1917,
       adsurl = {https://ui.adsabs.harvard.edu/abs/1917ktvm.book.....E}
}

@BOOK{Chapman1991,
       author = {{Chapman}, Sydney and {Cowling}, T.~G.},
        title = "{The Mathematical Theory of Non-uniform Gases}",
         year = 1991,
       adsurl = {https://ui.adsabs.harvard.edu/abs/1991mtnu.book.....C}
}

@ARTICLE{Anderson2014,
       author = {{Anderson}, D.~R. and {Collier Cameron}, A. and {Delrez}, L. and {Doyle}, A.~P. and {Faedi}, F. and {Fumel}, A. and {Gillon}, M. and {G{\'o}mez Maqueo Chew}, Y. and {Hellier}, C. and {Jehin}, E. and {Lendl}, M. and {Maxted}, P.~F.~L. and {Pepe}, F. and {Pollacco}, D. and {Queloz}, D. and {S{\'e}gransan}, D. and {Skillen}, I. and {Smalley}, B. and {Smith}, A.~M.~S. and {Southworth}, J. and {Triaud}, A.~H.~M.~J. and {Turner}, O.~D. and {Udry}, S. and {West}, R.~G.},
        title = "{Three newly discovered sub-Jupiter-mass planets: WASP-69b and WASP-84b transit active K dwarfs and WASP-70Ab transits the evolved primary of a G4+K3 binary}",
      journal = {\mnras},
         year = 2014,
        month = dec,
       volume = {445},
       number = {2},
        pages = {1114-1129},
          doi = {10.1093/mnras/stu1737},
archivePrefix = {arXiv},
       eprint = {1310.5654},
 primaryClass = {astro-ph.EP},
       adsurl = {https://ui.adsabs.harvard.edu/abs/2014MNRAS.445.1114A}
}

@ARTICLE{Hobbs2021,
       author = {{Hobbs}, Richard and {Rimmer}, Paul B. and {Shorttle}, Oliver and {Madhusudhan}, Nikku},
        title = "{Sulfur chemistry in the atmospheres of warm and hot Jupiters}",
      journal = {\mnras},
         year = 2021,
        month = sep,
       volume = {506},
       number = {3},
        pages = {3186-3204},
          doi = {10.1093/mnras/stab1839},
archivePrefix = {arXiv},
       eprint = {2101.08327},
 primaryClass = {astro-ph.EP},
       adsurl = {https://ui.adsabs.harvard.edu/abs/2021MNRAS.506.3186H}
}

@ARTICLE{Woitke2018,
       author = {{Woitke}, P. and {Helling}, Ch. and {Hunter}, G.~H. and {Millard}, J.~D. and {Turner}, G.~E. and {Worters}, M. and {Blecic}, J. and {Stock}, J.~W.},
        title = "{Equilibrium chemistry down to 100 K. Impact of silicates and phyllosilicates on the carbon to oxygen ratio}",
      journal = {\aap},
         year = 2018,
        month = jun,
       volume = {614},
          eid = {A1},
        pages = {A1},
          doi = {10.1051/0004-6361/201732193},
archivePrefix = {arXiv},
       eprint = {1712.01010},
 primaryClass = {astro-ph.EP},
       adsurl = {https://ui.adsabs.harvard.edu/abs/2018A&A...614A...1W}
}

@ARTICLE{Tsai2023,
       author = {{Tsai}, Shang-Min and {Lee}, Elspeth K.~H. and {Powell}, Diana and {Gao}, Peter and {Zhang}, Xi and {Moses}, Julianne and {H{\'e}brard}, Eric and {Venot}, Olivia and {Parmentier}, Vivien and {Jordan}, Sean and {Hu}, Renyu and {Alam}, Munazza K. and {Alderson}, Lili and {Batalha}, Natalie M. and {Bean}, Jacob L. and {Benneke}, Bj{\"o}rn and {Bierson}, Carver J. and {Brady}, Ryan P. and {Carone}, Ludmila and {Carter}, Aarynn L. and {Chubb}, Katy L. and {Inglis}, Julie and {Leconte}, J{\'e}r{\'e}my and {Line}, Michael and {L{\'o}pez-Morales}, Mercedes and {Miguel}, Yamila and {Molaverdikhani}, Karan and {Rustamkulov}, Zafar and {Sing}, David K. and {Stevenson}, Kevin B. and {Wakeford}, Hannah R. and {Yang}, Jeehyun and {Aggarwal}, Keshav and {Baeyens}, Robin and {Barat}, Saugata and {de Val-Borro}, Miguel and {Daylan}, Tansu and {Fortney}, Jonathan J. and {France}, Kevin and {Goyal}, Jayesh M. and {Grant}, David and {Kirk}, James and {Kreidberg}, Laura and {Louca}, Amy and {Moran}, Sarah E. and {Mukherjee}, Sagnick and {Nasedkin}, Evert and {Ohno}, Kazumasa and {Rackham}, Benjamin V. and {Redfield}, Seth and {Taylor}, Jake and {Tremblin}, Pascal and {Visscher}, Channon and {Wallack}, Nicole L. and {Welbanks}, Luis and {Youngblood}, Allison and {Ahrer}, Eva-Maria and {Batalha}, Natasha E. and {Behr}, Patrick and {Berta-Thompson}, Zachory K. and {Blecic}, Jasmina and {Casewell}, S.~L. and {Crossfield}, Ian J.~M. and {Crouzet}, Nicolas and {Cubillos}, Patricio E. and {Decin}, Leen and {D{\'e}sert}, Jean-Michel and {Feinstein}, Adina D. and {Gibson}, Neale P. and {Harrington}, Joseph and {Heng}, Kevin and {Henning}, Thomas and {Kempton}, Eliza M. -R. and {Krick}, Jessica and {Lagage}, Pierre-Olivier and {Lendl}, Monika and {Lothringer}, Joshua D. and {Mansfield}, Megan and {Mayne}, N.~J. and {Mikal-Evans}, Thomas and {Palle}, Enric and {Schlawin}, Everett and {Shorttle}, Oliver and {Wheatley}, Peter J. and {Yurchenko}, Sergei N.},
        title = "{Photochemically produced SO$_{2}$ in the atmosphere of WASP-39b}",
      journal = {\nat},
         year = 2023,
        month = may,
       volume = {617},
       number = {7961},
        pages = {483-487},
          doi = {10.1038/s41586-023-05902-2},
archivePrefix = {arXiv},
       eprint = {2211.10490},
 primaryClass = {astro-ph.EP},
       adsurl = {https://ui.adsabs.harvard.edu/abs/2023Natur.617..483T}
}

@ARTICLE{Faedi2011,
       author = {{Faedi}, F. and {Barros}, S.~C.~C. and {Anderson}, D.~R. and {Brown}, D.~J.~A. and {Collier Cameron}, A. and {Pollacco}, D. and {Boisse}, I. and {H{\'e}brard}, G. and {Lendl}, M. and {Lister}, T.~A. and {Smalley}, B. and {Street}, R.~A. and {Triaud}, A.~H.~M.~J. and {Bento}, J. and {Bouchy}, F. and {Butters}, O.~W. and {Enoch}, B. and {Haswell}, C.~A. and {Hellier}, C. and {Keenan}, F.~P. and {Miller}, G.~R.~M. and {Moulds}, V. and {Moutou}, C. and {Norton}, A.~J. and {Queloz}, D. and {Santerne}, A. and {Simpson}, E.~K. and {Skillen}, I. and {Smith}, A.~M.~S. and {Udry}, S. and {Watson}, C.~A. and {West}, R.~G. and {Wheatley}, P.~J.},
        title = "{WASP-39b: a highly inflated Saturn-mass planet orbiting a late G-type star}",
      journal = {\aap},
         year = 2011,
        month = jul,
       volume = {531},
          eid = {A40},
        pages = {A40},
          doi = {10.1051/0004-6361/201116671},
archivePrefix = {arXiv},
       eprint = {1102.1375},
 primaryClass = {astro-ph.EP},
       adsurl = {https://ui.adsabs.harvard.edu/abs/2011A&A...531A..40F}
}

@ARTICLE{Carone2020,
       author = {{Carone}, Ludmila and {Baeyens}, Robin and {Molli{\`e}re}, Paul and {Barth}, Patrick and {Vazan}, Allona and {Decin}, Leen and {Sarkis}, Paula and {Venot}, Olivia and {Henning}, Thomas},
        title = "{Equatorial retrograde flow in WASP-43b elicited by deep wind jets?}",
      journal = {\mnras},
         year = 2020,
        month = aug,
       volume = {496},
       number = {3},
        pages = {3582-3614},
          doi = {10.1093/mnras/staa1733},
archivePrefix = {arXiv},
       eprint = {1904.13334},
 primaryClass = {astro-ph.EP},
       adsurl = {https://ui.adsabs.harvard.edu/abs/2020MNRAS.496.3582C}
}

@ARTICLE{Baeyens2021,
       author = {{Baeyens}, Robin and {Decin}, Leen and {Carone}, Ludmila and {Venot}, Olivia and {Ag{\'u}ndez}, Marcelino and {Molli{\`e}re}, Paul},
        title = "{Grid of pseudo-2D chemistry models for tidally locked exoplanets - I. The role of vertical and horizontal mixing}",
      journal = {\mnras},
         year = 2021,
        month = aug,
       volume = {505},
       number = {4},
        pages = {5603-5653},
          doi = {10.1093/mnras/stab1310},
archivePrefix = {arXiv},
       eprint = {2105.02245},
 primaryClass = {astro-ph.EP},
       adsurl = {https://ui.adsabs.harvard.edu/abs/2021MNRAS.505.5603B}
}

@ARTICLE{Schlawin2024,
       author = {{Schlawin}, Everett and {Mukherjee}, Sagnick and {Ohno}, Kazumasa and {Bell}, Taylor J. and {Beatty}, Thomas G. and {Greene}, Thomas P. and {Line}, Michael and {Challener}, Ryan C. and {Parmentier}, Vivien and {Fortney}, Jonathan J. and {Rauscher}, Emily and {Wiser}, Lindsey and {Welbanks}, Luis and {Murphy}, Matthew and {Edelman}, Isaac and {Batalha}, Natasha and {Moran}, Sarah E. and {Mehta}, Nishil and {Rieke}, Marcia},
        title = "{Multiple Clues for Dayside Aerosols and Temperature Gradients in WASP-69 b from a Panchromatic JWST Emission Spectrum}",
      journal = {\aj},
         year = 2024,
        month = sep,
       volume = {168},
       number = {3},
          eid = {104},
        pages = {104},
          doi = {10.3847/1538-3881/ad58e0},
archivePrefix = {arXiv},
       eprint = {2406.15543},
 primaryClass = {astro-ph.EP},
       adsurl = {https://ui.adsabs.harvard.edu/abs/2024AJ....168..104S}
}

@ARTICLE{Madhusudhan2009,
       author = {{Madhusudhan}, N. and {Seager}, S.},
        title = "{A Temperature and Abundance Retrieval Method for Exoplanet Atmospheres}",
      journal = {\apj},
         year = 2009,
        month = dec,
       volume = {707},
       number = {1},
        pages = {24-39},
          doi = {10.1088/0004-637X/707/1/24},
archivePrefix = {arXiv},
       eprint = {0910.1347},
 primaryClass = {astro-ph.EP},
       adsurl = {https://ui.adsabs.harvard.edu/abs/2009ApJ...707...24M}
}

@ARTICLE{Dyrek2024,
       author = {{Dyrek}, Achr{\`e}ne and {Min}, Michiel and {Decin}, Leen and {Bouwman}, Jeroen and {Crouzet}, Nicolas and {Molli{\`e}re}, Paul and {Lagage}, Pierre-Olivier and {Konings}, Thomas and {Tremblin}, Pascal and {G{\"u}del}, Manuel and {Pye}, John and {Waters}, Rens and {Henning}, Thomas and {Vandenbussche}, Bart and {Ardevol Martinez}, Francisco and {Argyriou}, Ioannis and {Ducrot}, Elsa and {Heinke}, Linus and {van Looveren}, Gwenael and {Absil}, Olivier and {Barrado}, David and {Baudoz}, Pierre and {Boccaletti}, Anthony and {Cossou}, Christophe and {Coulais}, Alain and {Edwards}, Billy and {Gastaud}, Ren{\'e} and {Glasse}, Alistair and {Glauser}, Adrian and {Greene}, Thomas P. and {Kendrew}, Sarah and {Krause}, Oliver and {Lahuis}, Fred and {Mueller}, Michael and {Olofsson}, Goran and {Patapis}, Polychronis and {Rouan}, Daniel and {Royer}, Pierre and {Scheithauer}, Silvia and {Waldmann}, Ingo and {Whiteford}, Niall and {Colina}, Luis and {van Dishoeck}, Ewine F. and {{\"O}stlin}, G{\"o}ran and {Ray}, Tom P. and {Wright}, Gillian},
        title = "{SO$_{2}$, silicate clouds, but no CH$_{4}$ detected in a warm Neptune}",
      journal = {\nat},
         year = 2024,
        month = jan,
       volume = {625},
       number = {7993},
        pages = {51-54},
          doi = {10.1038/s41586-023-06849-0},
archivePrefix = {arXiv},
       eprint = {2311.12515},
 primaryClass = {astro-ph.EP},
       adsurl = {https://ui.adsabs.harvard.edu/abs/2024Natur.625...51D}
}

@ARTICLE{Turrini2021,
       author = {{Turrini}, D. and {Schisano}, E. and {Fonte}, S. and {Molinari}, S. and {Politi}, R. and {Fedele}, D. and {Pani{\'c}}, O. and {Kama}, M. and {Changeat}, Q. and {Tinetti}, G.},
        title = "{Tracing the Formation History of Giant Planets in Protoplanetary Disks with Carbon, Oxygen, Nitrogen, and Sulfur}",
      journal = {\apj},
         year = 2021,
        month = mar,
       volume = {909},
       number = {1},
          eid = {40},
        pages = {40},
          doi = {10.3847/1538-4357/abd6e5},
archivePrefix = {arXiv},
       eprint = {2012.14315},
 primaryClass = {astro-ph.EP},
       adsurl = {https://ui.adsabs.harvard.edu/abs/2021ApJ...909...40T}
}

@ARTICLE{Pacetti2022,
       author = {{Pacetti}, Elenia and {Turrini}, Diego and {Schisano}, Eugenio and {Molinari}, Sergio and {Fonte}, Sergio and {Politi}, Romolo and {Hennebelle}, Patrick and {Klessen}, Ralf and {Testi}, Leonardo and {Lebreuilly}, Ugo},
        title = "{Chemical Diversity in Protoplanetary Disks and Its Impact on the Formation History of Giant Planets}",
      journal = {\apj},
         year = 2022,
        month = sep,
       volume = {937},
       number = {1},
          eid = {36},
        pages = {36},
          doi = {10.3847/1538-4357/ac8b11},
archivePrefix = {arXiv},
       eprint = {2206.14685},
 primaryClass = {astro-ph.EP},
       adsurl = {https://ui.adsabs.harvard.edu/abs/2022ApJ...937...36P}
}

@ARTICLE{Guilluy2022,
       author = {{Guilluy}, G. and {Giacobbe}, P. and {Carleo}, I. and {Cubillos}, P.~E. and {Sozzetti}, A. and {Bonomo}, A.~S. and {Brogi}, M. and {Gandhi}, S. and {Fossati}, L. and {Nascimbeni}, V. and {Turrini}, D. and {Schisano}, E. and {Borsa}, F. and {Lanza}, A.~F. and {Mancini}, L. and {Maggio}, A. and {Malavolta}, L. and {Micela}, G. and {Pino}, L. and {Rainer}, M. and {Bignamini}, A. and {Claudi}, R. and {Cosentino}, R. and {Covino}, E. and {Desidera}, S. and {Fiorenzano}, A. and {Harutyunyan}, A. and {Lorenzi}, V. and {Knapic}, C. and {Molinari}, E. and {Pacetti}, E. and {Pagano}, I. and {Pedani}, M. and {Piotto}, G. and {Poretti}, E.},
        title = "{The GAPS Programme at TNG. XXXVIII. Five molecules in the atmosphere of the warm giant planet WASP-69b detected at high spectral resolution}",
      journal = {\aap},
         year = 2022,
        month = sep,
       volume = {665},
          eid = {A104},
        pages = {A104},
          doi = {10.1051/0004-6361/202243854},
archivePrefix = {arXiv},
       eprint = {2207.09760},
 primaryClass = {astro-ph.EP},
       adsurl = {https://ui.adsabs.harvard.edu/abs/2022A&A...665A.104G}
}

@ARTICLE{Murphy2025,
       author = {{Murphy}, Matthew M. and {Beatty}, Thomas G. and {Schlawin}, Everett and {Bell}, Taylor J. and {Radica}, Michael and {Kennedy}, Thomas D. and {Mehta}, Nishil and {Welbanks}, Luis and {Line}, Michael R. and {Parmentier}, Vivien and {Greene}, Thomas P. and {Mukherjee}, Sagnick and {Fortney}, Jonathan J. and {Ohno}, Kazumasa and {Wiser}, Lindsey and {Arnold}, Kenneth and {Rauscher}, Emily and {Edelman}, Isaac R. and {Rieke}, Marcia J.},
        title = "{A Panchromatic Characterization of the Evening and Morning Atmosphere of WASP-107 b: Composition and Cloud Variations, and Insight into the Effect of Stellar Contamination}",
      journal = {\aj},
         year = 2025,
        month = jul,
       volume = {170},
       number = {1},
          eid = {61},
        pages = {61},
          doi = {10.3847/1538-3881/addf38},
archivePrefix = {arXiv},
       eprint = {2505.13602},
 primaryClass = {astro-ph.EP},
       adsurl = {https://ui.adsabs.harvard.edu/abs/2025AJ....170...61M}
}

@ARTICLE{Mukherjee2025,
       author = {{Mukherjee}, Sagnick and {Fortney}, Jonathan J. and {Wogan}, Nicholas F. and {Sing}, David K. and {Ohno}, Kazumasa},
        title = "{Effects of Planetary Parameters on Disequilibrium Chemistry in Irradiated Planetary Atmospheres: From Gas Giants to Sub-Neptunes}",
      journal = {\apj},
         year = 2025,
        month = jun,
       volume = {985},
       number = {2},
          eid = {209},
        pages = {209},
          doi = {10.3847/1538-4357/adc7b3},
archivePrefix = {arXiv},
       eprint = {2410.17169},
 primaryClass = {astro-ph.EP},
       adsurl = {https://ui.adsabs.harvard.edu/abs/2025ApJ...985..209M}
}

@ARTICLE{Showman2009,
       author = {{Showman}, Adam P. and {Fortney}, Jonathan J. and {Lian}, Yuan and {Marley}, Mark S. and {Freedman}, Richard S. and {Knutson}, Heather A. and {Charbonneau}, David},
        title = "{Atmospheric Circulation of Hot Jupiters: Coupled Radiative-Dynamical General Circulation Model Simulations of HD 189733b and HD 209458b}",
      journal = {\apj},
         year = 2009,
        month = jul,
       volume = {699},
       number = {1},
        pages = {564-584},
          doi = {10.1088/0004-637X/699/1/564},
archivePrefix = {arXiv},
       eprint = {0809.2089},
 primaryClass = {astro-ph},
       adsurl = {https://ui.adsabs.harvard.edu/abs/2009ApJ...699..564S}
}

@ARTICLE{Showman2011,
       author = {{Showman}, Adam P. and {Polvani}, Lorenzo M.},
        title = "{Equatorial Superrotation on Tidally Locked Exoplanets}",
      journal = {\apj},
         year = 2011,
        month = sep,
       volume = {738},
       number = {1},
          eid = {71},
        pages = {71},
          doi = {10.1088/0004-637X/738/1/71},
archivePrefix = {arXiv},
       eprint = {1103.3101},
 primaryClass = {astro-ph.EP},
       adsurl = {https://ui.adsabs.harvard.edu/abs/2011ApJ...738...71S}
}

@ARTICLE{Parmentier2021,
       author = {{Parmentier}, Vivien and {Showman}, Adam P. and {Fortney}, Jonathan J.},
        title = "{The cloudy shape of hot Jupiter thermal phase curves}",
      journal = {\mnras},
         year = 2021,
        month = jan,
       volume = {501},
       number = {1},
        pages = {78-108},
          doi = {10.1093/mnras/staa3418},
archivePrefix = {arXiv},
       eprint = {2010.06934},
 primaryClass = {astro-ph.EP},
       adsurl = {https://ui.adsabs.harvard.edu/abs/2021MNRAS.501...78P}
}

@ARTICLE{Moses2022,
       author = {{Moses}, Julianne I. and {Tremblin}, Pascal and {Venot}, Olivia and {Miguel}, Yamila},
        title = "{Chemical variation with altitude and longitude on exo-Neptunes: Predictions for Ariel phase-curve observations}",
      journal = {Experimental Astronomy},
         year = 2022,
        month = apr,
       volume = {53},
       number = {2},
        pages = {279-322},
          doi = {10.1007/s10686-021-09749-1},
archivePrefix = {arXiv},
       eprint = {2103.07023},
 primaryClass = {astro-ph.EP},
       adsurl = {https://ui.adsabs.harvard.edu/abs/2022ExA....53..279M}
}

@ARTICLE{Parmentier2013,
       author = {{Parmentier}, Vivien and {Showman}, Adam P. and {Lian}, Yuan},
        title = "{3D mixing in hot Jupiters atmospheres. I. Application to the day/night cold trap in HD 209458b}",
      journal = {\aap},
         year = 2013,
        month = oct,
       volume = {558},
          eid = {A91},
        pages = {A91},
          doi = {10.1051/0004-6361/201321132},
archivePrefix = {arXiv},
       eprint = {1301.4522},
 primaryClass = {astro-ph.EP},
       adsurl = {https://ui.adsabs.harvard.edu/abs/2013A&A...558A..91P}
}

@ARTICLE{Bangera2025,
       author = {{Bangera}, Nidhi and {Helling}, Christiane and {Guilluy}, Gloria and {Cubillos}, Patricio and {Fossati}, Luca and {Giacobbe}, Paolo and {Rimmer}, Paul and {Kitzmann}, Daniel},
        title = "{Kinetic and Photochemical Disequilibrium in the Potentially Carbon-rich Atmosphere of the Warm Jupiter WASP-69 b}",
      journal = {\apj},
         year = 2025,
        month = feb,
       volume = {980},
       number = {1},
          eid = {147},
        pages = {147},
          doi = {10.3847/1538-4357/adaa7d},
archivePrefix = {arXiv},
       eprint = {2501.08463},
 primaryClass = {astro-ph.EP},
       adsurl = {https://ui.adsabs.harvard.edu/abs/2025ApJ...980..147B}
}

@ARTICLE{Husser2013,
       author = {{Husser}, T. -O. and {Wende-von Berg}, S. and {Dreizler}, S. and {Homeier}, D. and {Reiners}, A. and {Barman}, T. and {Hauschildt}, P.~H.},
        title = "{A new extensive library of PHOENIX stellar atmospheres and synthetic spectra}",
      journal = {\aap},
         year = 2013,
        month = may,
       volume = {553},
          eid = {A6},
        pages = {A6},
          doi = {10.1051/0004-6361/201219058},
archivePrefix = {arXiv},
       eprint = {1303.5632},
 primaryClass = {astro-ph.SR},
       adsurl = {https://ui.adsabs.harvard.edu/abs/2013A&A...553A...6H}
}

@INPROCEEDINGS{Claire2012,
       author = {{Claire}, M. and {Sheets}, J. and {Cohen}, M. and {Ribas}, I. and {Meadows}, V.~S. and {Catling}, D.~C.},
        title = "{The Evolution of Solar Flux: Quantitative Estimates for Planetary Studies}",
    booktitle = {AGU Fall Meeting Abstracts},
         year = 2012,
       volume = {2012},
        month = dec,
          eid = {P13B-1935},
        pages = {P13B-1935},
       adsurl = {https://ui.adsabs.harvard.edu/abs/2012AGUFM.P13B1935C}
}

@ARTICLE{Schneider2022a,
       author = {{Schneider}, Aaron David and {Carone}, Ludmila and {Decin}, Leen and {J{\o}rgensen}, Uffe Gr{\r{a}}e and {Molli{\`e}re}, Paul and {Baeyens}, Robin and {Kiefer}, Sven and {Helling}, Christiane},
        title = "{Exploring the deep atmospheres of HD 209458b and WASP-43b using a non-gray general circulation model}",
      journal = {\aap},
         year = 2022,
        month = aug,
       volume = {664},
          eid = {A56},
        pages = {A56},
          doi = {10.1051/0004-6361/202142728},
archivePrefix = {arXiv},
       eprint = {2202.09183},
 primaryClass = {astro-ph.EP},
       adsurl = {https://ui.adsabs.harvard.edu/abs/2022A&A...664A..56S}
}

@ARTICLE{Schneider2022b,
       author = {{Schneider}, Aaron David and {Carone}, Ludmila and {Decin}, Leen and {J{\o}rgensen}, Uffe Gr{\r{a}}e and {Helling}, Christiane},
        title = "{No evidence for radius inflation in hot Jupiters from vertical advection of heat}",
      journal = {\aap},
         year = 2022,
        month = oct,
       volume = {666},
          eid = {L11},
        pages = {L11},
          doi = {10.1051/0004-6361/202244797},
archivePrefix = {arXiv},
       eprint = {2210.01466},
 primaryClass = {astro-ph.EP},
       adsurl = {https://ui.adsabs.harvard.edu/abs/2022A&A...666L..11S}
}

@ARTICLE{Plaschzug2025,
       author = {{Plaschzug}, Alexander and {Reza}, Amit and {Carone}, Ludmila and {Gernjak}, Sebastian and {Helling}, Christiane},
        title = "{Accelerating exoplanet climate modelling: A machine learning approach to complement 3D GCM grid simulations}",
      journal = {arXiv e-prints},
         year = 2025,
        month = aug,
          eid = {arXiv:2508.10827},
        pages = {arXiv:2508.10827},
          doi = {10.48550/arXiv.2508.10827},
archivePrefix = {arXiv},
       eprint = {2508.10827},
 primaryClass = {astro-ph.EP},
       adsurl = {https://ui.adsabs.harvard.edu/abs/2025arXiv250810827P}
}

@article{Adcroft2004,
	title = {Implementation of an {Atmosphere}–{Ocean} {General} {Circulation} {Model} on the {Expanded} {Spherical} {Cube}},
	volume = {132},
	issn = {1520-0493, 0027-0644},
	url = {https://journals.ametsoc.org/view/journals/mwre/132/12/mwr2823.1.xml},
	doi = {10.1175/MWR2823.1},
	language = {EN},
	number = {12},
	urldate = {2025-03-18},
	journal = {Monthly Weather Review},
	author = {Adcroft, Alistair and Campin, Jean-Michel and Hill, Chris and Marshall, John},
	month = dec,
	year = {2004},
	note = {Publisher: American Meteorological Society
Section: Monthly Weather Review},
	pages = {2845--2863},
}

@incollection{ArakawaLamb1977,
title = {Computational Design of the Basic Dynamical Processes of the UCLA General Circulation Model},
editor = {JULIUS CHANG},
series = {Methods in Computational Physics: Advances in Research and Applications},
publisher = {Elsevier},
volume = {17},
pages = {173-265},
year = {1977},
booktitle = {General Circulation Models of the Atmosphere},
issn = {0076-6860},
doi = {https://doi.org/10.1016/B978-0-12-460817-7.50009-4},
url = {https://www.sciencedirect.com/science/article/pii/B9780124608177500094},
author = {Akio Arakawa and VIVIAN R. Lamb}
}

@ARTICLE{LiuShowman2013ApJ,
       author = {{Liu}, Beibei and {Showman}, Adam P.},
        title = "{Atmospheric Circulation of Hot Jupiters: Insensitivity to Initial Conditions}",
      journal = {\apj},
         year = 2013,
        month = jun,
       volume = {770},
       number = {1},
          eid = {42},
        pages = {42},
          doi = {10.1088/0004-637X/770/1/42},
archivePrefix = {arXiv},
       eprint = {1208.0126},
 primaryClass = {astro-ph.EP},
       adsurl = {https://ui.adsabs.harvard.edu/abs/2013ApJ...770...42L}
}

@article{TennysonEtal2020jqsrtExomol2020,
	adsurl = {https://ui.adsabs.harvard.edu/abs/2020JQSRT.25507228T},
	archiveprefix = {arXiv},
	author = {{Tennyson}, Jonathan and {Yurchenko}, Sergei N. and {Al-Refaie}, Ahmed F. and {Clark}, Victoria H.~J. and {Chubb}, Katy L. and {Conway}, Eamon K. and {Dewan}, Akhil and {Gorman}, Maire N. and {Hill}, Christian and {Lynas-Gray}, A.~E. and {Mellor}, Thomas and {McKemmish}, Laura K. and {Owens}, Alec and {Polyansky}, Oleg L. and {Semenov}, Mikhail and {Somogyi}, Wilfrid and {Tinetti}, Giovanna and {Upadhyay}, Apoorva and {Waldmann}, Ingo and {Wang}, Yixin and {Wright}, Samuel and {Yurchenko}, Olga P.},
	doi = {10.1016/j.jqsrt.2020.107228},
	eid = {107228},
	eprint = {2007.13022},
	journal = {\jqsrt},
	month = nov,
	pages = {107228},
	primaryclass = {astro-ph.SR},
	title = {{The 2020 release of the ExoMol database: Molecular line lists for exoplanet and other hot atmospheres}},
	volume = {255},
	year = 2020}

@article{Tennyson2016_ExoMol,
	adsurl = {https://ui.adsabs.harvard.edu/abs/2016JMoSp.327...73T},
	archiveprefix = {arXiv},
	author = {{Tennyson}, Jonathan and {Yurchenko}, Sergei N. and {Al-Refaie}, Ahmed F. and {Barton}, Emma J. and {Chubb}, Katy L. and {Coles}, Phillip A. and {Diamantopoulou}, S. and {Gorman}, Maire N. and {Hill}, Christian and {Lam}, Aden Z. and {Lodi}, Lorenzo and {McKemmish}, Laura K. and {Na}, Yueqi and {Owens}, Alec and {Polyansky}, Oleg L. and {Rivlin}, Tom and {Sousa-Silva}, Clara and {Underwood}, Daniel S. and {Yachmenev}, Andrey and {Zak}, Emil},
	doi = {10.1016/j.jms.2016.05.002},
	eprint = {1603.05890},
	journal = {Journal of Molecular Spectroscopy},
	month = sep,
	pages = {73-94},
	primaryclass = {astro-ph.GA},
	title = {{The ExoMol database: Molecular line lists for exoplanet and other hot atmospheres}},
	volume = {327},
	year = 2016}

@article{BorysowEtal2001jqsrtH2H2highT,
	adsurl = {https://ui.adsabs.harvard.edu/abs/2001JQSRT..68..235B},
	author = {{Borysow}, Aleksandra and {Jorgensen}, Uffe G. and {Fu}, Yi},
	doi = {10.1016/S0022-4073(00)00023-6},
	journal = {\jqsrt},
	month = feb,
	pages = {235-255},
	title = {{High-temperature (1000-7000 K) collision-induced absorption of H''2 pairs computed from the first principles, with application to cool and dense stellar atmospheres}},
	volume = {68},
	year = 2001}

@article{BorysowEtal1989apjH2HeRVRT,
	adsurl = {https://ui.adsabs.harvard.edu/abs/1989ApJ...336..495B},
	author = {{Borysow}, Aleksandra and {Frommhold}, Lothar and {Moraldi}, Massimo},
	doi = {10.1086/167027},
	journal = {\apj},
	month = jan,
	pages = {495},
	title = {{Collision-induced Infrared Spectra of H 2-He Pairs Involving 0 1 Vibrational Transitions and Temperatures from 18 to 7000 K}},
	volume = {336},
	year = 1989}

@article{BorysowEtal1988apjH2HeRT,
	adsurl = {https://ui.adsabs.harvard.edu/abs/1988ApJ...326..509B},
	author = {{Borysow}, Jacek and {Frommhold}, Lothar and {Birnbaum}, George},
	doi = {10.1086/166112},
	journal = {\apj},
	month = mar,
	pages = {509},
	title = {{Collision-induced Rototranslational Absorption Spectra of H 2-He Pairs at Temperatures from 40 to 3000 K}},
	volume = {326},
	year = 1988}

@article{BorysowFrommhold1989apjH2HeOvertones,
	adsurl = {https://ui.adsabs.harvard.edu/abs/1989ApJ...341..549B},
	author = {{Borysow}, Aleksandra and {Frommhold}, Lothar},
	doi = {10.1086/167515},
	journal = {\apj},
	month = jun,
	pages = {549},
	title = {{Collision-induced Infrared Spectra of H 2-He Pairs at Temperatures from 18 to 7000 K. II. Overtone and Hot Bands}},
	volume = {341},
	year = 1989}

@article{Borysow2002jqsrtH2H2lowT,
	adsurl = {https://ui.adsabs.harvard.edu/abs/2002A&A...390..779B},
	author = {{Borysow}, A.},
	doi = {10.1051/0004-6361:20020555},
	journal = {\aap},
	month = aug,
	pages = {779-782},
	title = {{Collision-induced absorption coefficients of H$_{2}$ pairs at temperatures from 60 K to 1000 K}},
	volume = {390},
	year = 2002}

@article{Allard19_Na_K,
	adsurl = {https://ui.adsabs.harvard.edu/abs/2019A&A...628A.120A},
	archiveprefix = {arXiv},
	author = {{Allard}, N.~F. and {Spiegelman}, F. and {Leininger}, T. and {Molliere}, P.},
	doi = {10.1051/0004-6361/201935593},
	eid = {A120},
	eprint = {1908.01989},
	journal = {\aap},
	month = aug,
	pages = {A120},
	primaryclass = {astro-ph.SR},
	title = {{New study of the line profiles of sodium perturbed by H$_{2}$}},
	volume = {628},
	year = 2019}

@ARTICLE{Dalgarno1962,
       author = {{Dalgarno}, A. and {Williams}, D.~A.},
        title = "{Rayleigh Scattering by Molecular Hydrogen.}",
      journal = {\apj},
         year = 1962,
        month = sep,
       volume = {136},
        pages = {690-692},
          doi = {10.1086/147428},
       adsurl = {https://ui.adsabs.harvard.edu/abs/1962ApJ...136..690D}
}

@ARTICLE{Chan1965,
       author = {{Chan}, Y.~M. and {Dalgarno}, A.},
        title = "{The refractive index of helium}",
      journal = {Proceedings of the Physical Society},
         year = 1965,
        month = feb,
       volume = {85},
       number = {2},
        pages = {227-230},
          doi = {10.1088/0370-1328/85/2/304},
       adsurl = {https://ui.adsabs.harvard.edu/abs/1965PPS....85..227C}
}

@ARTICLE{Richard2012,
       author = {{Richard}, C. and {Gordon}, I.~E. and {Rothman}, L.~S. and {Abel}, M. and {Frommhold}, L. and {Gustafsson}, M. and {Hartmann}, J. -M. and {Hermans}, C. and {Lafferty}, W.~J. and {Orton}, G.~S. and {Smith}, K.~M. and {Tran}, H.},
        title = "{New section of the HITRAN database: Collision-induced absorption (CIA)}",
      journal = {\jqsrt},
         year = 2012,
        month = jul,
       volume = {113},
       number = {11},
        pages = {1276-1285},
          doi = {10.1016/j.jqsrt.2011.11.004},
       adsurl = {https://ui.adsabs.harvard.edu/abs/2012JQSRT.113.1276R}
}

@ARTICLE{Goody1989,
       author = {{Goody}, Richard and {West}, Robert and {Chen}, Luke and {Crisp}, David},
        title = "{The correlated-k method for radiation calculations in nonhomogeneous atmospheres.}",
      journal = {\jqsrt},
         year = 1989,
        month = dec,
       volume = {42},
        pages = {539-550},
          doi = {10.1016/0022-4073(89)90044-7},
       adsurl = {https://ui.adsabs.harvard.edu/abs/1989JQSRT..42..539G}
}

@ARTICLE{2024ApJ...977...52S,
       author = {{Soni}, Vikas and {Acharyya}, Kinsuk},
        title = "{Signature of Vertical Mixing in Hydrogen-dominated Exoplanet Atmospheres}",
      journal = {\apj},
         year = 2024,
        month = dec,
       volume = {977},
       number = {1},
          eid = {52},
        pages = {52},
          doi = {10.3847/1538-4357/ad891f},
archivePrefix = {arXiv},
       eprint = {2410.12737},
 primaryClass = {astro-ph.EP},
       adsurl = {https://ui.adsabs.harvard.edu/abs/2024ApJ...977...52S}
}

@ARTICLE{Carone2025,
       author = {{Carone}, Ludmila and {Helling}, Christiane and {Gernjak}, Sebastian and {Leitner}, Hanna and {Janz}, Tamara},
        title = "{Exoplanet climate characterization with transit asymmetries -- A comprehensive population study from the optical to the infrared}",
      journal = {arXiv e-prints},
         year = 2025,
        month = nov,
          eid = {arXiv:2511.01548},
        pages = {arXiv:2511.01548},
          doi = {10.48550/arXiv.2511.01548},
archivePrefix = {arXiv},
       eprint = {2511.01548},
 primaryClass = {astro-ph.EP},
       adsurl = {https://ui.adsabs.harvard.edu/abs/2025arXiv251101548C}
}

@ARTICLE{Zamyatina2024,
       author = {{Zamyatina}, Maria and {Christie}, Duncan A. and {H{\'e}brard}, Eric and {Mayne}, Nathan J. and {Radica}, Michael and {Taylor}, Jake and {Baskett}, Harry and {Moore}, Ben and {Lils}, Craig and {Sergeev}, Denis E. and {Ahrer}, Eva-Maria and {Manners}, James and {Kohary}, Krisztian and {Feinstein}, Adina D.},
        title = "{Quenching-driven equatorial depletion and limb asymmetries in hot Jupiter atmospheres: WASP-96b example}",
      journal = {\mnras},
         year = 2024,
        month = apr,
       volume = {529},
       number = {2},
        pages = {1776-1801},
          doi = {10.1093/mnras/stae600},
archivePrefix = {arXiv},
       eprint = {2402.14535},
 primaryClass = {astro-ph.EP},
       adsurl = {https://ui.adsabs.harvard.edu/abs/2024MNRAS.529.1776Z}
}

@ARTICLE{Drummond,
       author = {{Drummond}, Benjamin and {Mayne}, Nathan J. and {Manners}, James and {Baraffe}, Isabelle and {Goyal}, Jayesh and {Tremblin}, Pascal and {Sing}, David K. and {Kohary}, Krisztian},
        title = "{The 3D Thermal, Dynamical, and Chemical Structure of the Atmosphere of HD 189733b: Implications of Wind-driven Chemistry for the Emission Phase Curve}",
      journal = {\apj},
         year = 2018,
        month = dec,
       volume = {869},
       number = {1},
          eid = {28},
        pages = {28},
          doi = {10.3847/1538-4357/aaeb28},
archivePrefix = {arXiv},
       eprint = {1810.09724},
 primaryClass = {astro-ph.EP},
       adsurl = {https://ui.adsabs.harvard.edu/abs/2018ApJ...869...28D}
}

@ARTICLE{Taylor2020,
       author = {{Taylor}, Jake and {Parmentier}, Vivien and {Irwin}, Patrick G.~J. and {Aigrain}, Suzanne and {Lee}, Elspeth K.~H. and {Krissansen-Totton}, Joshua},
        title = "{Understanding and mitigating biases when studying inhomogeneous emission spectra with JWST}",
      journal = {\mnras},
         year = 2020,
        month = apr,
       volume = {493},
       number = {3},
        pages = {4342-4354},
          doi = {10.1093/mnras/staa552},
archivePrefix = {arXiv},
       eprint = {2002.00773},
 primaryClass = {astro-ph.EP},
       adsurl = {https://ui.adsabs.harvard.edu/abs/2020MNRAS.493.4342T}
}

@ARTICLE{Challener2025,
       author = {{Challener}, Ryan C. and {Weiner Mansfield}, Megan and {Cubillos}, Patricio E. and {Piette}, Anjali A.~A. and {Coulombe}, Louis-Philippe and {Beltz}, Hayley and {Blecic}, Jasmina and {Rauscher}, Emily and {Bean}, Jacob L. and {Benneke}, Bj{\"o}rn and {Kempton}, Eliza M.-R. and {Harrington}, Joseph and {Komacek}, Thaddeus D. and {Parmentier}, Vivien and {Casewell}, S.~L. and {Iro}, Nicolas and {Mancini}, Luigi and {Nixon}, Matthew C. and {Radica}, Michael and {Steinrueck}, Maria E. and {Welbanks}, Luis and {Batalha}, Natalie M. and {Caceres}, Claudio and {Crossfield}, Ian J.~M. and {Crouzet}, Nicolas and {D{\'e}sert}, Jean-Michel and {Molaverdikhani}, Karan and {Nikolov}, Nikolay K. and {Palle}, Enric and {Rackham}, Benjamin V. and {Schlawin}, Everett and {Sing}, David K. and {Stevenson}, Kevin B. and {Tan}, Xianyu and {Turner}, Jake D. and {Zhang}, Xi},
        title = "{Horizontal and vertical exoplanet thermal structure from a JWST spectroscopic eclipse map}",
      journal = {arXiv e-prints},
         year = 2025,
        month = oct,
          eid = {arXiv:2510.24708},
        pages = {arXiv:2510.24708},
          doi = {10.48550/arXiv.2510.24708},
archivePrefix = {arXiv},
       eprint = {2510.24708},
 primaryClass = {astro-ph.EP},
       adsurl = {https://ui.adsabs.harvard.edu/abs/2025arXiv251024708C}
}

@ARTICLE{Murphy2024,
       author = {{Murphy}, Matthew M. and {Beatty}, Thomas G. and {Schlawin}, Everett and {Bell}, Taylor J. and {Line}, Michael R. and {Greene}, Thomas P. and {Parmentier}, Vivien and {Rauscher}, Emily and {Welbanks}, Luis and {Fortney}, Jonathan J. and {Rieke}, Marcia},
        title = "{Evidence for morning-to-evening limb asymmetry on the cool low-density exoplanet WASP-107 b}",
      journal = {Nature Astronomy},
         year = 2024,
        month = dec,
       volume = {8},
        pages = {1562-1574},
          doi = {10.1038/s41550-024-02367-9},
archivePrefix = {arXiv},
       eprint = {2406.09863},
 primaryClass = {astro-ph.EP},
       adsurl = {https://ui.adsabs.harvard.edu/abs/2024NatAs...8.1562M}
}

@ARTICLE{Baeyens2024,
       author = {{Baeyens}, Robin and {D{\'e}sert}, Jean-Michel and {Petrignani}, Annemieke and {Carone}, Ludmila and {Schneider}, Aaron David},
        title = "{Photodissociation and induced chemical asymmetries on ultra-hot gas giants. A case study of HCN on WASP-76 b}",
      journal = {\aap},
         year = 2024,
        month = jun,
       volume = {686},
          eid = {A24},
        pages = {A24},
          doi = {10.1051/0004-6361/202348022},
archivePrefix = {arXiv},
       eprint = {2309.00573},
 primaryClass = {astro-ph.EP},
       adsurl = {https://ui.adsabs.harvard.edu/abs/2024A&A...686A..24B}
}

@ARTICLE{Steinrueck2025,
       author = {{Steinrueck}, Maria E. and {Savel}, Arjun B. and {Christie}, Duncan A. and {Carone}, Ludmila and {Tsai}, Shang-Min and {Ak{\i}n}, Can and {Kennedy}, Thomas D. and {Kiefer}, Sven and {Lewis}, David A. and {Rauscher}, Emily and {Samra}, Dominic and {Zamyatina}, Maria and {Arnold}, Kenneth and {Baeyens}, Robin and {Gkouvelis}, Leonardos and {Haegele}, David and {Helling}, Christiane and {Mayne}, Nathan J. and {Powell}, Diana and {Roman}, Michael T. and {Beltz}, Hayley and {Espinoza}, N{\'e}stor and {Heng}, Kevin and {Iro}, Nicolas and {Kempton}, Eliza M. -R. and {Kreidberg}, Laura and {Kirk}, James and {Murphy}, Matthew M. and {Rackham}, Benjamin V. and {Tan}, Xianyu},
        title = "{Limb Asymmetries on WASP-39b: A Multi-GCM Comparison of Chemistry, Clouds, and Hazes}",
      journal = {arXiv e-prints},
         year = 2025,
        month = sep,
          eid = {arXiv:2509.21588},
        pages = {arXiv:2509.21588},
          doi = {10.48550/arXiv.2509.21588},
archivePrefix = {arXiv},
       eprint = {2509.21588},
 primaryClass = {astro-ph.EP},
       adsurl = {https://ui.adsabs.harvard.edu/abs/2025arXiv250921588S}
}

@ARTICLE{Thorngren2019,
       author = {{Thorngren}, Daniel and {Fortney}, Jonathan J.},
        title = "{Connecting Giant Planet Atmosphere and Interior Modeling: Constraints on Atmospheric Metal Enrichment}",
      journal = {\apjl},
         year = 2019,
        month = apr,
       volume = {874},
       number = {2},
          eid = {L31},
        pages = {L31},
          doi = {10.3847/2041-8213/ab1137},
archivePrefix = {arXiv},
       eprint = {1811.11859},
 primaryClass = {astro-ph.EP},
       adsurl = {https://ui.adsabs.harvard.edu/abs/2019ApJ...874L..31T}
}

@ARTICLE{Espinoza2024,
       author = {{Espinoza}, N{\'e}stor and {Steinrueck}, Maria E. and {Kirk}, James and {MacDonald}, Ryan J. and {Savel}, Arjun B. and {Arnold}, Kenneth and {Kempton}, Eliza M.-R. and {Murphy}, Matthew M. and {Carone}, Ludmila and {Zamyatina}, Maria and {Lewis}, David A. and {Samra}, Dominic and {Kiefer}, Sven and {Rauscher}, Emily and {Christie}, Duncan and {Mayne}, Nathan and {Helling}, Christiane and {Rustamkulov}, Zafar and {Parmentier}, Vivien and {May}, Erin M. and {Carter}, Aarynn L. and {Zhang}, Xi and {L{\'o}pez-Morales}, Mercedes and {Allen}, Natalie and {Blecic}, Jasmina and {Decin}, Leen and {Mancini}, Luigi and {Molaverdikhani}, Karan and {Rackham}, Benjamin V. and {Palle}, Enric and {Tsai}, Shang-Min and {Ahrer}, Eva-Maria and {Bean}, Jacob L. and {Crossfield}, Ian J.~M. and {Haegele}, David and {H{\'e}brard}, Eric and {Kreidberg}, Laura and {Powell}, Diana and {Schneider}, Aaron D. and {Welbanks}, Luis and {Wheatley}, Peter and {Brahm}, Rafael and {Crouzet}, Nicolas},
        title = "{Inhomogeneous terminators on the exoplanet WASP-39 b}",
      journal = {\nat},
         year = 2024,
        month = aug,
       volume = {632},
       number = {8027},
        pages = {1017-1020},
          doi = {10.1038/s41586-024-07768-4},
archivePrefix = {arXiv},
       eprint = {2407.10294},
 primaryClass = {astro-ph.EP},
       adsurl = {https://ui.adsabs.harvard.edu/abs/2024Natur.632.1017E}
}

@ARTICLE{Demangeon2024,
       author = {{Demangeon}, O.~D.~S. and {Cubillos}, P.~E. and {Singh}, V. and {Wilson}, T.~G. and {Carone}, L. and {Bekkelien}, A. and {Deline}, A. and {Ehrenreich}, D. and {Maxted}, P.~F.~L. and {Demory}, B.-O. and {Zingales}, T. and {Lendl}, M. and {Bonfanti}, A. and {Sousa}, S.~G. and {Brandeker}, A. and {Alibert}, Y. and {Alonso}, R. and {Asquier}, J. and {B{\'a}rczy}, T. and {Navascues}, D. Barrado and {Barros}, S.~C.~C. and {Baumjohann}, W. and {Beck}, M. and {Beck}, T. and {Benz}, W. and {Billot}, N. and {Biondi}, F. and {Borsato}, L. and {Broeg}, Ch. and {Buder}, M. and {Cameron}, A. Collier and {Csizmadia}, Sz. and {Davies}, M.~B. and {Deleuil}, M. and {Delrez}, L. and {Erikson}, A. and {Fortier}, A. and {Fossati}, L. and {Fridlund}, M. and {Gandolfi}, D. and {Gillon}, M. and {G{\"u}del}, M. and {G{\"u}nther}, M.~N. and {Heitzmann}, A. and {Helling}, Ch. and {Hoyer}, S. and {Isaak}, K.~G. and {Kiss}, L.~L. and {Lam}, K.~W.~F. and {Laskar}, J. and {des Etangs}, A. Lecavelier and {Magrin}, D. and {Mecina}, M. and {Mordasini}, Ch. and {Nascimbeni}, V. and {Olofsson}, G. and {Ottensamer}, R. and {Pagano}, I. and {Pall{\'e}}, E. and {Peter}, G. and {Piotto}, G. and {Pollacco}, D. and {Queloz}, D. and {Ragazzoni}, R. and {Rando}, N. and {Rauer}, H. and {Ribas}, I. and {Rieder}, M. and {Salmon}, S. and {Santos}, N.~C. and {Scandariato}, G. and {S{\'e}gransan}, D. and {Simon}, A.~E. and {Smith}, A.~M.~S. and {Stalport}, M. and {Szab{\'o}}, Gy. M. and {Thomas}, N. and {Udry}, S. and {Van Grootel}, V. and {Venturini}, J. and {Villaver}, E. and {Walton}, N.~A.},
        title = "{Asymmetry in the atmosphere of the ultra-hot Jupiter WASP-76 b}",
      journal = {\aap},
         year = 2024,
        month = apr,
       volume = {684},
          eid = {A27},
        pages = {A27},
          doi = {10.1051/0004-6361/202348270},
       adsurl = {https://ui.adsabs.harvard.edu/abs/2024A&A...684A..27D}
}

@ARTICLE{Eherenreich2020,
       author = {{Ehrenreich}, David and {Lovis}, Christophe and {Allart}, Romain and {Zapatero Osorio}, Mar{\'\i}a Rosa and {Pepe}, Francesco and {Cristiani}, Stefano and {Rebolo}, Rafael and {Santos}, Nuno C. and {Borsa}, Francesco and {Demangeon}, Olivier and {Dumusque}, Xavier and {Gonz{\'a}lez Hern{\'a}ndez}, Jonay I. and {Casasayas-Barris}, N{\'u}ria and {S{\'e}gransan}, Damien and {Sousa}, S{\'e}rgio and {Abreu}, Manuel and {Adibekyan}, Vardan and {Affolter}, Michael and {Allende Prieto}, Carlos and {Alibert}, Yann and {Aliverti}, Matteo and {Alves}, David and {Amate}, Manuel and {Avila}, Gerardo and {Baldini}, Veronica and {Bandy}, Timothy and {Benz}, Willy and {Bianco}, Andrea and {Bolmont}, {\'E}meline and {Bouchy}, Fran{\c{c}}ois and {Bourrier}, Vincent and {Broeg}, Christopher and {Cabral}, Alexandre and {Calderone}, Giorgio and {Pall{\'e}}, Enric and {Cegla}, H.~M. and {Cirami}, Roberto and {Coelho}, Jo{\~a}o M.~P. and {Conconi}, Paolo and {Coretti}, Igor and {Cumani}, Claudio and {Cupani}, Guido and {Dekker}, Hans and {Delabre}, Bernard and {Deiries}, Sebastian and {D'Odorico}, Valentina and {Di Marcantonio}, Paolo and {Figueira}, Pedro and {Fragoso}, Ana and {Genolet}, Ludovic and {Genoni}, Matteo and {G{\'e}nova Santos}, Ricardo and {Hara}, Nathan and {Hughes}, Ian and {Iwert}, Olaf and {Kerber}, Florian and {Knudstrup}, Jens and {Landoni}, Marco and {Lavie}, Baptiste and {Lizon}, Jean-Louis and {Lendl}, Monika and {Lo Curto}, Gaspare and {Maire}, Charles and {Manescau}, Antonio and {Martins}, C.~J.~A.~P. and {M{\'e}gevand}, Denis and {Mehner}, Andrea and {Micela}, Giusi and {Modigliani}, Andrea and {Molaro}, Paolo and {Monteiro}, Manuel and {Monteiro}, Mario and {Moschetti}, Manuele and {M{\"u}ller}, Eric and {Nunes}, Nelson and {Oggioni}, Luca and {Oliveira}, Ant{\'o}nio and {Pariani}, Giorgio and {Pasquini}, Luca and {Poretti}, Ennio and {Rasilla}, Jos{\'e} Luis and {Redaelli}, Edoardo and {Riva}, Marco and {Santana Tschudi}, Samuel and {Santin}, Paolo and {Santos}, Pedro and {Segovia Milla}, Alex and {Seidel}, Julia V. and {Sosnowska}, Danuta and {Sozzetti}, Alessandro and {Span{\`o}}, Paolo and {Su{\'a}rez Mascare{\~n}o}, Alejandro and {Tabernero}, Hugo and {Tenegi}, Fabio and {Udry}, St{\'e}phane and {Zanutta}, Alessio and {Zerbi}, Filippo},
        title = "{Nightside condensation of iron in an ultrahot giant exoplanet}",
      journal = {\nat},
         year = 2020,
        month = apr,
       volume = {580},
       number = {7805},
        pages = {597-601},
          doi = {10.1038/s41586-020-2107-1},
archivePrefix = {arXiv},
       eprint = {2003.05528},
 primaryClass = {astro-ph.EP},
       adsurl = {https://ui.adsabs.harvard.edu/abs/2020Natur.580..597E}
}

@ARTICLE{JWST2023,
       author = {{JWST Transiting Exoplanet Community Early Release Science Team} and {Ahrer}, Eva-Maria and {Alderson}, Lili and {Batalha}, Natalie M. and {Batalha}, Natasha E. and {Bean}, Jacob L. and {Beatty}, Thomas G. and {Bell}, Taylor J. and {Benneke}, Bj{\"o}rn and {Berta-Thompson}, Zachory K. and {Carter}, Aarynn L. and {Crossfield}, Ian J.~M. and {Espinoza}, N{\'e}stor and {Feinstein}, Adina D. and {Fortney}, Jonathan J. and {Gibson}, Neale P. and {Goyal}, Jayesh M. and {Kempton}, Eliza M.-R. and {Kirk}, James and {Kreidberg}, Laura and {L{\'o}pez-Morales}, Mercedes and {Line}, Michael R. and {Lothringer}, Joshua D. and {Moran}, Sarah E. and {Mukherjee}, Sagnick and {Ohno}, Kazumasa and {Parmentier}, Vivien and {Piaulet}, Caroline and {Rustamkulov}, Zafar and {Schlawin}, Everett and {Sing}, David K. and {Stevenson}, Kevin B. and {Wakeford}, Hannah R. and {Allen}, Natalie H. and {Birkmann}, Stephan M. and {Brande}, Jonathan and {Crouzet}, Nicolas and {Cubillos}, Patricio E. and {Damiano}, Mario and {D{\'e}sert}, Jean-Michel and {Gao}, Peter and {Harrington}, Joseph and {Hu}, Renyu and {Kendrew}, Sarah and {Knutson}, Heather A. and {Lagage}, Pierre-Olivier and {Leconte}, J{\'e}r{\'e}my and {Lendl}, Monika and {MacDonald}, Ryan J. and {May}, E.~M. and {Miguel}, Yamila and {Molaverdikhani}, Karan and {Moses}, Julianne I. and {Murray}, Catriona Anne and {Nehring}, Molly and {Nikolov}, Nikolay K. and {Petit dit de la Roche}, D.~J.~M. and {Radica}, Michael and {Roy}, Pierre-Alexis and {Stassun}, Keivan G. and {Taylor}, Jake and {Waalkes}, William C. and {Wachiraphan}, Patcharapol and {Welbanks}, Luis and {Wheatley}, Peter J. and {Aggarwal}, Keshav and {Alam}, Munazza K. and {Banerjee}, Agnibha and {Barstow}, Joanna K. and {Blecic}, Jasmina and {Casewell}, S.~L. and {Changeat}, Quentin and {Chubb}, K.~L. and {Col{\'o}n}, Knicole D. and {Coulombe}, Louis-Philippe and {Daylan}, Tansu and {de Val-Borro}, Miguel and {Decin}, Leen and {Dos Santos}, Leonardo A. and {Flagg}, Laura and {France}, Kevin and {Fu}, Guangwei and {Garc{\'\i}a Mu{\~n}oz}, A. and {Gizis}, John E. and {Glidden}, Ana and {Grant}, David and {Heng}, Kevin and {Henning}, Thomas and {Hong}, Yu-Cian and {Inglis}, Julie and {Iro}, Nicolas and {Kataria}, Tiffany and {Komacek}, Thaddeus D. and {Krick}, Jessica E. and {Lee}, Elspeth K.~H. and {Lewis}, Nikole K. and {Lillo-Box}, Jorge and {Lustig-Yaeger}, Jacob and {Mancini}, Luigi and {Mandell}, Avi M. and {Mansfield}, Megan and {Marley}, Mark S. and {Mikal-Evans}, Thomas and {Morello}, Giuseppe and {Nixon}, Matthew C. and {Ortiz Ceballos}, Kevin and {Piette}, Anjali A.~A. and {Powell}, Diana and {Rackham}, Benjamin V. and {Ramos-Rosado}, Lakeisha and {Rauscher}, Emily and {Redfield}, Seth and {Rogers}, Laura K. and {Roman}, Michael T. and {Roudier}, Gael M. and {Scarsdale}, Nicholas and {Shkolnik}, Evgenya L. and {Southworth}, John and {Spake}, Jessica J. and {Steinrueck}, Maria E. and {Tan}, Xianyu and {Teske}, Johanna K. and {Tremblin}, Pascal and {Tsai}, Shang-Min and {Tucker}, Gregory S. and {Turner}, Jake D. and {Valenti}, Jeff A. and {Venot}, Olivia and {Waldmann}, Ingo P. and {Wallack}, Nicole L. and {Zhang}, Xi and {Zieba}, Sebastian},
        title = "{Identification of carbon dioxide in an exoplanet atmosphere}",
      journal = {\nat},
         year = 2023,
        month = feb,
       volume = {614},
       number = {7949},
        pages = {649-652},
          doi = {10.1038/s41586-022-05269-w},
archivePrefix = {arXiv},
       eprint = {2208.11692},
 primaryClass = {astro-ph.EP},
       adsurl = {https://ui.adsabs.harvard.edu/abs/2023Natur.614..649J}
}

@ARTICLE{Brogi2018,
       author = {{Brogi}, M. and {Giacobbe}, P. and {Guilluy}, G. and {de Kok}, R.~J. and {Sozzetti}, A. and {Mancini}, L. and {Bonomo}, A.~S.},
        title = "{Exoplanet atmospheres with GIANO. I. Water in the transmission spectrum of HD 189 733 b}",
      journal = {\aap},
         year = 2018,
        month = jul,
       volume = {615},
          eid = {A16},
        pages = {A16},
          doi = {10.1051/0004-6361/201732189},
archivePrefix = {arXiv},
       eprint = {1801.09569},
 primaryClass = {astro-ph.EP},
       adsurl = {https://ui.adsabs.harvard.edu/abs/2018A&A...615A..16B}
}

@ARTICLE{Nakazawa2026,
       author = {{Nakazawa}, Kanon and {Kazumasa}, Ohno},
        title = "{Sulfur Enrichment in Close-in Exoplanet Atmospheres Induced by Pebble Drift across the Salt Line}",
      journal = {arXiv e-prints},
         year = 2026,
        month = feb,
          eid = {arXiv:2602.05300},
        pages = {arXiv:2602.05300},
          doi = {10.48550/arXiv.2602.05300},
archivePrefix = {arXiv},
       eprint = {2602.05300},
 primaryClass = {astro-ph.EP},
       adsurl = {https://ui.adsabs.harvard.edu/abs/2026arXiv260205300N}
}

@ARTICLE{Moses2011,
       author = {{Moses}, Julianne I. and {Visscher}, C. and {Fortney}, J.~J. and {Showman}, A.~P. and {Lewis}, N.~K. and {Griffith}, C.~A. and {Klippenstein}, S.~J. and {Shabram}, M. and {Friedson}, A.~J. and {Marley}, M.~S. and {Freedman}, R.~S.},
        title = "{Disequilibrium Carbon, Oxygen, and Nitrogen Chemistry in the Atmospheres of HD 189733b and HD 209458b}",
      journal = {\apj},
         year = 2011,
        month = aug,
       volume = {737},
       number = {1},
          eid = {15},
        pages = {15},
          doi = {10.1088/0004-637X/737/1/15},
archivePrefix = {arXiv},
       eprint = {1102.0063},
 primaryClass = {astro-ph.EP},
       adsurl = {https://ui.adsabs.harvard.edu/abs/2011ApJ...737...15M}
}

@ARTICLE{SommervilleThomas2026,
       author = {{Sommerville-Thomas}, Anna and {Kama}, Mihkel and {Shorttle}, Oliver and {Ran}, Jason},
        title = "{sponchpop II: Population Synthesis to Investigate Volatile Sulfur as a Fingerprint of Gas Giant Formation Histories}",
      journal = {arXiv e-prints},
         year = 2026,
        month = jan,
          eid = {arXiv:2601.10508},
        pages = {arXiv:2601.10508},
          doi = {10.48550/arXiv.2601.10508},
archivePrefix = {arXiv},
       eprint = {2601.10508},
 primaryClass = {astro-ph.EP},
       adsurl = {https://ui.adsabs.harvard.edu/abs/2026arXiv260110508S}
}

@article{barstow2020,
  title={Outstanding challenges of exoplanet atmospheric retrievals},
  author={Barstow, Joanna K and Heng, Kevin},
  journal={Space science reviews},
  volume={216},
  number={5},
  pages={82},
  year={2020},
  publisher={Springer}
}

@ARTICLE{Kempton2017,
       author = {{Kempton}, Eliza M.-R. and {Bean}, Jacob L. and {Parmentier}, Vivien},
        title = "{An Observational Diagnostic for Distinguishing between Clouds and Haze in Hot Exoplanet Atmospheres}",
      journal = {\apjl},
         year = 2017,
        month = aug,
       volume = {845},
       number = {2},
          eid = {L20},
        pages = {L20},
          doi = {10.3847/2041-8213/aa84ac},
archivePrefix = {arXiv},
       eprint = {1705.05847},
 primaryClass = {astro-ph.EP},
       adsurl = {https://ui.adsabs.harvard.edu/abs/2017ApJ...845L..20K}
}

@ARTICLE{Powell2019,
       author = {{Powell}, Diana and {Louden}, Tom and {Kreidberg}, Laura and {Zhang}, Xi and {Gao}, Peter and {Parmentier}, Vivien},
        title = "{Transit Signatures of Inhomogeneous Clouds on Hot Jupiters: Insights from Microphysical Cloud Modeling}",
      journal = {\apj},
         year = 2019,
        month = dec,
       volume = {887},
       number = {2},
          eid = {170},
        pages = {170},
          doi = {10.3847/1538-4357/ab55d9},
archivePrefix = {arXiv},
       eprint = {1910.07527},
 primaryClass = {astro-ph.EP},
       adsurl = {https://ui.adsabs.harvard.edu/abs/2019ApJ...887..170P}
}

@ARTICLE{Fortney2010,
       author = {{Fortney}, J.~J. and {Shabram}, M. and {Showman}, A.~P. and {Lian}, Y. and {Freedman}, R.~S. and {Marley}, M.~S. and {Lewis}, N.~K.},
        title = "{Transmission Spectra of Three-Dimensional Hot Jupiter Model Atmospheres}",
      journal = {\apj},
         year = 2010,
        month = feb,
       volume = {709},
       number = {2},
        pages = {1396-1406},
          doi = {10.1088/0004-637X/709/2/1396},
archivePrefix = {arXiv},
       eprint = {0912.2350},
 primaryClass = {astro-ph.EP},
       adsurl = {https://ui.adsabs.harvard.edu/abs/2010ApJ...709.1396F}
}

@ARTICLE{Prinoth2022,
       author = {{Prinoth}, Bibiana and {Hoeijmakers}, H. Jens and {Kitzmann}, Daniel and {Sandvik}, Elin and {Seidel}, Julia V. and {Lendl}, Monika and {Borsato}, Nicholas W. and {Thorsbro}, Brian and {Anderson}, David R. and {Barrado}, David and {Kravchenko}, Kateryna and {Allart}, Romain and {Bourrier}, Vincent and {Cegla}, Heather M. and {Ehrenreich}, David and {Fisher}, Chloe and {Lovis}, Christophe and {Guzm{\'a}n-Mesa}, Andrea and {Grimm}, Simon and {Hooton}, Matthew and {Morris}, Brett M. and {Oreshenko}, Maria and {Pino}, Lorenzo and {Heng}, Kevin},
        title = "{Titanium oxide and chemical inhomogeneity in the atmosphere of the exoplanet WASP-189 b}",
      journal = {Nature Astronomy},
         year = 2022,
        month = jan,
       volume = {6},
        pages = {449-457},
          doi = {10.1038/s41550-021-01581-z},
archivePrefix = {arXiv},
       eprint = {2111.12732},
 primaryClass = {astro-ph.EP},
       adsurl = {https://ui.adsabs.harvard.edu/abs/2022NatAs...6..449P}
}

@ARTICLE{molliere2019,
       author = {{Molli{\`e}re}, P. and {Wardenier}, J.~P. and {van Boekel}, R. and {Henning}, Th. and {Molaverdikhani}, K. and {Snellen}, I.~A.~G.},
        title = "{petitRADTRANS. A Python radiative transfer package for exoplanet characterization and retrieval}",
      journal = {\aap},
         year = 2019,
        month = jul,
       volume = {627},
          eid = {A67},
        pages = {A67},
          doi = {10.1051/0004-6361/201935470},
archivePrefix = {arXiv},
       eprint = {1904.11504},
 primaryClass = {astro-ph.EP},
       adsurl = {https://ui.adsabs.harvard.edu/abs/2019A&A...627A..67M}
}

@ARTICLE{rackham2018,
       author = {{Rackham}, Benjamin V. and {Apai}, D{\'a}niel and {Giampapa}, Mark S.},
        title = "{The Transit Light Source Effect: False Spectral Features and Incorrect Densities for M-dwarf Transiting Planets}",
      journal = {\apj},
         year = 2018,
        month = feb,
       volume = {853},
       number = {2},
          eid = {122},
        pages = {122},
          doi = {10.3847/1538-4357/aaa08c},
archivePrefix = {arXiv},
       eprint = {1711.05691},
 primaryClass = {astro-ph.EP},
       adsurl = {https://ui.adsabs.harvard.edu/abs/2018ApJ...853..122R}
}

@ARTICLE{Venot2014,
       author = {{Venot}, Olivia and {Ag{\'u}ndez}, Marcelino and {Selsis}, Franck and {Tessenyi}, Marcell and {Iro}, Nicolas},
        title = "{The atmospheric chemistry of the warm Neptune GJ 3470b: Influence of metallicity and temperature on the CH$_{4}$/CO ratio}",
      journal = {\aap},
         year = 2014,
        month = feb,
       volume = {562},
          eid = {A51},
        pages = {A51},
          doi = {10.1051/0004-6361/201322485},
archivePrefix = {arXiv},
       eprint = {1312.5163},
 primaryClass = {astro-ph.EP},
       adsurl = {https://ui.adsabs.harvard.edu/abs/2014A&A...562A..51V}
}

@ARTICLE{Marley2015,
       author = {{Marley}, M.~S. and {Robinson}, T.~D.},
        title = "{On the Cool Side: Modeling the Atmospheres of Brown Dwarfs and Giant Planets}",
      journal = {\araa},
         year = 2015,
        month = aug,
       volume = {53},
        pages = {279-323},
          doi = {10.1146/annurev-astro-082214-122522},
archivePrefix = {arXiv},
       eprint = {1410.6512},
 primaryClass = {astro-ph.EP},
       adsurl = {https://ui.adsabs.harvard.edu/abs/2015ARA&A..53..279M}
}

@ARTICLE{Visscher2011,
       author = {{Visscher}, Channon and {Moses}, Julianne I.},
        title = "{Quenching of Carbon Monoxide and Methane in the Atmospheres of Cool Brown Dwarfs and Hot Jupiters}",
      journal = {\apj},
         year = 2011,
        month = sep,
       volume = {738},
       number = {1},
          eid = {72},
        pages = {72},
          doi = {10.1088/0004-637X/738/1/72},
archivePrefix = {arXiv},
       eprint = {1106.3525},
 primaryClass = {astro-ph.EP},
       adsurl = {https://ui.adsabs.harvard.edu/abs/2011ApJ...738...72V}
}

@ARTICLE{2016MNRAS.460..855H,
       author = {{Helling}, Ch. and {Lee}, E. and {Dobbs-Dixon}, I. and {Mayne}, N. and {Amundsen}, D.~S. and {Khaimova}, J. and {Unger}, A.~A. and {Manners}, J. and {Acreman}, D. and {Smith}, C.},
        title = "{The mineral clouds on HD 209458b and HD 189733b}",
      journal = {\mnras},
         year = 2016,
        month = jul,
       volume = {460},
       number = {1},
        pages = {855-883},
          doi = {10.1093/mnras/stw662},
archivePrefix = {arXiv},
       eprint = {1603.04022},
 primaryClass = {astro-ph.EP},
       adsurl = {https://ui.adsabs.harvard.edu/abs/2016MNRAS.460..855H}
}

@ARTICLE{2023A&A...671A.122H,
       author = {{Helling}, Christiane and {Samra}, Dominic and {Lewis}, David and {Calder}, Robb and {Hirst}, Georgina and {Woitke}, Peter and {Baeyens}, Robin and {Carone}, Ludmila and {Herbort}, Oliver and {Chubb}, Katy L.},
        title = "{Exoplanet weather and climate regimes with clouds and thermal ionospheres. A model grid study in support of large-scale observational campaigns}",
      journal = {\aap},
         year = 2023,
        month = mar,
       volume = {671},
          eid = {A122},
        pages = {A122},
          doi = {10.1051/0004-6361/202243956},
archivePrefix = {arXiv},
       eprint = {2208.05562},
 primaryClass = {astro-ph.EP},
       adsurl = {https://ui.adsabs.harvard.edu/abs/2023A&A...671A.122H}
}
\bibliographystyle{aasjournalv7}

\end{document}